\documentclass[aps,twocolumn,letterpaper,superscriptaddress,pra,10pt]{revtex4-1}

\usepackage{physics}

\usepackage{CJK}
\usepackage{amsmath}
\usepackage{comment}
\usepackage{eufrak}
\usepackage{natbib}
\usepackage{hyperref}
\usepackage[left=0.75in, right=0.7in, top=0.7in, bottom=1in]{geometry}
\usepackage{xcolor}
\usepackage{graphicx}
\usepackage[font=small,labelfont=bf,justification=justified,format=plain]{caption}
\usepackage[font=small,labelfont=bf,justification=justified,format=plain]{subcaption}
\usepackage{placeins}

\begin{document}
\begin{CJK*}{GB}{}
\title{Frequency Dependent Magnetic Susceptibility and the \texorpdfstring{$q^2$}{q2} effective conductivity}
\author{Alistair H. Duff$^{1}$ , and J. E. Sipe}
\affiliation{Department of Physics, University of Toronto, Ontario M5S 1A7, Canada}

\begin{abstract} 
We apply a microscopic formalism for the calculation of material response properties to the problem of the generalization of a \textit{first-principles}, i.e based on the energy spectrum and geometric properties of the Bloch functions, derivation of the AC magnetic susceptibility. We find that the AC susceptibility forms only a part of the $q^2$ -- where $q$ is the wavevector of the applied field -- effective conductivity tensor, and many additional response tensors characterizing \textit{both} electric and magnetic multipole moments response to electromagnetic fields and their derivatives must be included to create the full gauge-invariant response. As was seen with the DC magnetic susceptibility and optical activity (characterized by the linear in \textit{q} contribution to the conductivity) one must be careful with the diagonal elements of the Berry connection. To our knowledge this is the only derivation of such a result general for crystalline insulators, with both `atomic like' contributions and `itinerant contributions' due to overlap of atomic orbitals and non-flat bands. Additionally, quantities familiar from quantum geometry like the Berry connection, curvature, and quantum metric appear extensively.  
\end{abstract}

\maketitle
\end{CJK*}

\section{Introduction}\label{Introduction}

The propagation of light in matter exhibits a rich range of phenomena even in the linear regime \cite{WootenOptical}, including linear and circular birefringence \cite{OpticalConductivityIvo,OpticalRotaryPower,JohnandKranendonk,OpticalActivityTheory,GyrotropicBirefringence}, and linear and circular dichroism. All of these can have a non-trivial frequency dependence. Very generally, we can describe the linear response of matter to radiation by relating the induced charge current density to the electromagnetic fields, which for a bulk medium can be written in the frequency and wavevector domain as  \cite{OpticalConductivityIvo, MultipoleTheoryOpticalActivity,PerryOptical} 
\begin{equation}
\label{CurrentDensity-qw}
    J^i(\textbf{q},\omega) = \sigma^{il}(\textbf{q},\omega) E^l(\textbf{q},\omega). 
\end{equation}
For media with weak spatial dispersion at the frequencies of interest, a power series expansion of the conductivity tensor $\sigma^{il}(\textbf{q},\omega)$ in orders of the wavevector $\textbf{q}$ of light can be employed,
\begin{equation}
\begin{split}
\label{CurrentDensity-Expansion}
    J^i(\textbf{q},\omega) = \sigma^{il}(\omega) E^l(\textbf{q},\omega) + \sigma^{ilj}(\omega) E^l(\textbf{q},\omega)q^j
    \\
    + \sigma^{iljk}(\omega) E^l(\textbf{q},\omega) q^j q^k + ...
\end{split}
\end{equation}
where the tensors that arise are related to the general $\textbf{q}$-dependent conductivity: $\sigma^{il}(\omega) = \sigma^{il}(0,\omega)$, $\sigma^{ilj}(\omega) = ( \partial \sigma^{il}(\textbf{q},\omega)/\partial q^j)_{\textbf{q}\rightarrow \textbf{0}}$, and so on. Various features of the optical response are associated with the contributions to the conductivity tensor at different orders of \textbf{q}: for example, the long-wavelength conductivity is related to $\sigma^{il}(\omega)$, which in turn determines the dominant contribution to the index of refraction; and \textit{natural} optical activity, whereby left- and right-circularly polarized light experience a different effective index of refraction in the absence of a magnetic field, is related to the tensor $\sigma^{ilj}(\omega)$ \cite{PerryOptical,MultipoleTheoryOpticalActivity,LandauElectrodynamics,ElectromagneticDispersiveMedia,GyrotropicBirefringence,DuffOpticalActivity}. Then, depending on the types of phenomena with which one is concerned, the series expansion can be truncated to capture the lowest order non-trivial contribution. For crystalline media, which are our focus of interest here, the contributions to the current due to each order of $q$ are generally a factor of $aq$ smaller than the contribution of the previous order, where $a$ is the lattice spacing of the material. Thus as $aq$ becomes significant we can expect these spatially dispersive effects to become relevant. In addition, symmetry constraints can cause the contributions to the conductivity at a given order of $q$ to vanish, necessarily requiring going to the next order to see any effects.

Within the framework of the macroscopic Maxwell equations in material media \cite{LorentzLectures,Fokker,Griffiths,MaxwellEquations,RosenfeldElectrons,JacksonElectrodynamics}, polarization fields \textbf{P} and magnetization fields \textbf{M} are traditionally introduced associated with the so-called bound charges in the medium. In addition, a free charge density $\varrho_\text{F}$ and current density $\textbf{J}_\text{F}$ can also be included. The macroscopic charge and current densities are then written as
\begin{equation}
\label{MacroJ}
\begin{split}
    \varrho(\textbf{x},t) &= -\nabla\cdot \textbf{P}(\textbf{x},t) + \varrho_\text{F}(\textbf{x},t),
    \\
    \textbf{J}(\textbf{x},t) &= \frac{\partial \textbf{P}(\textbf{x},t)}{\partial t} + c\nabla\times\textbf{M}(\textbf{x},t) + \textbf{J}_\text{F}(\textbf{x},t). 
\end{split}
\end{equation}
With equation (\ref{MacroJ}) in hand one can then relate the induced charge and current density to the induced multipole moments of the material \cite{MaxwellEquations}. Care must be taken when truncating the multipole expansions of the fields to ensure all contributions at a given order $q$ of the conductivity tensor are accounted for. In Section \ref{MultipolarContibutions} we outline how to do so. 

In a crystal there are complications to evaluating the multipole moment operators, since the position operator is unbounded in the Bloch function basis \cite{SeitzSolid}. The ``modern theory of polarization and magnetization" allows for the determination of the polarization instead from an adiabatic variation of the Hamiltonian, identifying a change in the polarization with the induced current, $\textbf{J} = d\textbf{P}/dt$ \cite{BerryVanderbilt,Macro_Polar,Elec_Polar,Beginner_Guide}. Additionally, the modern theory provides the insight that in a crystal there is both the ``atomic magnetization" associated with orbital currents identified with particular unit cells, and also an ``itinerant magnetization" -- even in an insulator -- due to the sites not being isolated from each other in a crystal \cite{BerryVanderbilt}. 

However, the ``modern theory" mainly treats adiabatic variations to the Hamiltonian. More recently, a ``microscopic" approach to defining polarization and magnetization fields in a crystal has been introduced \cite{Perry_Sipe,PerryOptical,PerryMagneto,PerryChern,PerryMetals,DuffOpticalActivity,DuffMagneticSusceptibility}. Here microscopic analogs of the fields of equation (\ref{MacroJ}) are introduced, and one can identify contributions to these fields associated with the lattice sites of the crystal. Moment expansions of the microscopic polarization and magnetization fields about the lattice sites can be performed, and the spatial average of these microscopic fields results in the macroscopic polarization $\textbf{P}(\textbf{x},t)$ and magnetization $\textbf{M}(\textbf{x},t)$ fields and their moment expansions. For a topologically trivial crystal in its ground state, the expressions for the polarization and magnetization agree with those of the ``modern theory." But with the microscopic theory the response to spatially varying and time dependent fields can also be considered. For a crystal treated in the independent particle approximation, and with the neglect of local field corrections, both $\sigma^{il}(\omega)$ and $\sigma^{ilj}(\omega)$ have been investigated, and expressions derived that are suitable for evaluation using the results of band structure calculations \cite{PerryOptical,DuffOpticalActivity}. 

The goal of this paper is a derivation of an expression for the response tensor $\sigma^{iljk}(\omega)$ characterizing topologically trivial crystals, treated in the independent particle approximation and with the neglect of local field corrections. One of the contributions to $\sigma^{iljk}(\omega)$ is the frequency dependent magnetic susceptibility, which has an interesting story in its own right. After all, the frequency dependent magnetic susceptibility might seem at first to describe a measurable response \cite{MagneticMeasurementDCI, MagneticMeasurementI, MagneticMeasurementII}. A first-principles expressions proves to be incredibly elusive, with the current literature appearing to treat the frequency dependence only phenomenologically \cite{MagneticMeasurementII,JonscherDielectricRelaxation,ColeCole,CasimirDuPres}. While the zero frequency magnetic susceptibility was found to be a gauge-invariant quantity \cite{DuffMagneticSusceptibility,RothMagSus} -- gauge-invariant in a sense that will become clear in Section \ref{MultipolarContibutions}, but for now this can be read as a proxy for a quantity being independent of the phases chosen for the Bloch functions used to make the calculation --  we find this is not so for its finite frequency extension. This is because the frequency dependent magnetic susceptibility does not describe a measureable quantity, it is only the \textit{total} magnetic field -- created by the magnetization current, the displacement current, and the free current -- that can be measured. Only if the latter two currents vanished or were negligible could one claim to be actually measuring the magnetization. Indeed at low frequencies this situation arises, but in general the total response requires considering the various tensors that arise at order $q^2$ in the response of the current density to the electric field, eq. (\ref{CurrentDensity-Expansion}), of which the magnetic susceptibility is only a part. And the individual constituents need not be gauge-invariant; only the full $\sigma^{iljk}(\omega)$ must be. 

Given that only the total conductivity corresponds to a ``physical quantity," one could ask why one would choose to introduce the macroscopic multipole moment fields at all, for they are individually gauge dependent. Why not just calculate the macroscopic charge and current densities directly to extract the response tensors from eq. (\ref{CurrentDensity-Expansion})? One practical reason is that in existing calculations that start from the minimal coupling Hamiltonian and compute the induced current density directly artificial divergences arise \cite{AversaSipe,NonlinearGhahramani,NonlinearOptics_VelocityGauge} . It is only with the identification of various sum rules that the divergences are shown to not be connected to any physical phenomena but instead from the need to truncate the basis used in calculations. Such a multipolar calculation does not exhibit these same non-physical divergences. Another reason is that a focus on the macroscopic multipole moment fields allows one to easily identify the ``molecular crystal limit," where the sites of the lattice are taken to be isolated; and with this in hand one can study the consequences of site-site dynamics in the optical response. A third reason is that the symmetry of the different multipole moment contributions can be easily identified. And a final reason we identify is the hope that the gauge dependence of the macroscopic moments can be utilized to choose local moments in a way that would simplify the inclusion of electron-electron interaction effects and the study of surface phenomena, effects that we neglect in this first treatment. 

The outline of this paper is as follows. In Section \ref{MultipolarContibutions} we introduce the macroscopic polarization and magnetization fields that are used to determine the charge and current densities in eq. (\ref{MacroJ}). We then introduce a collection of tensors that describe the response of the macroscopic polarization and magnetization fields to the Maxwell electromagnetic fields. Four of which are associated with the response to order $q$, characterized by $\sigma^{ilj}(\omega)$, these have been introduced before so we review the results from earlier work \cite{DuffMagneticSusceptibility,PerryOptical}. We then show how eight of these tensors can be combined to determine the response of the current to the Maxwell electromagnetic fields at order $q^2$. This effective conductivity tensor $\sigma^{iljk}(\omega)$ is divided into contributions associated with coupling to derivatives of the magnetic field $\sigma^{ilj}_L(\omega)$, and to double derivatives of the electric field $\sigma^{iljk}_K(\omega)$. These two quantities are independently gauge-invariant, and can be understood as identifying the magnetic and electric components of the $q^2$ response respectively. 

The macroscopic polarization and magnetization fields are constructed from the spatial average of microscopic polarization and magnetization fields that are themselves decomposed into site contributions in Section \ref{MicroMacro}. This site decomposition allows for a multipole expansion of the microscopic and hence macroscopic fields. We then demonstrate how the collection of macroscopic response tensors introduced in Section \ref{MultipolarContibutions} follows from a linear perturbative expansion of the microscopic fields. 

In Section \ref{sec:SPDM} we introduce the single particle density matrix (SPDM) expanded in the adjusted Wannier function basis; the details of such a basis are given in Appendix \ref{AppendixA}. Site quantities can then be expressed as the trace of a matrix multiplication of the SPDM and a ``site quantity matrix element," and in considering the linear response of a site quantity one needs to consider how both the SPDM and site matrix elements depend on the fields and their derivatives. 

In Section \ref{Sec:MultipoleMoments} we introduce expressions for the unperturbed electric dipole, quadrupole, and octupole, as well as the magnetic dipole and quadrupole, as single integrals over the BZ. These quantities depend on the Berry connection and band energies and their derivatives. We then introduce general $k$-dependent matrix elements for these quantities that will appear in the linear response such that the trace over filled states integrated over the Brillouin zone produces the unperturbed values. 

In Section \ref{Sec:ResponseTensors} we give the functional form of all eight response tensors outlined in Section \ref{MultipolarContibutions} that are used to determine the $q^2$ contribution to the effective conductivity tensor, and discuss the various contributions and symmetries of the tensors.

In Section \ref{sec:q2conductivity} we combine the multipole response tensors and demonstrate that the total induced current described by the tensors $\sigma^{iljk}(\omega)$, $\sigma^{ilj}_L(\omega)$, and $\sigma^{iljk}_K(\omega)$ are gauge-invariant within the limits we take. We point out that one could use the gauge dependent response tensors as they are, or one could construct and use explicitly gauge-invariant versions; the latter requires adding extra terms that are dependent on the quantum metric and the Berry curvature, but it is the approach we take. This is because the gauge-invariant versions are potentially more amenable for numerical evaluations, for only off-diagonal elements of the Berry connection are required, and they can be related to the velocity matrix elements.  

In Section \ref{sec:HaldaneModel} we apply the formulas we derive to the Haldane model to determine the frequency dependent magnetic susceptibility. Lastly, in Section \ref{sec:Conclusion} we conclude. 

\section{Multipolar contributions to the current density}\label{MultipolarContibutions}

The macroscopic polarization and magnetization fields can be expanded in terms of the multipole contributions,
\begin{equation}
\label{MacroscopicFields}
\begin{split}
    &P^i(\textbf{x},t) = \mathcal{P}^i(\textbf{x},t) - \frac{\partial \mathcal{Q}^{ij}_\mathcal{P}(\textbf{x},t)}{\partial x^j} + \frac{\partial^2 \mathcal{O}^{ijl}_\mathcal{P}(\textbf{x},t)}{\partial x^j \partial x^l} + ... 
    \\
    &M^i(\textbf{x},t) = \mathcal{M}^i(\textbf{x},t) - \frac{\partial \mathcal{Q}^{ij}_\mathcal{M}(\textbf{x},t)}{\partial x^j} + ... 
\end{split}
\end{equation}
where the uppercase script letters indicate macroscopic quantities: $\mathcal{P}^i(\textbf{x},t)$ is the electric dipole moment per unit volume, $\mathcal{Q}^{ij}_\mathcal{P}(\textbf{x},t)$ the electric quadrupole moment per unit volume, $\mathcal{O}^{ijl}_\mathcal{P}(\textbf{x},t)$ the electric octupole moment per unit volume, $\mathcal{M}^i(\textbf{x},t)$ the magnetic dipole moment per unit volume, and $\mathcal{Q}^{ij}_\mathcal{M}(\textbf{x},t)$ the magnetic quadrupole moment per unit volume \cite{MaxwellEquations}. The polarization and magnetization fields are spatial averages of corresponding microscopic quantities, which are themselves constructed from the microscopic charge and current densities using certain ``relators" \cite{HealyQuantum}. The particular expansion above results when the relators employ ``straight line paths" \cite{PerryMagneto,PerryOptical}, which we discuss further in Section \ref{MicroMacro}. Lastly, `...' indicates higher order electric and magnetic moments that do not appear when considering up to the $q^2$ contribution to the conductivity tensor. Table \ref{tab:1} provides a list of the different tensors we introduce that characterize the response of the different multipoles to the electric field and its derivatives. The tensors are grouped according to the order of $q$ at which they appear in eq. (\ref{CurrentDensity-Expansion}). Expressions for the induced multipole tensors that appear at order $q^2$ are given in Section \ref{Sec:ResponseTensors}. 

When grouping the multipole response contributions it is a matter of tracking the total number of spatial derivatives that appear. For example, the induced current depends on spatial derivatives of the magnetization field, therefore any response associated with the magnetization carries with it at least one spatial derivative. The induced current also depends on the time derivative of the polarization, therefore to obtain spatial derivatives another way is to consider the higher moments like the quadrupolarization, see eq. (\ref{MacroscopicFields}). The next way to obtain spatial derivatives is by directly coupling one of the multipole moments to a derivative of the electric field. Additionally, any multipole moment response to the magnetic field implicitly carries a spatial derivative since we can relate the time-derivative of the magnetic field to the curl of the electric field via Faraday's law,
\begin{equation}
\label{FaradaysLaw}
    \nabla\times\textbf{E}(\textbf{x},t) = -\frac{1}{c} \frac{\partial \textbf{B}(\textbf{x},t)}{\partial t}. 
\end{equation}
Then we see that the response of the magnetization to the magnetic field carries a double spatial derivative -- in other words it is part of the $q^2$ response -- since it picks up one derivative from how the magnetization is associated with the induced current, and another from describing the response to the magnetic field. 

\begin{table}
    \centering
    \begin{tabular}{|c|c|c|}
         \hline
         Multipole Field  & Tensor & EM Field 
         \\
         \hline \hline 
         $\mathcal{P}^i$ & $\chi^{il}_\mathcal{P}$ & $E^l$ 
         \\
         \hline \hline
         $\mathcal{P}^i$  & $\alpha^{il}_\mathcal{P}$ & $B^l$
         \\
         $\mathcal{P}^i$ & $\gamma_\mathcal{P}^{ijl}$ & $F^{jl}$
         \\
         $\mathcal{Q}_\mathcal{P}^{ij}$ & $\chi^{ijl}_\mathcal{Q}$  & $E^l$
         \\
         $\mathcal{M}^{i}$ & $\alpha^{li}_\mathcal{M}$ & $E^{l}$
         \\
         \hline \hline 
         $\mathcal{P}^{i}$ & $\Lambda^{ijl}$ & $L^{jl}$
         \\
         $\mathcal{P}^{i}$ & $\Pi^{ijlk}$ & $K^{jlk}$ 
         \\
         $\mathcal{Q}^{ij}_\mathcal{P}$ & $\Gamma^{ijl}$ & $B^l$
         \\
         $\mathcal{Q}^{ij}_\mathcal{P}$ & $\Sigma^{ijlk}$  & $F^{lk}$
         \\
         $\mathcal{M}^{i}$ & $\chi^{il}_\mathcal{M}$ & $B^{l}$
         \\
         $\mathcal{M}^{i}$  & $\gamma^{ijl}_\mathcal{M}$ & $F^{jl}$
         \\
         $\mathcal{Q}^{ij}_\mathcal{M}$  & $\beta^{ijl}_\mathcal{M}$ & $E^l$
         \\
         $\mathcal{O}^{ijl}_\mathcal{P}$  & $\Omega^{ijlk}$ & $E^{k}$
         \\
         \hline
    \end{tabular}
    \caption{List of response tensors that describe the multipole moments induced by the applied electromagnetic field. Note that the multipoles, tensors, and fields and their derivatives have different rank. \textbf{F} is a shorthand notation for the symmetric derivatives of the electric field introduced in equation (\ref{F-field}), \textbf{K} is a shorthand notation for the symmetric double derivatives of the electric field introduced in equation (\ref{K-field}), and \textbf{L} are the derivatives of the magnetic field introduced in equation (\ref{L-field}). The three different boxes group the tensors based on at what order of $q$ they appear in describing the total conductivity.}
    \label{tab:1}
\end{table}

For the topologically trivial insulators that we consider here, there is no free current response at linear order in the electromagnetic field amplitude, and the usual long wavelength response of the induced current, described by $\sigma^{il}(\omega)$, is due to the oscillation of induced dipole moments in the material \cite{AversaSipe}. This is simply

\begin{equation}
\label{conductivity0}
\begin{split}
    \sigma^{il}(\omega) = -i\omega \chi^{il}_\mathcal{P}(\omega),
\end{split}
\end{equation}
where $\chi^{il}_\mathcal{P}(\omega)$ is the usual electric susceptibility.

\subsection{Contributions to order \texorpdfstring{$q$}{q}} 

To describe spatially dispersive effects requires considering additional multipole moments beyond the electric dipole. Here we review the contributions to the conductivity tensor $\sigma^{il}(\textbf{q},\omega)$ that arise at $\mathcal{O}(q)$, and have been identified in previous work \cite{PerryOptical,DuffOpticalActivity}. These include the response of the electric dipole moment to the magnetic field, and the response of the magnetic dipole moment to the electric field.
The response of the Fourier components $\textbf{M}(\omega)$ and $\textbf{P}(\omega)$ to $\textbf{E}(\omega)$ and $\textbf{B}(\omega)$ respectively are given by
\begin{equation}
\label{MPS}
\begin{split}
    \alpha^{li}_{\mathcal{M}}(\omega) &\equiv \frac{\partial M^i}{\partial E^l},
    \\
    \alpha^{il}_\mathcal{P}(\omega) &\equiv \frac{\partial P^i}{\partial B^l},
\end{split}
\end{equation}
where $\alpha^{il}_{\mathcal{M}}(\omega)$ and $\alpha^{il}_\mathcal{P}(\omega)$ are not in general equal at an arbitrary frequency \cite{DuffOpticalActivity,PerryOptical,VanderbiltOMP,OpticalConductivityIvo}.  

While there are other contributions at $\mathcal{O}(q)$ to $\sigma^{il}(\textbf{q},\omega)$, we first focus on these two because at zero frequency they are the only ones that survive, and in fact are characterized by a single magnetoelectric polarizability tensor \cite{BerryVanderbilt,OMP1,OMP2,OMP3,VanderbiltOMP}
\begin{equation}\label{MP0}
    \alpha^{il}_{\mathcal{M}}(0) = \alpha^{il}_\mathcal{P}(0) \equiv \alpha^{il},
\end{equation}
which describes the so-called ``magnetoelectric effect" (ME); for it to be non-vanishing both time-reversal and inversion symmetry must be broken. The magnetoelectric polarizability tensor $\alpha^{il}$ can be divided into three contributions \cite{DuffOpticalActivity}: a spin dependent contribution and two ``orbital parts," termed the ``Chern-Simons contribution" and the ``cross-gap contribution." The Chern-Simons contribution depends only on the occupied subspace and is isotropic \cite{PerryMagneto}. 

The natures of the two orbital parts are linked to the gauge freedom associated with the Bloch bundle. One can introduce at first a simple and then more complicated version of this gauge freedom. First, there is the simple and familiar $\textbf{k}$-dependent phase indeterminacy of the Bloch function solutions, whereby the multiplication of a complex phase does not alter the physical observables such as the charge and current densities. Second, more generally one can perform a so-called ``multiband gauge-transformation of the Bloch bundle" -- of which the phase rotations are only a subset. The multiband gauge-transformations are employed, for example, to create maximally localized Wannier functions \cite{MaxLocWannier,ExponentialLocWannier}. Here one forms new superpositions of the filled Bloch states at each \textbf{k}; they are typically not energy eigenstates. While the Chern-Simons term is \textit{not} sensitive to simple phase rotations, it is sensitive to the larger class of multiband gauge-transformations of the Bloch bundle \cite{BerryVanderbilt,PerryMagneto}. On the other hand, the cross-gap contribution depends on matrix elements that connect the occupied and unoccupied bands, is in general not isotropic, and is gauge-invariant in the larger multiband sense.  

There are other quantities of interest that are either sensitive or insensitive to the different versions of this gauge freedom. For example, the Abelian Berry curvature is insensitive to the simple version of this gauge freedom, and thus is sometimes said to be gauge-invariant \cite{BerryVanderbilt}. However, the diagonal elements of the Berry connection have a sensitivity to it. The ground state polarization of an infinite crystal inherits a ``quantum of ambiguity" from its dependence on the diagonal elements of the Berry connection, and it can be linked to how one chooses the unit cell \cite{BerryVanderbilt}. 

As we move beyond considering just the static response this whole perspective must be expanded. At finite frequency, only inversion symmetry need be broken for magnetoelectric effects to appear, the two response tensors of eq. (\ref{MPS}) need not be equal, and additional tensors must be considered to describe the response due to other multipoles: There is the response of the electric dipole moment to symmetric derivatives of the electric field, and the response of the electric quadrupole moment to the electric field \cite{PerryOptical,JohnandKranendonk}. In all, the contributions to the conductivity at $\mathcal{O}(q)$ can be written as the combination of the aforementioned response tensors (shown in the second box of Table \ref{tab:1}) 
\begin{equation}
\label{sigmaq}
\begin{split}
    \sigma^{ilj}(\omega) = 
    \omega \gamma^{ijl}_{\mathcal{P}}(\omega) - \omega \chi^{ijl}_\mathcal{Q}(\omega) 
    \\
    -ic \alpha^{ia}_\mathcal{P}(\omega) \epsilon^{ajl} + ic \epsilon^{ijb} \alpha^{lb}_{\mathcal{M}}(\omega).
\end{split}
\end{equation}

Upon closer inspection of the second line of equation (\ref{sigmaq}) it is clear that the contributions to $\alpha^{il}_\mathcal{P}(\omega)$ and $\alpha^{il}_\mathcal{M}(\omega)$ that are frequency independent and isotropic -- which is exactly how the Chern-Simons contribution enters -- make no contribution to $\sigma^{ilj}(\omega)$ \cite{PerryOptical}. This is expected since any bulk observable property must be gauge-invariant, and since the Chern-Simons contribution is not gauge-invariant, it should not -- and does not -- contribute to the bulk induced current \cite{PerryMagneto}. The two additional response tensors are defined as follows: $\gamma^{ijl}_\mathcal{P}(\omega)$ is the response of the electric dipole to the symmetric derivatives of the electric field; and $\chi^{ijl}_\mathcal{Q}(\omega)$ is the response of the electric quadrupole to the electric field. For their functional form the reader is directed to earlier work \cite{PerryOptical}, as well as to Appendix \ref{AppendixE}. The symmetric derivatives $F^{jl}(\textbf{x},\omega)$ of the electric field are defined as 
\begin{equation}
\label{F-field}
\begin{split}
    F^{jl}(\textbf{x},\omega) = \frac{1}{2} \Big( \frac{\partial E^j(\textbf{x},\omega)}{\partial x^l} + \frac{\partial E^l(\textbf{x},\omega)}{\partial x^j}\Big).
\end{split}
\end{equation}
At finite frequency the $\mathcal{O}(q)$ contributions to the conductivity tensor can be non zero in crystals that do not break time-reversal symmetry if the point group of the crystal is chiral \cite{MultipoleTheoryOpticalActivity}; then $\sigma^{ilj}(\omega)$ describes natural optical activity, where the plane of polarization of light rotates as the light propagates, due to the difference in the phase velocities of left and right circularly polarized light, and circular dichroism if the frequency of light is above the band gap, where there is a difference in the absorption of the different helicities of light \cite{OpticalActivityTheory}. 

\subsection{Presuppositions}\label{sec:Presuppositions}

The validity of the calculations leading to eq. (\ref{sigmaq}), and the identification of the expressions for the different terms that contribute, are based on three presuppositions. The first is the presupposition that the insulator is \textit{topologically trivial}; a consequence of this used in the calculation is that localized ``valence" Wannier functions can be constructed from the valence bands, and that localized ``conduction" Wannier functions can be constructed from the conduction bands \cite{WannierFunc_Insulators}. A second is that there is no difficulty in constructing Bloch functions over the Brillouin zone that vary sufficiently smoothly such that the partial integrations required in performing the kind of integrals necessary to reduce the expressions to terms amenable to band structure calculations (see Appendix \ref{AppendixD}) are not problematic; we refer to this as the \textit{smoothness} presupposition. We note that work on Chern insulators indicates that even there this would be valid \cite{PerryChern,FreedMath}. The third presupposition is that demonstrating that our expression for eq. (\ref{sigmaq}) is invariant to changes of the phase of the Bloch functions -- which can be done by demonstrating that no diagonal components of the Berry connection contribute to the total expression -- is sufficient to guarantee that the expression can be taken to be gauge-invariant in the larger sense defined above. We refer to this as the \textit{sufficiency} presupposition. In fact, the larger multiband gauge invariance was explicitly proven as well in earlier work \cite{PerryOptical,DuffOpticalActivity}, but we will only prove the former in the following sections.

\subsection{Contributions to order \texorpdfstring{$q^2$}{q2}}

When inversion symmetry is unbroken the first non-zero spatially dispersive contribution to the conductivity tensor $\sigma^{il}(\textbf{q},\omega)$ requires going beyond the linear in $q$ contribution identified by $\sigma^{ilj}(\omega)$. It is then helpful to consider the two ways in which the electromagnetic fields enter the second line of equation (\ref{CurrentDensity-Expansion}): directly as a double derivative of the electric field; or through the derivatives of the magnetic field. In the multipole expansion of the current we find that there are a total of eight contributions to the $\mathcal{O}(q^2)$ conductivity tensor, $\sigma^{iljk}(\omega)$. They are: the response of the electric dipole to derivatives of the magnetic field $\Lambda^{ijl}(\omega)$ and to the symmetric double derivatives of the electric field $\Pi^{ijlk}(\omega)$; the response of the electric quadrupole to the magnetic field $\Gamma^{ijl}(\omega)$ and to symmetric derivatives of the electric field $\Sigma^{ijlk}(\omega)$; the response of the electric octupole to the electric field $\Omega^{ijlk}(\omega)$; the response of the magnetic dipole to the magnetic field $\chi^{il}_\mathcal{M}(\omega)$ and to the symmetric derivatives of the electric field $\gamma^{ijl}_\mathcal{M}(\omega)$; and lastly the response of the magnetic quadrupole to the electric field $\beta^{ijl}_\mathcal{M}(\omega)$. These tensors, by definition, have various symmetries one would expect based on the associated multipole moments, discussed further in Section \ref{Sec:ResponseTensors}. They are not necessarily gauge-invariant individually, since they do not alone describe a measureable quantity like the current or charge density, but their combination, which describes spatially dispersive effects, must be. 

Over-viewing the collection of these terms, we see that this multipole decomposition of the contributions to the induced current requires one tensor in the long wavelength limit (see eq. (\ref{conductivity0})), and $4n$ tensors for $\mathcal{O}(q^n)$ for $n>1$. At each subsequent order an additional electric and magnetic multipole moment coupled to the electric field must be added; the electric dipole moment couples to the new higher derivatives of the electric and magnetic fields; and the other multipole moments that were present at the previous order of $q$ now couple to one higher derivative of the electromagnetic fields. A schematic of the couplings for $\mathcal{O}(q^n)$ from $n=0$ to $n=2$ is shown in Figure \ref{fig:Oq2}. 

Working through this scheme, we first convert eq. (\ref{CurrentDensity-Expansion}) to the real space and frequency domain
\begin{widetext}

\begin{equation}
\begin{split}
    J^{i(1)}(\textbf{x},\omega) = \sigma^{il}(\omega) E^l(\textbf{x},\omega) - i\sigma^{ilj}(\omega) \frac{\partial E^l(\textbf{x},\omega)}{\partial x^j} - \sigma^{iljk}(\omega) \frac{\partial^2 E^l(\textbf{x},\omega)}{\partial x^j \partial x^k} + ...,
\end{split}
\end{equation}
and identify a natural partitioning of $\sigma^{iljk}(\omega)$ into the response associated with the derivatives of the magnetic field and that associated with symmetric double derivatives of the electric field, 

\begin{equation}
\begin{split}
    \sigma^{iljk}(\omega) \frac{\partial^2 E^l(\textbf{x},\omega)}{\partial x^j \partial x^k} = \sigma^{iljk}_{K}(\omega) K^{ljk}(\textbf{x},\omega) + \sigma^{ilj}_L(\omega) L^{jl}(\textbf{x},\omega),
\end{split}
\end{equation}
where
\begin{equation}
\begin{split}\label{K-field}
    K^{jlk}(\textbf{x},\omega) = \frac{1}{3}\Big(\frac{\partial^2 E^j(\textbf{x},\omega)}{\partial x^l \partial x^k} + \frac{\partial^2 E^l(\textbf{x},\omega)}{\partial x^j \partial x^k} + \frac{\partial^2 E^k(\textbf{x},\omega)}{\partial x^l \partial x^j} \Big),
\end{split}
\end{equation}
and
\begin{equation}
\begin{split}\label{L-field}
    L^{jl}(\textbf{x},\omega) = \frac{\partial B^l(\textbf{x},\omega)}{\partial x^j}.
\end{split} 
\end{equation}

\begin{figure}[h]
\includegraphics[width=0.85\textwidth]{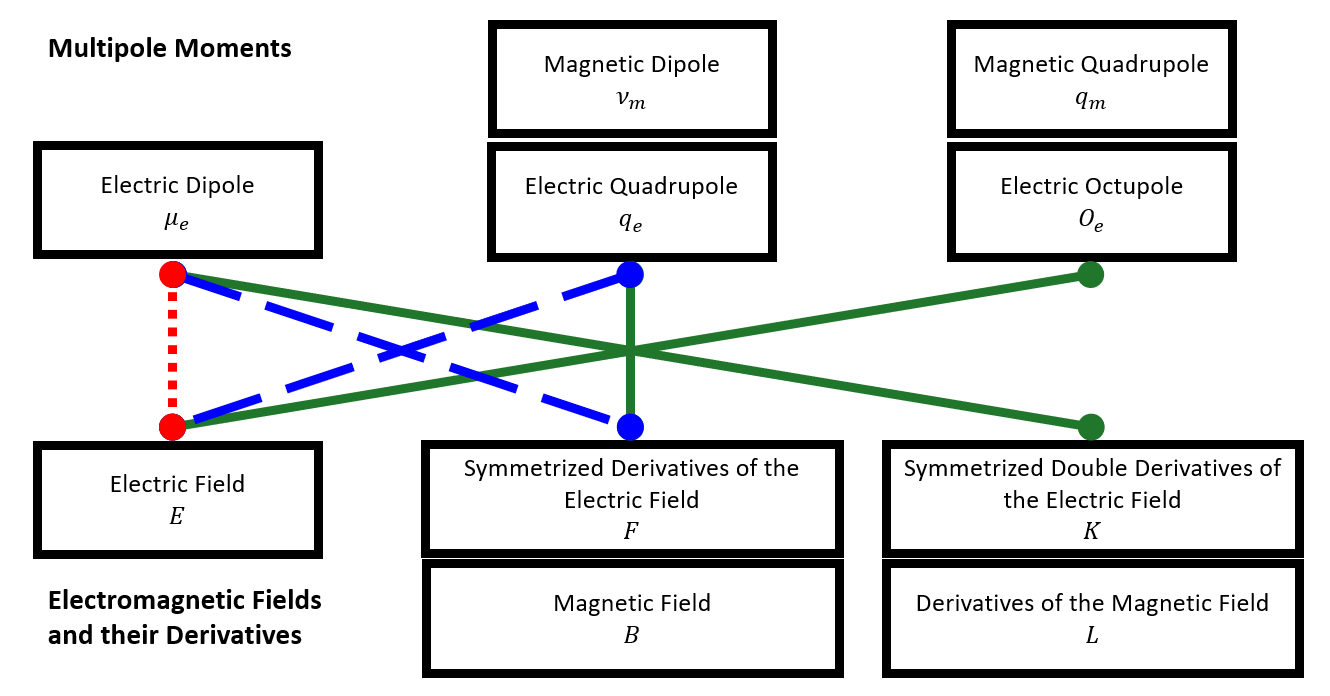}
\caption{ In the long wavelength limit for a topologically trivial insulator it is only the electric dipole coupling to the electric field, indicated by a dotted red line, that is required to determine the conductivity tensor. To consider optical activity, which is described by the tensor $\sigma^{ilj}(\omega)$, requires including the magnetic dipole moment and the electric quadrupole moment coupling to the electric field. Additionally we must consider the associated fields, the magnetic field and symmetrized derivatives of the electric field coupled to the electric dipole moment, see equation (\ref{sigmaq}). These 4 couplings are indicated by the dashed blue lines. Going beyond the regularly considered optical activity that includes only the linear in $q$ response requires adding the next order of multipole moments, the magnetic quadrupole and electric octupole, as well as higher derivatives of the fields, e.g. the double derivatives of the electric field and single derivatives of the magnetic field. The pattern continues, shifting all the preceding multipole couplings seen  to the next set of electromagnetic field derivatives, and coupling the new multipole moments to the electric field. These eight couplings are indicated by the three solid green lines.}
\label{fig:Oq2}
\end{figure}

We will see that the terms $\sigma^{ilj}_L(\omega)$ and $\sigma^{iljk}_K(\omega)$ are well behaved in the limit $\omega\rightarrow 0$, and are individually gauge-invariant. The latter can be expected because electric and magnetic fields satisfying Faraday's law and the vanishing of the divergence of the magnetic field can be constructed so that $L^{jl}(\textbf{x},\omega) = 0$ and $K^{ljk}(\textbf{x},\omega) \neq 0$, and alternately so that $L^{jl}(\textbf{x},\omega) \neq 0$ and $K^{ljk}(\textbf{x},\omega) = 0$. Thus $L^{jl}(\textbf{x},\omega)$ and $K^{ljk}(\textbf{x},\omega)$ lead to the magnetic and electric contributions to the $q^2$ response, and $\sigma^{ilj}_L(\omega)$ and $\sigma^{iljk}_K(\omega)$ respectively are the response coefficients that identify them. With the definitions above and Faraday's law, we can write

\begin{equation}
\label{eq:q2conductivity}
    \sigma^{iljk}(\omega) = \sigma^{iljk}_K(\omega)  + \frac{ic}{\omega}\epsilon^{akl}\sigma^{iaj}_{L}(\omega). 
\end{equation}

Using manipulations outlined in Appendix \ref{AppendixB} we can combine the eight multipole response tensors mentioned above to obtain the expressions

\begin{equation}\label{q2conductivityL}
\begin{split}
    \sigma^{ilj}_L(\omega) = -c\epsilon^{ijk}\Big( \chi^{kl}_{\mathcal{M}}(\omega) - \frac{i\omega}{2c} \epsilon^{lab} \beta^{kab}_{\mathcal{M}}(\omega) \Big) + \epsilon^{iak} \epsilon^{lab} \frac{i\omega}{3} \Big( \frac{1}{2} \Big( \beta^{kjb}_{\mathcal{M}}(\omega) + \beta^{kbj}_{\mathcal{M}}(\omega)\Big)-\gamma^{kjb}_{\mathcal{M}}(\omega) \Big) 
    \\
    +i\omega \Big(\Lambda^{ijl}(\omega)-\Gamma^{ijl}(\omega) \Big) 
    - \frac{\omega^2}{3c} \epsilon^{lab} \Big( \Sigma^{ibaj}(\omega) + 2\Omega^{ijab}(\omega) \Big) ,
\end{split}    
\end{equation}
and 
\begin{equation}\label{q2conductivityK}
\begin{split}
    \sigma^{iljk}_K(\omega) = \frac{1}{6}\sum_{\{jlk\} }\Bigg[c\epsilon^{ijn}\Big( \beta_{\mathcal{M}}^{nlk}(\omega) - \gamma^{nlk}_{\mathcal{M}}(\omega)  \Big)
    +i\omega \Big(\Pi^{ijlk}(\omega)+\Omega^{ijlk}(\omega)-\Sigma^{ijlk}(\omega) \Big)\Bigg], 
\end{split}    
\end{equation}
where $\sum_{ \{jlk\} }$ indicates a summation over all $3!=6$ permutations of the Cartesian indices $j$, $l$, and $k$. This can be done with impunity to construct a $\sigma^{iljk}_{K}(\omega)$ that is symmetric in the last three indices, since the symmetry of $K^{ljk}(\omega)$ under exchange of any of the Cartesian indices guarantees that any part of $\sigma^{iljk}_K(\omega)$ that is antisymmetric upon exchange of any of the last three indices will not contribute to the induced current. 

\end{widetext}

Turning to the special limit of zero frequency, we find that only the magnetization tensors $\chi^{il}_\mathcal{M}(\omega)$, $\gamma^{ijl}_\mathcal{M}(\omega)$, and $\beta^{ijl}_\mathcal{M}(\omega)$ contribute to the total induced current and determine $\sigma^{ilj}_L(0)$ and $\sigma^{ijlk}_K(0)$. The only contribution that remains for $\sigma^{ilj}_L(0)$ is due to the familiar magnetic susceptibility tensor, $\chi^{il}_\mathcal{M}(0)$ \cite{DuffMagneticSusceptibility}. This result, therefore, must be gauge-invariant -- and indeed it is -- as it describes a physically observable response. If one instead considers the response to only symmetric double derivatives of the electric field, still at zero frequency, the only contributions are due to $\gamma_\mathcal{M}$ and $\beta_{\mathcal{M}}$. Thus $\beta^{ijl}_\mathcal{M}(0) - \gamma_\mathcal{M}^{ijl}(0)$ must be gauge-invariant; see eq. (\ref{q2conductivityK}). 

As we have shown in earlier work \cite{DuffOpticalActivity,DuffMagneticSusceptibility}, to put expressions such as those for $\sigma^{ilj}_L(\omega)$ and $\sigma^{iljk}_K(\omega)$ in a computationally convenient form -- such that all the contributions appearing can be written in terms of the velocity matrix elements, the gauge-covariant non-Abelian Berry curvature, and the quantum metric -- is not trivial. This stems from how seemingly problematic integrands can, upon using sum rules and integration by parts, be written as gauge-invariant ``geometric" quantities after a certain amount of ``repackaging;" this involves the appearance of additional terms. Some of the confusion surrounding the earlier studies of the DC magnetic susceptibility is due to this difficulty \cite{BlountMagSus,HebbornSondheimerEarlyDiamag,RothMagSus,GaoGeometricalSus,OgataMagSusI,OgataMagSusII,OgataMagSusIII,OgataMagSus2017,FukuyamaBismuth,FukuyamaEarlyWork,PiechonMagSus}; in comparing with other expressions in the literature it can be necessary to use sum rules, integration by parts, and careful treatments of the diagonal elements of the Berry connection. To move to computationally convenient forms of $\sigma^{ilj}_L(\omega)$ and $\sigma^{iljk}_K(\omega)$ we can do such ``repackaging" as well. In section \ref{sec:q2conductivity} we do this, and give the explicit form of the additional terms that arise. 

\section{Microscopic to Macroscopic Fields}\label{MicroMacro}

For a detailed description of the microscopic theory we employ, the interested reader is directed to earlier work \cite{Perry_Sipe,PerryMagneto,PerryOptical,DuffOpticalActivity,DuffMagneticSusceptibility}. The essence is that we do not want to directly define the various induced multipole moments outlined in Section \ref{MultipolarContibutions} in the Bloch basis, since expectation values of the position operator are ill-defined \cite{SeitzSolid}. While the Bloch energy eigenfunctions $\psi_{n\textbf{k}}(\textbf{x}) \equiv \langle \textbf{x}|\psi_{n\textbf{k}}\rangle$ with associated energy bands $E_{n\textbf{k}}$ are a natural basis for a crystal, where $n$ labels a band index and $\hbar \textbf{k}$ the crystal momentum, one can equally define a set of exponentially localized Wannier functions (ELWFs) to serve as a basis for computations \cite{ExponentialLocWannier}. For topologically trivial insulators one can construct a set of filled (empty) Wannier functions from the set of filled (empty) bands. In general this band mixing is described by a unitary matrix U(\textbf{k}) with components $U_{n\alpha}(\textbf{k})$ that relate the Bloch and Wannier function bases $(W_{\alpha\textbf{R}}(\textbf{x}) \equiv \langle \textbf{x}| \alpha\textbf{R}\rangle)$ via a Fourier transform

\begin{equation}
\label{Wannier}
\begin{split}
    W_{\alpha\textbf{R}}(\textbf{x}) = \sqrt{\frac{\mathcal{V}_{uc}}{(2\pi)^3}} \int_{BZ} d\textbf{k} e^{-i\textbf{k}\cdot\textbf{R}} \sum_{n} U_{n\alpha}(\textbf{k}) \psi_{n\textbf{k}}(\textbf{x}),
\end{split}
\end{equation}
where $\alpha$ labels an orbital type and \textbf{R} the associated lattice site, and $\mathcal{V}_{uc}$ is the volume of the unit cell. 

The unitary matrix U(\textbf{k}) and the Bloch eigenvectors $|\psi_{n\textbf{k}}\rangle$ are chosen to be periodic over the first Brillouin zone. Associated with the $|\psi_{n\textbf{k}}\rangle$ are cell-periodic Bloch functions $u_{n\textbf{k}}(\textbf{x}) \equiv \langle \textbf{x}|n\textbf{k}\rangle$. Since we consider topologically trivial insulators, $U_{n\alpha}(\textbf{k})$ is only non-zero when $n$ and $\alpha$ both label occupied (or unoccupied) bands or Wannier functions respectively. The orthogonality conditions on the above set of functions are as follows: $\langle \psi_{m\textbf{k}'}|\psi_{n\textbf{k}}\rangle = \delta_{nm} \delta(\textbf{k}-\textbf{k}')$, $\langle \alpha\textbf{R}|\beta\textbf{R}'\rangle = \delta_{\alpha\beta} \delta_{\textbf{R}\textbf{R}'}$, and $(m\textbf{k}|n\textbf{k}) = \delta_{nm}$; the cell periodic functions are only integrated over a single unit cell due to their periodicity, so we have introduced the notation
\begin{equation}
    (g|h) \equiv \frac{1}{\mathcal{V}_{uc}} \int_{\mathcal{V}_{uc}} g^*(\textbf{x}) h(\textbf{x}) d\textbf{x}.
\end{equation}
Likewise, one can introduce a set of cell-periodic functions associated with the Wannier basis $u_{\alpha\textbf{k}}(\textbf{x}) = \sum_{n} U_{n\alpha} u_{n\textbf{k}}(\textbf{x})$. 

We then begin a calculation by writing the microscopic analogues of the fields in eq. (\ref{MacroJ}) in terms of operators on the equal time lesser electronic single particle Green function \cite{DuffOpticalActivity}
\begin{equation}
    G(\textbf{x},\textbf{y};t) = i\langle \hat{\psi}^\dag(\textbf{y},t) \hat{\psi}(\textbf{x},t) \rangle,
\end{equation}
where $\hat{\psi}(\textbf{x},t)$ ($\hat{\psi}^\dag(\textbf{x},t)$) is the electron annihilation (creation) operator, and the angled brackets $\langle ... \rangle$ indicate the ground state expectation value. To decompose the microscopic quantities into contributions associated with each site \textbf{R} we expand the fermionic field operators in an ``adjusted Wannier functions" basis defined in Appendix \ref{AppendixA}. The expansion coefficients are then just the adjusted Wannier functions $\hat{\psi}(\textbf{x},t) = \sum_{\alpha\textbf{R}} \hat{a}_{\alpha\textbf{R}}(t) \bar{W}_{\alpha\textbf{R}}(\textbf{x},t)$ and $\hat{a}_{\alpha\textbf{R}}(t)$ $(\hat{a}^\dag_{\alpha\textbf{R}}(t)$) the associated fermionic annihilation (creation) operator of an electron in a state described by an ``orbital type" $\alpha$ associated with a lattice vector $\textbf{R}$. These operators obey the standard fermionic anti-commutation relations. Following this procedure the polarization and magnetization fields are decomposed into contributions associated with each site $\textbf{R}$, 
\begin{equation}\label{pr}
    \textbf{p}(\textbf{x},t) = \sum_\textbf{R} \textbf{p}_\textbf{R}(\textbf{x},t),
\end{equation}
and
\begin{equation}\label{mr}
    \textbf{m}(\textbf{x},t) = \sum_\textbf{R} \textbf{m}_\textbf{R}(\textbf{x},t).
\end{equation}
The magnetization fields are decomposed further into `atomic', `itinerant', and `spin' contributions,
\begin{equation}\label{magnetizationfields}
    \textbf{m}_\textbf{R}(\textbf{x},t) = \bar{\textbf{m}}_\textbf{R}(\textbf{x},t) + \tilde{\textbf{m}}_\textbf{R}(\textbf{x},t) + \breve{\textbf{m}}_\textbf{R}(\textbf{x},t).
\end{equation}
The lowercase $\textbf{p}(\textbf{x},t)$ and $\textbf{m}(\textbf{x},t)$ are the microscopic analogs of the macroscopic polarization and magnetization fields that are present in eq. (\ref{MacroJ}). 

The site polarization fields $\textbf{p}_\textbf{R}(\textbf{x},t)$ are obtained from 
\begin{equation}
\label{site_pr}
\begin{split}
    p^i_\textbf{R}(\textbf{x},t) = \int d\textbf{w} s^i(\textbf{x};\textbf{w},\textbf{R}) \rho_\textbf{R}(\textbf{w},t),
\end{split}
\end{equation}
where $\rho_\textbf{R}(\textbf{x},t)$ is the microscopic charge density associated with the lattice site \textbf{R} \cite{notes_on_microscopic_rho}
\begin{equation}
\begin{split}\label{microchargedensity}
    \rho_\textbf{R}(\textbf{x},t) = \frac{e}{2} \sum_{\alpha\beta\textbf{R}'\textbf{R}''} \Big( \delta_{\textbf{R}\textbf{R}'} + \delta_{\textbf{R}\textbf{R}''} \Big)\langle \hat{a}^\dag_{\beta\textbf{R}'}(t) \hat{a}_{\alpha\textbf{R}''}(t) \rangle
    \\ 
    \times \bar{W}^\dag_{\beta\textbf{R}'}(\textbf{x},t) \bar{W}_{\alpha\textbf{R}''}(\textbf{x},t)
    + \sum_{i} q_i \delta(\textbf{x}-\textbf{R}-\textbf{d}_i),
\end{split}
\end{equation}
where the nuclei in the unit cell associated with site \textbf{R} have charges $q_i$ and are located at positions $\textbf{d}_i$ with respect to $\textbf{R}$. The `relator' $s^i(\textbf{w};\textbf{x},\textbf{R})$ is defined as
\begin{equation}\label{s-relator}
\begin{split}
    s^i(\textbf{x};\textbf{w},\textbf{R}) = \int_{C(\textbf{w},\textbf{R})} dz^i \delta(\textbf{x}-\textbf{z}),
\end{split}
\end{equation}
where $C(\textbf{w},\textbf{R})$ specifies a path from \textbf{R} to \textbf{w}. We then can perform a multipole expansion of the site polarization fields of eq. (\ref{site_pr}) about the site location $\textbf{R}$. An example of this procedure is given in Appendix \ref{AppendixC} to obtain the electric and magnetic multipole moments. More details on this, and how the site magnetization fields $\textbf{m}_\textbf{R}(\textbf{x},t)$ are constructed, can be found in earlier work \cite{notes_on_microscopic_rho}. Taking the path $C(\textbf{w},\textbf{R})$ to be a straight line, the multipole expansion of the `site' polarization and magnetization fields about their associated lattice sites produces
\begin{equation}
\label{polarizationMultipoles}
\begin{split}
    p^i_\textbf{R}(\textbf{x},t) = \delta(\textbf{x}-\textbf{R}) \mu^i_\textbf{R}(t) - \frac{\partial \delta(\textbf{x}-\textbf{R})}{\partial x^j} q^{ij}_{\mathcal{P},\textbf{R}}(t)
    \\
    + \frac{\partial^2 \delta(\textbf{x}-\textbf{R})}{\partial x^j \partial x^l} o^{ijl}_{\mathcal{P},\textbf{R}}(t) + ...,
\end{split}
\end{equation}
and
\begin{equation}
\label{magnetizationMultipoles}
\begin{split}
    m^i_\textbf{R}(\textbf{x},t) = \delta(\textbf{x}-\textbf{R}) \nu^i_\textbf{R}(t) - \frac{\partial \delta(\textbf{x}-\textbf{R})}{\partial x^j} q^{ij}_{\mathcal{M},\textbf{R}}(t) + ...,
\end{split}
\end{equation}
where $\mu^i_\textbf{R}(t)$, $q^{ij}_{\mathcal{P},\textbf{R}}(t)$, $o^{ijl}_{\mathcal{P},\textbf{R}}(t)$, $\nu^i_\textbf{R}(t)$, and $q^{ij}_{\mathcal{M},\textbf{R}}(t)$ are the electric dipole, quadrupole, and octupole moments and the magnetic dipole and quadrupole moments, respectively, associated with the lattice site \textbf{R}. The electric dipole moment, for example, is given by
\begin{equation}
\label{site_electricdipole}
    \mu^i_\textbf{R}(t) = \int d\textbf{x} (x^i-R^i) \rho_\textbf{R}(\textbf{x},t).
\end{equation}
See section II of an earlier work \cite{DuffMagneticSusceptibility} for the form of the microscopic atomic, itinerant, and spin magnetic moments. 

We then perform the usual Fourier series analysis of any time dependent quantities 
\begin{equation}
    f(t) = \sum_{\omega} e^{-i\omega t} f(\omega),
\end{equation}
to consider the frequency components of the induced multipoles that appear at order $q^2$ in the effective conductivity tensor. Within the independent particle approximation we expect the multipole moments appearing in eq. (\ref{polarizationMultipoles}) and eq. (\ref{magnetizationMultipoles}) to depend on the microscopic electric and magnetic fields and their derivatives in the neighborhood of \textbf{R}. We then introduce microscopic response tensors with a tilde accent that relate the induced multipole moments with the microscopic fields and their derivatives evaluated at the site \textbf{R}. We neglect so-called ``local field corrections" and take the fields to be the macroscopic Maxwell fields $\textbf{E}(\textbf{x},t)$ and $\textbf{B}(\textbf{x},t)$ \cite{MaxwellEquations}, and thus obtain
\begin{widetext}
\begin{equation}
\label{ElectricMultipoleResponseTensors}
\begin{split}
    &\frac{ \mu^{i}_\textbf{R}(\omega)}{\mathcal{V}_\text{uc}} = \tilde{\chi}^{il}_{\mathcal{P}}(\omega) E^l(\textbf{R},\omega) + \tilde{\alpha}^{il}_{\mathcal{P}}(\omega) B^l(\textbf{R},\omega) + \tilde{\gamma}^{ijl}_\mathcal{P}(\omega) F^{jl}(\textbf{R},\omega) + \tilde{\Lambda}^{ijl}(\omega) L^{jl}(\textbf{R},\omega) + \tilde{\Pi}^{ijlk}(\omega) K^{jlk}(\textbf{R},\omega), 
    \\
    &\frac{ q^{ij}_{\mathcal{P},\textbf{R}}(\omega)}{\mathcal{V}_\text{uc}} = \tilde{\chi}^{ijl}_\mathcal{Q}(\omega) E^l(\textbf{R},\omega)+ \tilde{\Gamma}^{ijl}(\omega) B^l(\textbf{R},\omega) + \tilde{\Sigma}^{ijlk}(\omega) F^{lk}(\textbf{R},\omega) , 
    \\
    &\frac{ o^{ijl}_{\mathcal{P},\textbf{R}}(\omega)}{\mathcal{V}_\text{uc}} = \tilde{\Omega}^{ijlk}(\omega) E^k(\textbf{R},\omega),
\end{split}
\end{equation}
and
\begin{equation}
\label{MagneticMultipoleResponseTensors}
\begin{split}
    &\frac{\nu^{i}_\textbf{R}(\omega)}{\mathcal{V}_\text{uc}} = \tilde{\alpha}^{li}_{\mathcal{M}}(\omega)E^l(\textbf{R},\omega)  + \tilde{\chi}^{il}_{\mathcal{M}}(\omega)B^l(\textbf{R},\omega) + \tilde{\gamma}^{ijl}_{\mathcal{M}}(\omega)F^{jl}(\textbf{R},\omega) ,
    \\
    &\frac{ q^{ij}_{\mathcal{M},\textbf{R}}(\omega)}{\mathcal{V}_\text{uc}} = \tilde{\beta}^{ijl}_{\mathcal{M}}(\omega) E^l(\textbf{R},\omega).
\end{split}
\end{equation} 
We consider only the induced moments that linearly depend upon the electromagnetic fields; time-independent moments could exist if the crystal breaks the appropriate symmetries in the ground state. In eq. (\ref{ElectricMultipoleResponseTensors}) and eq. (\ref{MagneticMultipoleResponseTensors}) we identify the one response tensor at $\mathcal{O}(q^0)$, the four that appear at order $\mathcal{O}(q)$ and the  remaining eight contributions at order $\mathcal{O}(q^2)$.

The macroscopic fields are identified as the spatial average of the corresponding microscopic fields, using a weighting function $w(\textbf{x})$; its integral over all space is unity and its range is much greater than the lattice spacing but much less than the wavelength of light. The macroscopic polarization field, for example, is given by
\begin{equation}
\label{spatialaveraging}
\begin{split}
    \textbf{P}(\textbf{x},\omega) = \int d\textbf{x}' w(\textbf{x}-\textbf{x}')\textbf{p}(\textbf{x}',\omega)= \sum_\textbf{R} \int d\textbf{x}' w(\textbf{x}-\textbf{x}') \textbf{p}_\textbf{R}(\textbf{x}',\omega).
\end{split}
\end{equation}
Additionally, we assume that the macroscopic fields are not affected by further averaging; any corrections that would arise from this approximation can be considered part of the local field corrections that we neglect.  For example, we can then write
\begin{equation}
\begin{split}
    \mathcal{V}_{uc} \sum_\textbf{R} w(\textbf{x}-\textbf{R}) \textbf{E}(\textbf{R},\omega) \rightarrow \textbf{E}(\textbf{x},\omega)  . 
\end{split}
\end{equation}
For details of this approach, see Appendix B of Mahon and Sipe \cite{PerryOptical} and references cited therein.

As an example, for the macroscopic electric dipole moment per unit volume we then find 
\begin{equation}
\begin{split}
    \mathcal{P}^i(\textbf{x},\omega) &= \sum_\textbf{R} \int d\textbf{x}' w(\textbf{x}-\textbf{x}') \delta(\textbf{x}'-\textbf{R})
    \mu^i_\textbf{R}(\omega),
    \\
    &= \mathcal{V}_{uc} \sum_\textbf{R} w(\textbf{x}-\textbf{R}) \Bigg[
    \tilde{\chi}^{il}_\mathcal{P}(\omega) E^l(\textbf{R},\omega) + \tilde{\alpha}^{il}_\mathcal{P}(\omega) B^l(\textbf{R},\omega) + ... 
    \Bigg] 
    \\
    &= \tilde{\chi}^{il}_{\mathcal{P}}(\omega) E^l(\textbf{x},\omega) + \tilde{\alpha}^{il}_{\mathcal{P}}(\omega) B^l(\textbf{x},\omega) + ...
\end{split}
\end{equation}
Thus, within the neglect of local field corrections as discussed above, we can identify the macroscopic response tensors outlined in Table \ref{tab:1} with the microscopic multipole response tensors,  $\chi^{il}_\mathcal{P}(\omega)= \tilde{\chi}^{il}_\mathcal{P}(\omega),$ and so on.  

Applying this averaging procedure for all the induced multipole moments that are relevant up to order $q^2$, in linear response we obtain the following result for the macroscopic \textbf{P} and \textbf{M} fields, 
\begin{equation}
\label{PolarizationExpansion}
\begin{split}
    P^i(\textbf{x},\omega) = ... + \Big( \Lambda^{ijl}(\omega) - \Gamma^{ijl}(\omega) \Big) L^{jl}(\textbf{x},\omega) 
    + \Pi^{ijlk}(\omega) K^{jlk}(\textbf{x},\omega)
    \\
    - \Sigma^{ijlk}(\omega) \frac{\partial F^{lk}(\textbf{x},\omega)}{\partial x^j} 
    + \Omega^{ijlk}(\omega) \frac{\partial E^k(\textbf{x},\omega)}{\partial x^j \partial x^l},
\end{split}
\end{equation}
and
\begin{equation}
\label{MagnetizationExpansion}
\begin{split}
    M^k(\textbf{x},\omega) = ... + \chi^{kl}_\mathcal{M}(\omega) B^l(\textbf{x},\omega) + \gamma^{kjl}_\mathcal{M}(\omega) F^{jl}(\textbf{x},\omega) 
    - \beta^{kjl}_\mathcal{M}(\omega) \frac{\partial E^l(\textbf{x},\omega)}{\partial x^j}, 
\end{split}
\end{equation}

\end{widetext}
where `...' indicate the unperturbed and up to $\mathcal{O}(q)$ contributions. We have thus identified the eight response tensors outlined in Section \ref{MultipolarContibutions}. Using eq. (\ref{PolarizationExpansion}) and eq. (\ref{MagnetizationExpansion}) in the expression for the current density $\textbf{J}$ (the second line of eq. (\ref{MacroJ})), these eight tensors can be combined to obtain the two tensors $\sigma^{ilj}_L(\omega)$ and $\sigma^{ijlk}_K(\omega)$ that describe the $q^2$ contributions to the induced current; see Appendix \ref{AppendixB}.

\section{Linear Response}\label{sec:SPDM}

In constructing the response calculation it will be useful to introduce the single particle density matrix (SPDM) in the adjusted Wannier function basis, 
\begin{equation}\label{spdm}
\begin{split}
    \eta_{\alpha\textbf{R};\beta\textbf{R}'}(t) \equiv \langle \hat{a}^\dag_{\beta\textbf{R}'}(t) \hat{a}_{\alpha\textbf{R}}(t) \rangle e^{i\Phi(\textbf{R}',\textbf{R};t)}, 
\end{split}
\end{equation}
where $e^{i\Phi(\textbf{R}',\textbf{R};t)}$ is a generalized Peierls phase factor, and where the phase is a line integral of the electromagnetic vector potential $\textbf{A}(\textbf{x},t)$ of the applied fields from $\textbf{R}$ to $\textbf{R}'$ \cite{SylviaGLF,Perry_Sipe}, 

\begin{equation}\label{phiphase}
    \Phi(\textbf{R}',\textbf{R};t) \equiv \int d\textbf{x} s^i(\textbf{x};\textbf{R}',\textbf{R}) A^i(\textbf{x},t).
\end{equation}
Implementing eq. (\ref{s-relator}) in eq. (\ref{phiphase}) confirms that the phase is indeed a line integral of the vector potential. The use of the relator $s^i(\textbf{x};\textbf{R}',\textbf{R})$ allows for a systematic expansion of the multipole contributions; see for example Appendix C of Mahon and Sipe \cite{PerryMagneto}. For the extension to spatially varying fields see Appendix \ref{AppendixC} of the present manuscript.

Any site quantity, such as
$\textbf{p}_\textbf{R}(\textbf{x},t)$ or the three contributions to $\textbf{m}_\textbf{R}(\textbf{x},t)$, can be expressed as a trace of the `matrix multiplication' of the SPDM and the respective site quantity matrix element, 
\begin{equation}
\label{GeneralSiteQuantity}
\begin{split}
    \Lambda_\textbf{R}(\textbf{x},t) = \sum_{\alpha\beta\textbf{R}'\textbf{R}''} \Lambda_{\alpha\textbf{R}';\beta\textbf{R}''}(\textbf{x},\textbf{R};t) \eta_{\beta\textbf{R}'';\alpha\textbf{R}'}(t),
\end{split}
\end{equation}
where $\Lambda_\textbf{R}(\textbf{x},t)$ is a stand in for some arbitrary site quantity with $\Lambda_{\alpha\textbf{R}';\beta\textbf{R}''}(\textbf{x},\textbf{R};t)$ the corresponding ``site quantity matrix element." See section IV of Duff et al. for the example of the site spin magnetization \cite{DuffMagneticSusceptibility}.

The zeroth order SPDM for a topologically trivial insulator, our focus in this paper, is
\begin{equation}\label{SPDM0}
    \eta^{(0)}_{\alpha\textbf{R};\beta\textbf{R}'} = f_{\alpha} \delta_{\alpha\beta}\delta_{\textbf{R}\textbf{R}'},
\end{equation}
where $f_{\alpha}$ is the occupation of an orbital type $\alpha$ which is either 0 or 1 for zero temperature insulators. The response of the SPDM to an electric field, symmetrized single and double derivatives of the electric field, the magnetic field, and the derivatives of the magnetic field are all obtained from solving the dynamical equation for the SPDM; these results are given in Appendix \ref{AppendixD}.

When performing the perturbation expansion both the site quantity matrix element and the SPDM can depend on the electromagnetic fields. Thus in linear response one contribution is obtained from taking the site quantity matrix element to zeroth order and the SPDM to first order,
\begin{equation}
\begin{split}
    \Lambda^{(I)}_\textbf{R}(\textbf{x},t) = \sum_{\alpha\beta\textbf{R}'\textbf{R}''} \Lambda^{(0)}_{\alpha\textbf{R}';\beta\textbf{R}''}(\textbf{x},\textbf{R}) \eta^{(1)}_{\beta\textbf{R}'';\alpha\textbf{R}'}(t);
\end{split}
\end{equation}
it is dubbed the \textit{dynamical} contribution. The second contribution is obtained from taking the SPDM to zeroth order and the site quantity matrix element to first order,
\begin{equation}
\begin{split}
    \Lambda^{(II)}_{\textbf{R}}(\textbf{x},t) = \sum_{\alpha\beta\textbf{R}'\textbf{R}''}  \Lambda^{(1)}_{\alpha\textbf{R}';\beta\textbf{R}''}(\textbf{x},\textbf{R};t)\eta^{(0)}_{\beta\textbf{R}'';\alpha\textbf{R}'}(t);
\end{split}
\end{equation}
it is dubbed the \textit{compositional} contribution. The superscript $(0)$ indicates this is the ground state expression and the superscript $(1)$ indicates a quantity that is linear in the Maxwell fields. Every multipole response tensor will have a \textit{dynamical} contribution, but only some will have an additional \textit{compositional} contribution. 

As a simple example of the nature of these contributions and how they will arise, we consider the response of an atom to an electromagnetic field. Here there is only one site \textbf{R}, which can be identified with some center-point used to specify the position of the atom, and instead of Wannier functions we have the atomic orbitals \cite{HealyQuantum,HealyPZW,PZWNewPaper}. We take the Hamiltonian to be
\begin{equation}
\label{AtomicHamiltonian}
\begin{split}
    \hat{H}(t) = &\hat{H}^{(0)} -\hat{\mu}^{i}_E \cdot E^i(\textbf{R},t) - \hat{q}^{ij}_E F^{ij}(\textbf{R},t) 
    \\
    &- \hat{\nu}_P^i B^i(\textbf{R},t) - \frac{1}{2} \hat{\nu}_D^i(t) B^i(\textbf{R},t) + ... 
\end{split}
\end{equation}
where $\hat{H}^{(0)}$ is the atomic Hamiltonian before any external fields are applied, $\hat{\mu}^{i}_E$ is the electric dipole operator, $\hat{q}^{ij}_E$ the electric quadrupole operator, $\hat{\nu}_P^i$ the paramagnetic dipole operator, and $\hat{\nu}_D^i(t)$ the diamagnetic dipole operator. The last of these depends on time since it depends explicitly on the magnetic field $\textbf{B}(\textbf{R},t)$. 

Moving into the interaction picture, the expectation values of these multipole operators are of a very similar form to the general site quantities in the problem we address in this paper, 
\begin{equation}
    \langle \hat{\mathcal{O}}(t)\rangle = \sum_{\alpha\beta} \mathcal{O}_{\alpha\beta}(t) \langle \hat{a}_\alpha^\dag(t) \hat{a}_{\beta}(t) \rangle,
\end{equation}
where $\hat{a}^\dag_\alpha(t)$ ($\hat{a}_\alpha(t)$) is the fermionic creation (annihilation) operator for an orbital type $\alpha$. We have in general allowed for time dependent matrix elements $\mathcal{O}_{\alpha\beta}(t)$, however at this level of approximation it is only the diamagnetic dipole matrix elements $\boldsymbol\nu_{\text{D}:\alpha\beta}(t)$ that have any time dependence through their dependence on the magnetic field.

We can then perform a straightforward perturbative calculation to determine how the single particle density matrix elements $\langle \hat{a}^\dag_{\alpha}(t) \hat{a}_{\beta}(t)\rangle$ depend on the electromagnetic fields. Then considering for example the linear response of the electric dipole, since the matrix elements $\boldsymbol\mu_{\alpha\beta}$ have no explicit field dependence, the result only depends on how $\langle \hat{a}_\alpha^\dag(t) \hat{a}_{\beta}(t) \rangle$ depends on the fields. The electric dipole response to an electric field is
\begin{equation}
\label{Atomic:EdipoleE}
\begin{split}
    \langle \mu^{i,(E)}(\omega) \rangle = \sum_{\alpha\beta} f_{\beta\alpha} \frac{ \mu^i_{\beta \alpha} \mu^j_{\alpha\beta} }{E_{\beta}-E_{\alpha} - \hbar(\omega+i0^+)} E^j(\textbf{R},\omega).
\end{split}
\end{equation}
Eq. (\ref{Atomic:EdipoleE}) is an example of a \textit{dynamical} contribution. There are two key features to highlight. First, since we are considering the response of the $i$'th Cartesian component of the electric dipole, the respective multipole matrix element $\mu^i_{\beta\alpha}$ appears. If we were instead considering the response of the $ij$'th component of the quadrupole then $q^{ij}_{E:\beta\alpha}$ would appear in its place. Second, the matrix element of the operator that couples to the electromagnetic field of interest (or its derivative) in the interaction Hamiltonian is included. Since eq. (\ref{Atomic:EdipoleE}) describes the response to the $j$'th component of the electric field $\mu^j_{\alpha\beta}$ appears. If instead we considered the response to symmetric derivatives of the electric field $F^{jl}(\textbf{R},\omega)$ it would be the electric quadrupole matrix elements $q^{jl}_{E:\alpha\beta}$ that would appear in its place. When we generalize to crystalline systems, expressions that follow the general form of eq. (\ref{Atomic:EdipoleE}) will be referred to as the ``expected result." 

Even within the atomic limit a deviation from the above expected results arises due to the explicit dependence of the diamagnetic dipole operator matrix elements to the magnetic field. Notably, the linear response of the diamagnetic dipole can only be due to the magnetic field, since any changes to the density matrix $\langle \hat{a}^\dag_\alpha(t) \hat{a}_\beta(t)\rangle$ would necessarily enter at second order in the total response. This inherent dependence of the operator on the magnetic field leads to what we call the \textit{compositional} contribution. This produces the atomic diamagnetism   
\begin{equation}
\begin{split}
    \langle \nu_D^{i,(B)}(\omega)\rangle = \frac{e^2}{2mc^2} \sum_{\alpha} f_{\alpha} \epsilon^{iab} \epsilon^{lbj} q^{aj}_{E:\alpha\alpha}  B^l(\textbf{R},\omega). 
\end{split}
\end{equation}
Furthermore, when considering the linear response of an operator to the magnetic field (other than that of the diamagnetic dipole), it is the paramagnetic dipole matrix elements $\nu^j_{P:\alpha\beta}$ that appear. This is because in the Hamiltonian of eq. (\ref{AtomicHamiltonian}) the paramagnetic dipole energy is first order in the magnetic field but the diamagnetic dipole energy is second order.   

In a crystal additional deviations from these expected results arise since sites are not isolated from each other. Additionally, in determining analogs of the multipole matrix elements for a crystal the so called ``intraband position matrix elements" \cite{AversaSipe} must be treated carefully and produce extra contributions with no atomic analogs. This is because in a crystal one cannot merely make an identification of position matrix elements in the Bloch basis with the Berry connection.

\section{Multipole Moments in the ground state}\label{Sec:MultipoleMoments}

Returning to a topologically trivial insulator, we examine the multipole moments expanded about the lattice sites, and the associated macroscopic fields, in the ground state.  This will allow us to identify matrix elements
that will be useful in constructing the expressions for linear response. To construct these ground state quantities we use the ground state SPDM (\ref{SPDM0}) and the appropriate site quantity matrix element in the absence of any fields, $\Lambda^{(0)}_{\alpha\textbf{R}';\alpha\textbf{R}'}(\textbf{x},\textbf{R})$. Depending on the quantity of interest we will rely on one of the `relators' borrowed from atomic physics \cite{HealyQuantum}. The polarization depends on the `s-relator' that we have already introduced in eq. (\ref{s-relator}), and the magnetization depends on the `$\alpha$-relator'
\begin{equation}\label{alpha-relator}
\begin{split}
    \alpha^{jk}(\textbf{x};\textbf{w},\textbf{R}) = \epsilon^{jmn} \int_{C(\textbf{w},\textbf{R})} dz^m \frac{\partial z^n}{\partial w^k} \delta(\textbf{x}-\textbf{z}), 
\end{split}
\end{equation}
which `relates' the magnetization to the current density as $\textbf{s}(\textbf{x};\textbf{w},\textbf{R})$ relates the polarization to the charge density;
see \cite{DuffOpticalActivity} and Appendix \ref{AppendixC}. 
Note that the site quantity matrix element and SPDM are defined in the ``adjusted Wannier function" basis, not the Bloch function basis. This is important since, by construction, the Wannier functions are chosen to be localized about the sites \textbf{R} of the lattice and so the multipole moments are well defined in this basis.

We begin with some known results. The macroscopic ground state polarization is obtained by substituting eq. (\ref{polarizationMultipoles}) in eq. (\ref{spatialaveraging}),
and taking the applied fields to zero. Since we assume a homogeneous crystal the unperturbed charge density associated with each site satisfies $\rho_{\textbf{R}}(\textbf{x}) = \rho_{\textbf{0}}(\textbf{x}-\textbf{R}$). Thus, the dipole moment per unit volume is $\mathcal{P}^{(0)i} = {\mu^{i(0)}_\textbf{R}}/{\mathcal{V}_{uc}}$, where the superscript $(0)$ indicates this is the ground state expression before any fields are applied, and any $\textbf{R}$ can be chosen since the electric dipole moment associated with each site (\ref{site_electricdipole}) is identical. The macroscopic ground state polarization is then simply $P^{(0)i}=\mathcal{P}^{(0)i}$, since when the crystal is unperturbed the higher order electric multipole moments per unit volume are uniform and do not contribute to $P^{(0)i}$ (see eq. (\ref{MacroscopicFields})). The result can be expressed as the sum of the dipole moments of the filled Wannier functions divided by the unit cell volume,       
\begin{equation}
\begin{split}
    P^{(0)i} = \frac{e}{\mathcal{V}_{uc}} \sum_{\alpha} f_{\alpha} \int d\textbf{x} W^\dag_{\alpha\textbf{R}}(\textbf{x}) (x^i-R^i) W_{\alpha\textbf{R}}(\textbf{x}),
\end{split}
\end{equation}
where again any $\textbf{R}$ can be chosen, and we have omitted the ionic contribution, ${\mathcal{V}_{uc}^{-1}} \sum_{a} q_a d^i_a$. We can then use eq. (\ref{Wannier}), which relates the Wannier functions and Bloch functions, to write the real-space integral of the dipole moment as a single BZ-integral \cite{BerryVanderbilt}
\begin{equation}
\label{Polarization0_W}
\begin{split}
    P^{(0)i} = e\sum_{\alpha} f_{\alpha} \int_{BZ} \frac{d\textbf{k}}{(2\pi)^3} \tilde{\xi}^i_{\alpha\alpha},
\end{split}
\end{equation}
where the matrix elements of the Berry connection in the Wannier cell-periodic basis are
\begin{equation}
\begin{split}
    \tilde{\xi}^i_{\alpha\beta} = i(\alpha\textbf{k}| \partial_i \beta\textbf{k}).
\end{split}
\end{equation}
These matrix elements are smooth functions of \textbf{k}, which is not necessarily true for the Berry connection matrix elements defined in the eigenstate Bloch cell-periodic basis \cite{WannierInterpolation},
\begin{equation}
\begin{split}
    \xi^i_{nm} = i(n\textbf{k}|\partial_i m\textbf{k}).
\end{split}
\end{equation}
The Berry connection matrix elements in the Bloch and Wannier cell-periodic bases are related by
\begin{equation}
\label{BerryTransformation}
\begin{split}
    \tilde{\xi}^i_{\alpha \beta} = \sum_{nm} U^\dag_{\alpha n}\Big( \xi^i_{nm} + \mathcal{W}^i_{nm} \Big) U_{m \beta},
\end{split}
\end{equation}
where $\mathcal{W}^i_{nm} = i\sum_{\alpha} \partial_i U_{n\alpha} U^\dag_{\alpha m}$ depends on the unitary matrix that describes the transformation between the Bloch and Wannier bases. Because the Wannier cell-periodic functions are constructed such that they are smooth everywhere over the BZ \cite{WannierInterpolation}, the singularities that can arise in the Berry connection matrix elements defined in the eigenstate Bloch cell-periodic basis are canceled in eq. (\ref{BerryTransformation}) by singularities in the $\mathcal{W}_{nm}$.

Finally, using eq. (\ref{BerryTransformation}) in eq. (\ref{Polarization0_W}) we can write
\begin{equation}
\label{GroundStateP}
\begin{split}
    P^{(0)i} = e \sum_{n} f_{n} \int_{BZ} \frac{d\textbf{k}}{(2\pi)^3} \Big( \xi^i_{nn} + \mathcal{W}^i_{nn} \Big),
\end{split}
\end{equation}
in agreement with the modern theory \cite{BerryVanderbilt}. The presence of the $\mathcal{W}^i_{nn}$ explicitly shows that this quantity is gauge dependent; only changes in the polarization are physically well defined, as there is a quantum of ambiguity in the computationally determined values \cite{BerryVanderbilt}. 

We next turn to the ground state magnetization. As indicated in eq. (\ref{magnetizationfields}) the magnetization can be decomposed into `atomic', `itinerant', and `spin' contributions. Setting the fields to zero in the expressions for the site magnetization fields, and taking the spatial average, we obtain contributions expressed as real space integrals in the Wannier function basis. Analogous to the situation for the ground state polarization, for the ground state magnetization there is only a contribution from the magnetic dipole moment per unit volume. For the ground state atomic magnetization we find 
\begin{widetext}
\begin{equation}
\label{AtomicMagWannier}
\begin{split}
    \bar{M}^{(0)i} = \frac{e}{4 \mathcal{V}_{uc} c} \epsilon^{iab} \sum_{\alpha} \int d\textbf{x} W^\dag_{\alpha\textbf{R}}(\textbf{x}) \{ (x^a-R^a) , \hat{v}^b(\textbf{x}) \} W_{\alpha\textbf{R}}(\textbf{x}), 
\end{split}
\end{equation}
where $\hat{v}^b(\textbf{x})$ is the velocity operator that follows from the definition of the Hamiltonian \cite{DuffOpticalActivity}, and $\{ \cdot , \cdot \}$ is the anti-commutator. The itinerant magnetization is
\begin{equation}
\label{ItinerantMagWannier}
\begin{split}
    \tilde{M}^{(0)i} = \frac{e}{2\hbar \mathcal{V}_{uc} c} \epsilon^{iab} \sum_{\alpha\beta\textbf{R}'} f_{\alpha} \text{Im}\Bigg[ 
    R' \int d\textbf{x} W^\dag_{\beta\textbf{R}'}(\textbf{x}) W_{\alpha\textbf{0}}(\textbf{x}) H^{(0)}_{\alpha\textbf{0};\beta\textbf{R}'} 
    \Bigg], 
\end{split}
\end{equation}
where $H^{(0)}_{\alpha\textbf{0};\beta\textbf{R}'}$ are the Hamiltonian matrix elements in the absence of any fields, given by eq. (\ref{HamiltonianMatrixElements}). Lastly, the spin magnetization is
\begin{equation}
\label{SpinMagWannier}
\begin{split}
    \breve{M}^{(0)i} = \frac{e}{m \mathcal{V}_{uc} c} \sum_{\alpha} f_{\alpha} \int d\textbf{x} W^\dag_{\alpha\textbf{R}}(\textbf{x}) \frac{\hbar}{2} \sigma^i W_{\alpha\textbf{R}}(\textbf{x}), 
\end{split}
\end{equation}
where $\boldsymbol\sigma$ is the vector of Pauli matrices. Detailed derivations of these expressions can be found in earlier work \cite{Perry_Sipe,DuffOpticalActivity}.
\end{widetext}

Individually, the expressions for the atomic and itinerant contributions are gauge dependent. That is, when the real space integrals are converted to integrals over the BZ written with matrix elements defined in the cell-periodic Bloch basis, the individual results depend on how those cell-periodic Bloch basis functions are chosen; this is clear because there are diagonal elements of the Berry connection appearing, as well as additional terms arising 
that depend on the $\mathcal{W}$ matrix elements. However, as we show in Appendix \ref{ExtraAppendix}, the sum of the expressions for the atomic and itinerant contributions does not depend on diagonal elements of the Berry connection \textit{after} integration by parts, and the dependence on the $\mathcal{W}_{nm}$ matrix elements vanishes if we assume topologically trivial insulators (i.e., $f_{nm} \mathcal{W}^i_{nm} = 0$). When the expression (\ref{SpinMagWannier}) for the spin magnetization is written in terms of Bloch function quantities it \textit{is} gauge-invariant, and so we arrive at an expression for the total ground state magnetization that is gauge-invariant, 
\begin{equation}
\label{GroundStateMag}
\begin{split}
    M^{(0)i} = &\frac{ie}{2\hbar c} \epsilon^{iab} \sum_{n, s\neq n} f_{n} \int_{BZ} \frac{d\textbf{k}}{(2\pi)^3} (E_{n\textbf{k}}+E_{s\textbf{k}}) \xi^a_{ns}\xi^b_{sn} 
    \\
    + &\frac{e}{mc} \sum_{n} f_{n} \int_{BZ} \frac{d\textbf{k}}{(2\pi)^3} S^i_{nn},
\end{split}
\end{equation}
where $S^i_{nm} = \hbar/2 (u_{n\textbf{k}}|\sigma^i|u_{m\textbf{k}})$; this is in agreement with the `modern theory of magnetization' \cite{BerryVanderbilt}.

It will be useful to introduce general matrix elements $M^l_{nm}$ associated with the ground state magnetization, such that eq. (\ref{GroundStateMag}) can be written as a sum over the diagonal components of the matrix associated with the occupied states of the crystal, i.e.,
\begin{equation}
\label{GMagTrace}
    M^{(0)i} = \sum_{n} f_{n} \int_{BZ} \frac{d\textbf{k}}{(2\pi)^3} M^l_{nn}. 
\end{equation}
This is because general multipole matrix elements like $M^l_{nm}$ will appear when considering the linear response in Section \ref{Sec:ResponseTensors}, acting as optical transition matrix elements. We choose a Hermitian form, which we earlier identified as the ``spontaneous magnetization matrix element" \cite{DuffMagneticSusceptibility},
\begin{widetext}
\begin{equation}
\label{SpontaneousMagnetization}
\begin{split}
    M^l_{nm} = \frac{e}{4c} \epsilon^{lab} \Big( \sum_{s} \Big(\xi^a_{ns}v^b_{sm} + v^b_{ns}\xi^a_{sm}\Big) + \frac{1}{\hbar} \partial_b (E_{n\textbf{k}}+E_{m\textbf{k}}) \xi^a_{nm}\Big) + \frac{e}{mc} S^l_{nm},
\end{split}
\end{equation}
\end{widetext}
where $v^b_{sm}$ is the velocity matrix element evaluated in the Bloch basis
\begin{equation}
\label{velocityEq}
\begin{split}
    v^b_{sm} = \frac{i}{\hbar} (E_{s\textbf{k}}-E_{m\textbf{k}}) \xi^b_{sm} + \frac{1}{\hbar} \partial_b E_{m\textbf{k}} \delta_{sm}. 
\end{split}
\end{equation}
Eq. (\ref{SpontaneousMagnetization}) is not gauge covariant -- in that if one applies a change of phase to the Bloch functions $M^l_{nm}$ changes by more than just a phase -- but one can show that its use in eq. (\ref{GMagTrace}) leads to the correct expression (\ref{GroundStateMag}) for $M^{(0)i}$ using integration by parts (see \ref{sec:Presuppositions}).  

In a similar manner, based on the expression (\ref{GroundStateP}) for the ground state polarization we can identify an electric dipole matrix element as $P^i_{nm} = e\xi^i_{nm}$. Note that here we make this identification independent of the term involving $\mathcal{W}$, the appearance of which is related to the ground state polarization being gauge dependent, unlike the ground state magnetization. In identifying corresponding matrix elements that will be useful from the ground state expressions for other macroscopic multipole moments, we will also neglect such ``basis dependent terms." Of course, the matrix elements so identified -- such as $P^i_{nm}$ here -- will themselves still be gauge dependent. We now turn to those other macroscopic multipole moments.

The electric quadrupolarization is obtained by expanding the `s-relator' to obtain a term proportional to single derivatives of $\delta(\textbf{x}-\textbf{R})$ in eq. (\ref{site_pr}). The ground state quadrupole moment evaluated in the Wannier basis is then
\begin{equation}
\begin{split}
    \mathcal{Q}^{(0)ij}_{\mathcal{P}} = &\frac{e}{2 \mathcal{V}_{uc}} \sum_{\alpha} f_{\alpha} 
    \int d\textbf{x} W^\dag_{\alpha\textbf{0}}(\textbf{x}) x^i x^j W_{\alpha\textbf{0}}(\textbf{x}). 
\end{split}
\end{equation}
We may then convert to the Bloch basis to obtain
\begin{equation}
\label{quadrupole_moment}
\begin{split}
    \mathcal{Q}^{(0)ij}_\mathcal{P} = \frac{e}{4} \sum_{ns} f_{n} \int_{BZ} \frac{d\textbf{k}}{(2\pi)^3} \Big( \xi^i_{ns}\xi^j_{sn} + \xi^j_{ns}\xi^i_{sn} \Big) + f_{\mathcal{Q}}(\mathcal{W}),
\end{split}
\end{equation}
where a function $f_{\mathcal{Q}}(\mathcal{W})$ carries the $\mathcal{W}$ dependence that arises when transforming the non gauge covariant electric quadrupole from the Wannier to the Bloch basis. It is an example of the ``basis dependent terms" we neglect. The Bloch basis expression for the quadrupole moment has a striking resemblance to the quantum metric of \textbf{k}-space; however, the sum over $s$ in eq. (\ref{quadrupole_moment}) is not restricted to $s \neq n$ \cite{QuantumGeometry}. This is part of the reason the quadrupole moment as written above is gauge dependent; in the atomic limit the gauge dependence reflects the origin dependence of the quadrupole moment, which arises if there is a nonzero dipole moment. One could try to identify a gauge independent definition of the electric quadrupole moment by removing the diagonal elements of the Berry connection from eq. (\ref{quadrupole_moment}), which has been considered when looking at the multipolar contributions to the optical activity of a crystal \cite{MultipoleTheoryOpticalActivity}. We do not choose to make this substitution, since it is important to keep track of these pieces systematically to ensure gauge invariance of the total optical response tensors determined at the end of our calculations. It is then convenient to define the electric quadrupole matrix elements as
\begin{equation}
\begin{split}
    \mathcal{Q}^{ij}_{\mathcal{P}:nm} = \frac{e}{4} \sum_{s} \Big( \xi^i_{ns}\xi^j_{sm} + \xi^j_{ns}\xi^i_{sm} \Big).
\end{split}
\end{equation}

The electric octupole moment is obtained by identifying the term proportional to double derivatives of $\delta(\textbf{x}-\textbf{R})$ when expanding the `s-relator' in eq. (\ref{site_pr}). As was done for the electric quadrupole moment, the octupole is first defined in the Wannier basis, then transformed to the Bloch function basis, and the basis transformation dependent terms discarded. Following that, we can define Hermitian octupole matrix elements as 
\begin{equation}
\label{octupole_moment}
\begin{split}
    \mathcal{O}^{abc}_{\mathcal{P}:nm} = &\frac{e}{36}  \sum_{ \{abc\} } \Bigg[ \sum_{s} \Bigg( 
    \sum_{l} \xi^a_{ns}\xi^b_{sl}\xi^c_{lm} 
    \\
    &+ \frac{i}{2}\Big( \xi^a_{ns} \partial_b \xi^c_{sm} - \partial_b \xi^a_{ns} \xi^c_{sm} \Big) \Bigg) - \partial_a \partial_b \xi^c_{nm} 
    \Bigg],
\end{split}
\end{equation}
since the unperturbed macroscopic electric octupole moment, minus its $f_{\mathcal{O}}(\mathcal{W})$ contribution, is obtained by integrating the matrix elements of eq. (\ref{octupole_moment}) over the BZ and taking the trace over the filled states.  

There are two notable features of eq. (\ref{octupole_moment}) for $\mathcal{O}^{abc}_{\mathcal{P}:nm}$: First, it is gauge dependent due to the diagonal elements of the Berry connection included in the sums over band indices and the non gauge covariant $\textbf{k}$-derivatives on the Berry connection matrix elements. Second, the second line with $\textbf{k}$-derivatives would not be present in the atomic analogue, but is present here for a crystal due to Wannier functions at different sites in general having common support. In the atomic limit the gauge dependence can be understood as an origin dependence of the electric octupole moment.

The ground state magnetic quadrupolarization has three contributions, as illustrated in equation (\ref{magnetizationfields}). By expanding the `$\alpha$ relator' up to terms proportional to single derivatives of $\delta(\textbf{x}-\textbf{R})$ and inserting this result into the site quantity expressions for the magnetization (see Duff et al. \cite{DuffMagneticSusceptibility}) the atomic and itinerant contributions can be identified. Converting to the Bloch basis and neglecting contributions of the form $f_{\mathcal{Q}_\mathcal{M}}(\mathcal{W})$, we find  
\begin{widetext}
\begin{equation}
\begin{split}
    \bar{Q}^{(0)ij}_{\mathcal{M}} = &\frac{e}{12 c} \epsilon^{iab} \sum_{nsl} f_n \int_{BZ} \frac{d\textbf{k}}{(2\pi)^3} \Bigg[ \xi^j_{ns} \Big( \xi^a_{sl}v^b_{ln} + v^b_{sl}\xi^a_{ln} \Big) 
    + \Big(\xi^a_{ns}v^b_{sl} + v^b_{ns}\xi^a_{sl}\Big) \xi^j_{ln} \Bigg] 
    \\
    +&\frac{ie}{12 c} \epsilon^{iab} \sum_{ns} f_n \int_{BZ} \frac{d\textbf{k}}{(2\pi)^3} \Big( v^b_{ns}\partial_a \xi^j_{sn} - \partial_a \xi^j_{ns} v^b_{sn} \Big), 
\end{split}
\end{equation}
and
\begin{equation}
\begin{split}
    \tilde{Q}^{(0)ij}_{\mathcal{M}} = -\frac{e}{12\hbar c} \epsilon^{iab} \sum_{ns} f_n \int_{BZ} \frac{d\textbf{k}}{(2\pi)^3} \partial_a E_{n\textbf{k}} \Big( \xi^b_{ns}\xi^j_{sn} + \xi^j_{ns}\xi^b_{sn}\Big). 
\end{split}
\end{equation}
\end{widetext}
The spin magnetic moment is not constructed by the use of the `$\alpha$-relator' so an ad-hoc moment expansion of $\breve{m}_\textbf{R}(\textbf{x})$ produces
\begin{equation}
\begin{split}
    \breve{Q}^{(0)ij}_{\mathcal{M}} = \frac{e}{2mc} \sum_{ns} f_n \int_{BZ} \frac{d\textbf{k}}{(2\pi)^3} \Big( S^i_{ns}\xi^j_{sn} + \xi^j_{ns}S^i_{sn} \Big).
\end{split}
\end{equation}
None of $\bar{Q}^{(0)ij}_{\mathcal{M}}$, $\tilde{Q}^{(0)ij}_{\mathcal{M}}$, nor $\breve{Q}^{(0)ij}_{\mathcal{M}}$ are gauge-invariant, nor is their sum. This is in contrast to the magnetization eq. (\ref{GroundStateMag}), where no contribution of the form $f_{\mathcal{\mathcal{M}}}(\mathcal{W})$ arose, and in the final result there was no term that depended on components of diagonal elements $\boldsymbol\xi_{nn}$. This is in line with our understanding of the atomic limit, where the total magnetic dipole moment is unique but where the magnetic quadrupole in fact does have an origin dependence if there is a non-zero magnetic dipole moment. Likewise the electric dipole, quadrupole, and octupole show a dependence on the origin of the coordinate system if their respective lower degree moment is non-zero, and in extending the consideration to crystalline systems these quantities are dependent on the diagonal elements of the Berry connection. 
\begin{widetext}
We then only introduce a combined atomic and spin magnetic quadrupole matrix element,
\begin{equation}
\begin{split}
    \mathcal{Q}^{ij}_{\mathcal{M}:nm} = \frac{e}{12 c} \epsilon^{iab} \Bigg[ 
    \xi^j_{nl} \Big( \xi^a_{ls}v^b_{sm} + v^b_{ls}\xi^a_{sm}\Big) + \Big(\xi^a_{ns}v^b_{sl} + v^b_{ns}\xi^a_{sl} \Big) \xi^j_{lm} + i \Big( v^b_{nl} \partial_a \xi^j_{lm} - \partial_a \xi^j_{nl} v^b_{lm} \Big)
    \Bigg]
    \\
    +\frac{ie}{12\hbar c} \epsilon^{iab} \partial_j \partial_a (E_{n\textbf{k}}-E_{m\textbf{k}}) \xi^b_{nm} 
    +\frac{e}{2mc} \Big( S^i_{nl}\xi^j_{lm} + \xi^j_{nl}S^i_{lm} \Big).
\end{split}
\end{equation}
The itinerant contributions are left out, since at finite frequency their effects on the optical response tensors are not naturally described by their inclusion in these matrix elements, in contrast to the situation when writing the spontaneous magnetization matrix element for the DC response. However, we have added a traceless contribution, since it arises in the optical response tensor expressions. As a word of caution, while we have used the ground state expressions for the macroscopic multipole moments as a starting point for identifying the matrix elements introduced here -- once the basis dependent terms of the form $f(\mathcal{W})$ have been neglected -- our strategy has been to define them in a way that simplifies our calculated results in the following sections, in particular by choosing them to be Hermitian. And sum rules can be used to re-express the matrix elements, so many other functional forms that are equivalent could exist \cite{DuffMagneticSusceptibility}.

\end{widetext}




\section{Induced Multipole Response Tensors}\label{Sec:ResponseTensors}

In this section we identify the analytic form of the response tensors introduced in Table \ref{tab:1}, each involving a single integral over the Brillouin zone and depending on both the spectral and geometric properties of the Hamiltonian. Here when we refer to a quantity as being ``geometric" in nature, we mean that it depends on quantities like the quantum metric or Berry curvature. The Hermitian multipole matrix elements introduced above will make frequent appearance in the response tensors. The results given below are the total response, i.e. the sum of the dynamical and compositional contributions. To produce these results the appropriate site quantity matrix element is identified by looking at the relator expansion of the polarization or magnetization.

Part of the results will depend only upon the quantities like the Berry connection, band energies, and their derivatives. However, there will also be a gauge dependence tracked by the derivatives of the matrix elements $U_{n\alpha}(\textbf{k})$, the quantities $\mathcal{W}^a_{nm} \equiv i\sum_{\alpha} \partial_a U_{n\alpha} U^\dag_{\alpha m}$ and their derivatives. These track the general multiband gauge-dependence that includes the indeterminacy of the phase of the Bloch functions if one chooses a transformation that makes $\mathcal{W}^a_{nm}$ purely diagonal. In the following we omit these ``basis dependent terms." In the case of a gauge-transformation described by $\mathcal{W}^a_{nm} = \delta_{nm} \partial_a \phi_{n}(\textbf{k})$ (i.e. this describes a different periodic $\textbf{k}$-dependent phase for the Bloch functions) the terms that depend on the $\mathcal{W}$'s vanish when all the multipole contributions are combined appropriately in the constitutive equation for the current, see equations (\ref{q2conductivityL}) and (\ref{q2conductivityK}). This cancellation requires the \textit{topologically trivial} and \textit{smoothness} presuppositions. Thus the expressions reported below are the ones that should be computed to determine the effective $q^2$ conductivity tensor. As we have seen in previous manuscripts, the basis dependent terms vanish when calculating any ``physical" response tensor. This was seen for optical activity \cite{DuffOpticalActivity} and the DC magnetic susceptibility \cite{DuffMagneticSusceptibility}.  

\subsection{Magnetization Response to B - Magnetic Susceptibility}

Previously we determined the static limit of the magnetic susceptibility of an insulator and partitioned it into three relatively simple contributions \cite{DuffMagneticSusceptibility},

\begin{equation}
\label{magstatic}
    \chi^{kl}_{\text{static}} = \chi^{kl}_{\text{VV}} + \chi^{kl}_\text{occ} + \chi^{kl}_{\text{geo}}.
\end{equation}
The first term is a generalization of Van Vleck paramagnetism to crystals, and can be written very compactly with the use of the spontaneous magnetization matrix elements defined in eq. (\ref{SpontaneousMagnetization}) 

\begin{equation}
    \chi^{kl}_\text{VV} = \sum_{mn} f_{nm} \int_{BZ} \frac{d\textbf{k}}{(2\pi)^3} \frac{M^k_{nm}M^l_{mn}}{\Delta_{mn}(\textbf{k})},
\end{equation}
where $\Delta_{mn}(\textbf{k}) = E_{m\textbf{k}}-E_{n\textbf{k}}$. The occupied term $\chi^{kl}_\text{occ}$ represents a generalization of the atomic diamagnetism to include the itinerant features of Bloch electrons, 
\begin{widetext}
\begin{equation}
\begin{split}
    \chi^{kl}_{\text{occ}} = \frac{e^2}{4\hbar^2 c^2} \epsilon^{kab} \epsilon^{lcd} \sum_{nm} f_{n} \int_{BZ} \frac{d\textbf{k}}{(2\pi)^3} \text{Re}\Bigg[ \Big( \frac{\hbar^2}{m}\delta_{bc} - \partial_b \partial_c E_{n\textbf{k}}  \Big) \xi^a_{nm}\xi^d_{mn} \Bigg]. 
\end{split}
\end{equation}
Note that $\frac{\hbar^2}{m}\delta_{bc}$ must be replaced by the Hessian matrix if one uses an effective tight-binding model; this is required due to sums over bands necessarily being truncated. Lastly, the geometric contribution, which has no atomic analogue, is

\begin{equation}
\begin{split}
    \chi^{kl}_\text{geo} = -\frac{e}{2\hbar c} \sum_{nm} f_{n} \int_{BZ} \frac{d\textbf{k}}{(2\pi)^3} \text{Re} \Bigg[ \Omega^k_{nm} \Big( M^l_{mn} + \frac{e}{8\hbar c} \Delta_{nm}(\textbf{k}) \Omega^l_{mn} \Big) 
    \\
    + \Big( M^k_{nm} + \frac{e}{8\hbar c} \Delta_{nm}(\textbf{k}) \Omega^k_{nm} \Big) \Omega^l_{mn} \Bigg],
\end{split}
\end{equation}
where $\Omega^l_{nm}$ is the curl of the non-Abelian Berry connection, $\Omega^l_{nm} = \epsilon^{lab} \partial_a \xi^b_{nm}$. As was shown previously, the static magnetic susceptibility is a gauge-invariant quantity and can be written in many different ways, but all related by sum rules. One particular form has all diagonal elements removed which makes the gauge invariance explicit (cf. \cite{DuffMagneticSusceptibility}). 

As we extend the calculation of $\chi^{kl}_\mathcal{M}$ to treat the finite frequency response, new contributions arise in addition to the static response. We can write the total frequency dependent susceptibility then as
\begin{equation}
\label{magsus}
\begin{split}
    \chi^{kl}_\mathcal{M}(\omega) = & \chi^{kl}_\text{occ} + \sum_{mn} f_{nm} \int_{BZ} \frac{d\textbf{k}}{(2\pi)^3} \frac{M^k_{nm}}{\Delta_{mn}(\textbf{k}) - \hbar(\omega+i0^+)}\Big( M^l_{mn}  - \frac{e\omega}{4c} \epsilon^{lab} \frac{\partial_a (E_{m\textbf{k}}+E_{n\textbf{k}}) \xi^b_{mn}}{\Delta_{mn}(\textbf{k})-\hbar(\omega+i0^+)} \Big)
    \\
    &-\frac{e}{4\hbar c} \sum_{mn} \int_{BZ} \frac{d\textbf{k}}{(2\pi)^3} \Bigg[ 
    \Big( f_{n} + f_{m} \Big) \Big( \Omega^k_{nm} M^l_{mn} + M^k_{nm} \Omega^l_{mn} \Big)
    +f_{nm} \frac{e}{4\hbar c}\Big(\Delta_{nm}(\textbf{k}) - \hbar\omega \Big) \Big(
    \Omega^k_{nm} \Omega^l_{mn} 
    \Big)
    \Bigg]
    \\
    &-\frac{ie^2 \omega}{16 c^2} \epsilon^{kcd} \epsilon^{lab} \sum_{mn} f_{nm} \int_{BZ} \frac{d\textbf{k}}{(2\pi)^3} \frac{\partial_a \partial_c v^d_{nm} \xi^b_{mn}}{\Delta_{mn}-\hbar(\omega+i0^+)}, 
\end{split}
\end{equation}
\end{widetext}
which reduces to $\chi^{kl}_\text{static}$ at $\omega=0$. 
The first term is the occupied contribution, which is the analog in insulators of the atomic diamagnetism. It is peculiar since it is \textit{frequency independent} and persists for all frequencies of the magnetic field. The next contribution to $\chi^{kl}_\mathcal{M}(\omega)$ is what one would naively expect as the extension of the Van Vleck paramagnetism to finite frequency, the ``expected result." However, we also find that the magnetic dipole matrix element $M^l_{mn}$ now gains a frequency dependent and ``itinerant" contribution, as it depends on the gradient of the band energies.

We identify the second line of eq. (\ref{magsus}) as a generalization of the geometric contribution to the magnetic susceptibility $\chi^{kl}_\text{geo}$. The change when going to finite frequency is to take $\Delta_{nm}(\textbf{k}) \rightarrow \Delta_{nm}(\textbf{k})-\hbar\omega$. The added frequency dependent contribution is purely an itinerant magnetization compositional contribution. 

To understand its existence, recall the need for defining an itinerant magnetization due to motion of electrons between the sites in the crystal. This itinerant magnetization requires the overlap of Wannier functions at different sites. Then when considering the interaction with the electric field we make the choice to couple to the average of the electric field evaluated at both sites involved in the expression. Therefore, spatial variation of the electric field will influence the coupling, which is exactly the case when one considers a time-dependent magnetic field since the curl of the electric field must be non-zero by Faraday's law. This can then be recast as a magnetic response of the magnetization that only appears at finite frequency. 

The third line comes from an addition to the atomic magnetic dipole beyond what one would naively expect from taking the simple atomic expression of position crossed with velocity. This is another example of how when extending to the crystal limit one must treat the \textit{intraband} position matrix elements carefully.

\begin{widetext}
\subsection{Magnetization Response to F}

\begin{equation}
\label{gammaMF}
\begin{split}
    \gamma^{ijl}_\mathcal{M}(\omega) =& \sum_{mn} f_{nm} \int_{BZ} \frac{d\textbf{k}}{(2\pi)^3} \frac{M^i_{nm}}{\Delta_{mn}(\textbf{k}) - \hbar(\omega+i0^+)} \Bigg( 
    \mathcal{Q}^{jl}_{\mathcal{P}:mn} + \frac{ie}{4} \frac{ \partial_j (E_{n\textbf{k}}+E_{m\textbf{k}}) \xi^l_{mn} + \partial_l (E_{n\textbf{k}}+E_{m\textbf{k}})\xi^j_{mn} }{\Delta_{mn}-\hbar(\omega+i0^+)}
    \Bigg)
    \\
    &-\frac{e}{2\hbar c} \sum_{mn} f_{n} 
    \int_{BZ} \frac{d\textbf{k}}{(2\pi)^3} \text{Re}\Bigg[\Omega^i_{nm} \Big( \mathcal{Q}^{jl}_{\mathcal{P}:mn} + \frac{ie}{4} \Big( \partial_j \xi^l_{mn} + \partial_l \xi^j_{mn} \Big) 
    \Bigg]
    \\
    &-\frac{e^2}{16 c} \epsilon^{iab} \sum_{mn} f_{nm} \int_{BZ} \frac{d\textbf{k}}{(2\pi)^3} \frac{ \partial_j \partial_a v^b_{nm} \xi^l_{mn} + \partial_l \partial_a v^b_{nm} \xi^j_{mn}  }{\Delta_{mn}-\hbar(\omega+i0^+)}
\end{split}
\end{equation} 
In the above response tensor we have the ``expected result" in the first line, the multiplication of the magnetization and the electric quadrupole matrix elements divided by the optical transition energy $\Delta_{mn}(\textbf{k})-\hbar(\omega+i0^+)$. However, we also have an ``itinerant" like addition to the quadrupole moment matrix elements $\mathcal{Q}^{jl}_{\mathcal{P}:mn}$. 

We then have a term that is very reminiscent of the geometric magnetic susceptibility that arises from the compositional contribution to the itinerant magnetic dipole. In fact, it has the same origin as the frequency dependent contribution to line two of equation (\ref{magsus}), except the symmetric derivatives of the electric field are used instead of the antisymmetric derivatives. Here we can identify the curl of the Berry connection ($\boldsymbol{\Omega}$) times a \textit{modified} quadrupole matrix element. The modification arises from carefully treating the \textit{intraband} position matrix elements.

The last line of equation (\ref{gammaMF}) is almost identical to the third line of equation (\ref{magsus}); however, in equation (\ref{gammaMF}) the integrand is not multiplied by the frequency and the tensor is symmetric upon exchange of the $j$ and $l$ Cartesian indices. This line comes from the terms in the SPDM response that connect different lattice sites $\textbf{R} \neq \textbf{R}'$. 

\subsection{Magnetic Quadrupolarization response to E}

\begin{equation}
\label{betaM}
\begin{split}
    \beta^{ijl}_\mathcal{M}(\omega) = &e\sum_{mn} f_{nm} \int_{BZ} \frac{d\textbf{k}}{(2\pi)^3} \frac{ \mathcal{Q}^{ij}_{\mathcal{M}:nm} \xi^l_{mn}}{\Delta_{mn}(\textbf{k})-\hbar(\omega+i0^+)}
    \\
    &-\frac{e^2}{24\hbar c} \epsilon^{iab} \sum_{mn} f_{nm} \int_{BZ} \frac{d\textbf{k}}{(2\pi)^3} \Bigg[
    \partial_a (E_{n\textbf{k}}+E_{m\textbf{k}})\Big(\xi^b_{nl}\xi^j_{lm} + \xi^j_{nl}\xi^b_{lm}\Big)
    \Bigg] \frac{ \xi^l_{mn}}{\Delta_{mn}(\textbf{k})-\hbar(\omega+i0^+)}  
    \\
    &-\frac{e^2}{8 c} \epsilon^{iab} \sum_{mn} f_{nm} \int_{BZ} \frac{d\textbf{k}}{(2\pi)^3} \frac{ \xi^l_{mn}}{\Delta_{mn}(\textbf{k})-\hbar(\omega+i0^+)} \Bigg[ 
    \partial_j \partial_a v^b_{nm} +\frac{i}{3\hbar} (E_{m\textbf{k}}-E_{n\textbf{k}}) \partial_j \partial_a \xi^b_{nm}
    \Bigg]
    \\
    &+\frac{e^2}{6\hbar c} \sum_{n} f_{n} \int_{BZ} \frac{d\textbf{k}}{(2\pi)^3} \text{Re}\Bigg[ 
    \epsilon^{iab} \partial_a \xi^l_{nm} \xi^j_{ml} \xi^b_{ln} + 2i \partial_j \xi^l_{nm} \Omega^i_{mn} 
    \Bigg].
\end{split}
\end{equation}
The first line is again the ``expected result" of the magnetic quadrupole matrix elements times the polarization matrix elements divided by the optical transition energy. The second line contains the itinerant contribution to the magnetic quadrupolarization matrix elements. The third line is an added term that comes from the dynamical response of the \textit{atomic} magnetic quadrupole, like the third line of equation (\ref{magsus}) and (\ref{gammaMF}) it contains a double derivative of velocity matrix elements. The fourth line of equation (\ref{betaM}) comes from the compositional response of the itinerant magnetic quadrupole, and appears to be of a geometric nature, and so has no atomic analogue. 
\vspace{20pt}
\subsection{Electric Dipole response to L}

\begin{equation}
\label{Lambda}
\begin{split}
    \Lambda^{ijl}(\omega) = &e \sum_{mn} f_{nm} \int_{BZ} \frac{d\textbf{k}}{(2\pi)^3} \frac{\xi^i_{nm} \mathcal{Q}^{lj}_{\mathcal{M}:mn}}{\Delta_{mn} - \hbar(\omega+i0^+)} 
    \\
    &+\frac{ie^2}{12 c} \epsilon^{lab} \sum_{lmn} f_{nm} \int_{BZ} \frac{d\textbf{k}}{(2\pi)^3} \xi^i_{nm} \Big( \xi^a_{ml}v^b_{ln} + v^b_{ml}\xi^a_{ln} - \frac{1}{\hbar} \partial_a (E_{m\textbf{k}}+E_{n\textbf{k}}) \xi^b_{mn} \Big) \frac{ \partial_j (E_{n\textbf{k}}+E_{m\textbf{k}})}{(\Delta_{mn}(\textbf{k})-\hbar(\omega+i0^+))^2} 
    \\
    &+\frac{ie^2}{2mc} \sum_{mn} f_{nm} \int_{BZ} \frac{d\textbf{k}}{(2\pi)^3} \xi^i_{nm} S^l_{mn} \frac{ \partial_j (E_{n\textbf{k}}+E_{m\textbf{k}})}{(\Delta_{mn}(\textbf{k})-\hbar(\omega+i0^+))^2} 
    \\
    &+\frac{ie^2}{12 c} \epsilon^{lab} \sum_{lmn} f_{nm} \int_{BZ} \frac{d\textbf{k}}{(2\pi)^3} \xi^i_{nm} \Big( \xi^j_{ml}v^b_{ln} + v^b_{ml}\xi^j_{ln} - \frac{1}{\hbar} \partial_j (E_{n\textbf{k}}+E_{m\textbf{k}}) \xi^b_{mn} \Big) \frac{ \partial_a (E_{n\textbf{k}}+E_{m\textbf{k}})}{(\Delta_{mn}(\textbf{k})-\hbar(\omega+i0^+))^2}
    \\
    &-\frac{i\omega e^2}{12 c} \epsilon^{lab} \sum_{mn} f_{nm} \int_{BZ} \frac{d\textbf{k}}{(2\pi)^3} \frac{ \partial_j \partial_a \xi^i_{nm} \xi^b_{mn}}{\Delta_{mn}(\textbf{k})-\hbar(\omega+i0^+)}
    \\
    &- \frac{ie^2 \omega}{12 c} \epsilon^{lab} \sum_{mn} f_{nm} \int_{BZ} \frac{d\textbf{k}}{(2\pi)^3} \xi^i_{nm} \xi^b_{mn} \Bigg[ 
    \frac{\partial_j \partial_a (E_{n\textbf{k}}-E_{m\textbf{k}})}{(\Delta_{mn}-\hbar(\omega+i0^+))^2} + 2 \frac{\partial_j (E_{m\textbf{k}}+E_{n\textbf{k}}) \partial_a (E_{m\textbf{k}}+E_{n\textbf{k}})}{(\Delta_{mn}-\hbar(\omega+i0^+))^3} 
    \Bigg] 
    \\
    &+\frac{e^2}{12\hbar c}  \sum_{mn} f_{n}  
    \int_{BZ} \frac{d\textbf{k}}{(2\pi)^3} \text{Re}\Bigg[ \epsilon^{lab}\sum_{l}\Big( \xi^j_{nl}\xi^b_{lm} + \xi^b_{nl}\xi^j_{lm}\Big) \partial_a \xi^i_{mn}  - 2i \Omega^l_{nm} \partial_j \xi^i_{mn} 
    \Bigg].
\end{split}
\end{equation}
Here we find the most complicated multipole response tensor we derive, there are many different contributions to be parsed. We begin with the first line. It is the ``expected result" of the polarization matrix element multiplied by the magnetic quadrupolarization matrix element. All the additional contributions beyond the expected result arise from non-zero intersite matrix elements (i.e. for $\textbf{R}\neq \textbf{R}'$).  

The second and third line together \textit{almost} have a very simple interpretation. The second line is $1/3$ of the orbital contributions to the magnetic dipole moment and the third line is $1/2$ the spin contribution to the magnetic dipole moment matrix elements. The fourth line is then a $j$ and $a$ Cartesian index swapped version of the second line, but as such cannot be identified with the orbital magnetization. 

The fifth and sixth lines are then contributions that vanish as $\omega\rightarrow 0$, and depend on double derivatives of the Berry connection or the band energies. 

The last line is geometric in nature, and comes from the compositional contribution to the response. It contains both the antisymmetric derivatives of the non-Abelian berry connection and a symmetric product of the Berry connection matrix elements. If not for the unconstrained sum, the latter could be identified with the quantum metric. 

\subsection{Electric Dipole response to K}

\begin{equation}
\label{Pi_Response}
\begin{split}
    \Pi^{ijlk}(\omega) &= \frac{e}{6} \sum_{mn} f_{nm} \sum_{ \{jlk\} } \int_{BZ} \frac{d\textbf{k}}{(2\pi)^3} \frac{\xi^i_{nm} }{\Delta_{mn}(\textbf{k}) - \hbar(\omega+i0^+)} \Bigg[ 
    \mathcal{O}^{jlk}_{\mathcal{P}:mn} + \frac{i}{2} \frac{ \partial_j (E_{n\textbf{k}}+E_{m\textbf{k}}) \mathcal{Q}^{lk}_{\mathcal{P}:mn}}{\Delta_{mn}-\hbar(\omega+i0^+)}
    \\
    &+ \frac{e}{4} \frac{\partial_j \partial_l (E_{m\textbf{k}}-E_{n\textbf{k}}) \xi^k_{mn} }{\Delta_{mn}-\hbar(\omega+i0^+)}
    + \frac{e}{4} \frac{\partial_j (E_{m\textbf{k}}-E_{n\textbf{k}}) \partial_l \xi^k_{mn}}{\Delta_{mn}-\hbar(\omega+i0^+)}
    - \frac{e}{2} \frac{(\partial_j E_{n\textbf{k}} \partial_l E_{n\textbf{k}} + \partial_j E_{m\textbf{k}} \partial_l E_{m\textbf{k}})\xi^k_{mn} }{ (\Delta_{mn}-\hbar(\omega+i0^+))^2 }
    \Bigg].
\end{split}
\end{equation}
The first term in the brackets of eq. (\ref{Pi_Response}) produces the ``expected result", the multiplication of the polarization and octupolarization matrix elements. The remaining contributions appear to be itinerant in nature, involving single and double derivatives of the band energies and derivatives of the Berry connection matrix elements. There are no compositional contributions; the above is purely due to how the SPDM is altered by the applied field. The entire tensor is made explicitly symmetric under exchange of any of the Cartesian indices $\{i,j,k\}$ with the sum over all permutations of the indices $\sum_{\{jlk\}}$ since this tensor is contracted with the symmetric double derivatives of the electric field $K^{jlk}(\textbf{x},\omega)$. 
\vspace{10pt}
\subsection{Quadrupolarization response to F}

\begin{equation}\label{SigmaF}
\begin{split}
    \Sigma^{ijlk}(\omega) &= \sum_{mn} f_{nm} \int_{BZ} \frac{d\textbf{k}}{(2\pi)^3} \frac{ \mathcal{Q}^{ij}_{\mathcal{P}:nm}  }{\Delta_{mn}(\textbf{k})-\hbar(\omega+i0^+)} \Bigg[ 
    \mathcal{Q}^{kl}_{\mathcal{P}:mn} + \frac{ie}{4}\frac{\partial_k (E_{n\textbf{k}}+E_{m\textbf{k}}) \xi^l_{mn} + \partial_l (E_{n\textbf{k}}+E_{m\textbf{k}}) \xi^k_{mn}}{\Delta_{mn}(\textbf{k})-\hbar(\omega+i0^+)}
    \Bigg]
    \\
    &-\frac{e^2}{16} \sum_{mn} f_{nm} \int_{BZ} \frac{d\textbf{k}}{(2\pi)^3} \frac{\partial_l\Big( \partial_i \xi^j_{nm} + \partial_j \xi^i_{nm} \Big)   \xi^k_{mn} + \partial_k \Big( \partial_i \xi^j_{nm} + \partial_j \xi^i_{nm}\Big)\xi^l_{mn}}{\Delta_{mn}(\textbf{k})-\hbar(\omega+i0^+)} .
\end{split}
\end{equation}
The first term in the brackets of eq. (\ref{SigmaF}) produces the ``expected result". We also have the ``itinerant"-like addition to the quadrupole moment matrix elements $Q^{kl}_{\mathcal{P}:mn}$ that appears in the first line of eq. (\ref{gammaMF}). The second line arises due to the matrix elements between orbitals at different lattice sites; hence we consider it another itinerant feature.  
\vspace{20pt}
\subsection{Quadrupolarization response to B}
\begin{equation}
\label{Gamma}
\begin{split}
    \Gamma^{ijl}(\omega) = &\sum_{mn} f_{nm} \int_{BZ} \frac{d\textbf{k}}{(2\pi)^3} \frac{ \mathcal{Q}^{ij}_{\mathcal{P}:nm} }{\Delta_{mn}(\textbf{k})-\hbar(\omega+i0^+)} \Bigg[ 
    M^l_{mn} -\frac{e\omega}{4c} \epsilon^{lab} \frac{ \partial_a (E_{n\textbf{k}}+E_{m\textbf{k}})\xi^b_{mn} }{\Delta_{mn}(\textbf{k})-\hbar(\omega+i0^+)} 
    \Bigg] 
    \\
    &-\frac{i\omega e^2}{16 c} \epsilon^{lab} \sum_{mn} f_{nm} \int_{BZ} \frac{d\textbf{k}}{(2\pi)^3} \frac{ \partial_a\Big(  \partial_j \xi^i_{nm} + \partial_i \xi^j_{nm}  \Big) \xi^b_{mn}}{\Delta_{mn}(\textbf{k})-\hbar(\omega+i0^+)}
    \\
    &-\frac{e}{2\hbar c} \sum_{nm} f_{n} \int_{BZ} \frac{d\textbf{k}}{(2\pi)^3} \text{Re}\Bigg[ \Omega^l_{nm} \Big( \mathcal{Q}^{ij}_{\mathcal{P}:mn} + \frac{ie}{4} \Big(\partial_i \xi^j_{mn} + \partial_j \xi^i_{mn}\Big) \Big) \Bigg]. 
\end{split}
\end{equation}
The first line contains the ``expected result" as well as the same frequency dependent modification to the magnetization matrix seen in eq. (\ref{magsus}). The second line vanishes at zero frequency and captures some of the effects of the itinerant nature of the Bloch electrons on the response. The last line appears to be geometric, coming from the compositional contribution to the response, containing the curl of the Berry connection and the same modified quadrupole matrix elements seen in eq. (\ref{gammaMF}).
\end{widetext}

\subsection{Octupolarization response to E}
Here we have simply the ``expected result" of the octupolarization times the polarization matrix elements,

\begin{equation}
\label{Omega_Response}
\begin{split}
    \Omega^{ijlk}(\omega) = e \sum_{mn} f_{nm} \int_{BZ} \frac{d\textbf{k}}{(2\pi)^3} \frac{ \mathcal{O}^{ijl}_{\mathcal{P}:nm}\xi^k_{mn}}{\Delta_{mn}(\textbf{k})-\hbar(\omega+i0^+)}.
\end{split}
\end{equation}

\subsection{Symmetries of the Response Tensors}
It is instructive to explicitly discuss the symmetries of the various response tensors and additional imposed constraints when the crystal satisfies time-reversal and/or inversion symmetry. Recall the perturbative expansion of the multipole moments in response to the applied electromagnetic fields, equations (\ref{ElectricMultipoleResponseTensors}) and (\ref{MagneticMultipoleResponseTensors}): $\Pi^{ijlk}(\omega)$ is unaffected by permutations of the last three indices; $\Sigma^{ijlk}(\omega)$ is unaffected by permutations of the first two indices or the last two indices; $\Gamma^{ijl}(\omega)$ is unaffected by permutations of the first two indices; $\Omega^{ijlk}(\omega)$ is unaffected by permutations of the first three indices; and $\gamma^{ijl}_{\mathcal{M}}(\omega)$ is unaffected by permutations of the last two indices. 

If one considers a system that does not break inversion symmetry then $E_{n\textbf{k}} = E_{n-\textbf{k}}$, $\xi^i_{nm}(\textbf{k}) = -\xi^i_{nm}(-\textbf{k})$, and $S^i_{nm}(\textbf{k}) = S^i_{nm}(-\textbf{k})$. Thus $M^l_{nm}$ and $\mathcal{Q}^{ij}_{\mathcal{P}:nm}$ are even functions of $\textbf{k}$, and $\mathcal{O}^{ijk}_{\mathcal{P}:nm}$ and $\mathcal{Q}^{ij}_{\mathcal{M}:nm}$ are odd functions of $\textbf{k}$. So all integrands of the response tensors at order $q^2$ are even functions of $\textbf{k}$, they do not trivially vanish. This is in contrast to the order $q$ response tensors that describe optical activity and require inversion symmetry breaking.  

If one considers a system that does not break time-reversal symmetry then $E_{n\textbf{k}} = E_{n'-\textbf{k}}$, where here the prime notation indicates a pair of eigenstates related by time-reversal symmetry, since we consider half-integer spinor states. In a particular gauge one can choose the following relationship for some of the base matrix elements
\begin{equation}
\begin{split}
    \xi^a_{nm}(\textbf{k}) = \xi^a_{m'n'}(-\textbf{k}), 
    \hspace{10pt}
    S^a_{nm}(\textbf{k}) = -S^a_{m'n'}(-\textbf{k}),
    \\
    \xi^a_{nm'}(\textbf{k}) = -\xi^a_{mn'}(-\textbf{k}),
    \hspace{10pt}
    S^a_{nm'}(\textbf{k}) = S^a_{mn'}(-\textbf{k}).
\end{split}
\end{equation}
Assuming a time-reversal symmetric ground state, if the state $n$ is filled then so too is the state $n'$. Assuming the energy $\hbar\omega$ remains below the band gap, with some relabeling of band indices one can then show the following: $\chi^{kl}(\omega) = \chi^{kl}(-\omega)$; $\gamma^{ijl}_{\mathcal{M}}(\omega) = -\gamma^{ijl}_{\mathcal{M}}(-\omega)$; $\beta^{ijl}_{\mathcal{M}}(\omega) = -\beta^{ijl}_{\mathcal{M}}(-\omega)$; $\Lambda^{ijl}(\omega) = -\Lambda^{ijl}(-\omega)$; $\Pi^{ijlk}(\omega) = \Pi^{ijlk}(-\omega)$; $\Sigma^{ijlk}(\omega) = \Sigma^{ijlk}(-\omega)$; $\Gamma^{ijl}(\omega) = -\Gamma^{ijl}(-\omega)$; and lastly $\Omega^{ijlk}(\omega) = \Omega^{ijlk}(-\omega)$. In total one finds that $\sigma^{ilj}_{L}(\omega) = \sigma^{ijl}_{L}(-\omega)$ and $\sigma^{ijlk}_K(\omega) = -\sigma^{ijlk}_K(-\omega)$, this follows from how the multipole contributions are combined in equation (\ref{q2conductivityL}) and (\ref{q2conductivityK}). In addition there are some properties that hold even if the energy is taken above the band gap or if one includes scattering phenomenologically $0^+ \rightarrow 1/\tau$, where $\tau$ is a mean lifetime: if not for the itinerant and frequency dependent contribution in line one and line three of equation (\ref{magsus}) one could say $\chi^{kl}(\omega) = \chi^{lk}(\omega)$; we also find many contributions vanish. These are: the frequency dependent part of the second line of equation (\ref{magsus}); the second line of equation (\ref{gammaMF}); the last line of equation (\ref{betaM}); the last line of equation (\ref{Lambda}); and the last line of equation (\ref{Gamma}). Thus, all the terms we have identified as `geometric' will vanish with time-reversal symmetry, except for those from $\chi^{kl}_\text{static}$.  
\vspace{10pt}
\section{The \texorpdfstring{$q^2$}{q2} effective conductivity tensor}\label{sec:q2conductivity}

The induced multipole moment response tensors introduced in Section \ref{Sec:ResponseTensors} are not gauge-invariant by themselves, but when we combine them all in the constitutive equation for the current to obtain the expressions for $\sigma^{ilj}_L(\omega)$ and $\sigma^{ijlk}_K(\omega)$, see eq. (\ref{q2conductivityL}) and eq. (\ref{q2conductivityK}), the result that describes the induced current density is gauge-invariant. This is an expected result, as the induced charge current at order $q^2$ should be a physically measureable quantity. 

In the case of optical activity the order $q$ contributions to the effective conductivity tensor could be written such that the diagonal elements of the Berry connection were all explicitly removed, making the gauge invariance explicit \cite{MultipoleTheoryOpticalActivity,DuffOpticalActivity}. In a similar fashion for the DC magnetic susceptibility, the diagonal elements of the Berry connection could be collected and repackaged by the use of sum rules and integration by parts into a geometric term that had no dependence on the diagonal elements, thus making the gauge invariance explicit \cite{DuffMagneticSusceptibility}. To do this we first rewrote equation (\ref{magstatic}) as the sum of three terms identified by Ogata \cite{OgataMagSus2017}, $\chi^{kl}_\text{inter}$, $\chi^{kl}_\text{occ}$, and $\chi^{kl}_\text{occ2}$. Then we labeled quantities with the diagonal elements removed with an over set ring - the so called `purified' terms - and the terms that were removed and collected up were indicated by an over bar accent,
\begin{equation}
\label{chiDC_ring}
\begin{split}
    \chi^{kl}_\text{static} = \mathring{\chi}^{kl}_\text{inter} + \mathring{\chi}^{kl}_\text{occ} + \mathring{\chi}^{kl}_\text{occ2} + \bar{\chi}^{kl}.
\end{split}
\end{equation}
In a perhaps surprising twist one can rewrite $\bar{\chi}^{il}$ as
\begin{equation}
\begin{split}
    \bar{\chi}^{kl} = \frac{1}{2} \mathring{\chi}^{kl}_\text{occ2:Orb} + \mathring{\chi}^{kl}_\text{occ2:Spin}.
\end{split}
\end{equation}
See section VII of Duff et al. \cite{DuffMagneticSusceptibility} for the details and distinction between the orbital and spin contributions to $\mathring{\chi}^{kl}_\text{occ2}$.

The story of the $q^2$ effective conductivity is naturally more complicated, since as we have demonstrated in the earlier sections the magnetic susceptibility is only one of eight response tensors that needs to be considered, whereas there were only four response tensors at order $q$. To create purified response tensors we take the diagonal elements out of the sums. For example,
\begin{widetext}
\begin{equation}
\label{purifiedtensorI}
\begin{split}
    \sum_{l} \xi^i_{nl}\xi^j_{lm} = \sum_{l\neq n,m} \xi^i_{nl}\xi^j_{lm} + \Big( \xi^i_{nn}\xi^j_{nm} + \xi^i_{nm}\xi^j_{mm}\Big),
\end{split}
\end{equation}
where we identify the first term on the right side of equation (\ref{purifiedtensorI}) as the purified term. The terms in brackets go into the over bar accent part of the response tensor. One also must replace the $\textbf{k}$ derivatives of the non-Abelian Berry connection with gauge covariant derivatives \cite{NonlinearOpticalSipe},
\begin{equation}
\label{gaugecovariantderivative}
    \partial_a \xi^b_{nm} = \partial_{;a} \xi^b_{nm} + i(\xi^a_{nn}-\xi^a_{mm}) \xi^b_{nm},
\end{equation}
where we employ a shorthand notation for the gauge-covariant derivatives as:
\begin{equation}\label{gaugecovariantderivative_eq}
    \partial_{;a} \xi^b_{nm} \equiv \Big( \partial_a - i\Big( \xi^a_{nn}-\xi^a_{mm}\Big) \Big) \xi^b_{nm} .
\end{equation}
From eq. (\ref{gaugecovariantderivative}) the gauge covariant derivative is included in the purified response tensor and the other terms collected separately. Note that under a phase transformation of the Bloch functions equation (\ref{gaugecovariantderivative_eq}) transforms covariantly, hence the name.  

An added complication arises in the multipole tensors involved in defining $\sigma^{iljk}(\omega)$; there are double \textbf{k} derivatives of matrix elements. The replacement to gauge covariant derivatives must be done with care since there is a choice of which derivative to perform first, or to take the derivatives symmetrically, and this leads to a different expression. If we evaluate the derivatives from right to left we find
\begin{equation}
\label{gaugecovariantdoublederivative_eq}
\begin{split}
    \partial_i \partial_j \xi^a_{nm} = \partial_{;i}\partial_{;j} \xi^a_{nm} + i(\xi^i_{nn}-\xi^i_{mm}) \partial_{;j} \xi^a_{nm} + i(\xi^j_{nn} - \xi^j_{mm}) \partial_{;i} \xi^a_{nm} 
    \\
    -(\xi^i_{nn}-\xi^i_{mm})(\xi^j_{nn}-\xi^j_{mm}) \xi^a_{nm} +i\partial_i (\xi^j_{nn}-\xi^j_{mm}) \xi^a_{nm}.
\end{split}
\end{equation}
With this choice all the contributions that one picks up to make the substitutions to gauge-covariant derivatives and pulling out diagonal elements of the Berry connection from the sums leads to the expression for $\bar{\sigma}_L^{ilj}(\omega)$ and $\bar{\sigma}_K^{ilj}(\omega)$. Thus, we can write
\begin{equation}
\label{sigmaL}
\begin{split}
    \sigma^{ilj}_L(\omega) = -c\epsilon^{ijk}\Big( \mathring{\chi}^{kl}(\omega) - \frac{i\omega}{2c} \epsilon^{lab} \mathring\beta^{kab}(\omega) \Big) + \epsilon^{iak}\epsilon^{lab} \frac{i\omega}{3} \Big( \frac{1}{2} \Big( \mathring{\beta}^{kjb}(\omega) + \mathring{\beta}^{kbj}(\omega) \Big)
    -\mathring{\gamma}^{kjb}(\omega)  \Big)
    \\
    +i\omega \Big(\mathring{\Lambda}^{ijl}(\omega) - \mathring{\Gamma}^{ijl}(\omega) \Big) - \frac{\omega^2}{3c} \epsilon^{lab} \Big( \mathring{\Sigma}^{ibaj}(\omega) + 2\mathring{\Omega}^{ijab}(\omega) \Big) + \bar{\sigma}^{ilj}_L(\omega),
\end{split}
\end{equation}
where the extra contribution can be written to be explicitly gauge-invariant as
\begin{equation}
\begin{split}
    \bar{\sigma}^{ilj}_L(\omega) &= \frac{i\omega e^2}{18\hbar c} \epsilon^{lab} \sum_{n,m\neq n} f_{n} \int_{BZ} \frac{d\textbf{k}}{(2\pi)^3} \Bigg[ \Big(\partial_a \xi^i_{nn} - \partial_i \xi^a_{nn} \Big)\Big(\xi^j_{nm}\xi^b_{mn} + \xi^b_{nm}\xi^j_{mn} \Big) 
    \\
    &-\frac{i\omega e^2}{12\hbar c} \epsilon^{lab} \sum_{n,m\neq n} f_{n} \int_{BZ} \frac{d\textbf{k}}{(2\pi)^3} \Bigg[ 
    \partial_a \xi^b_{nn} \Big( \xi^j_{nm}\xi^i_{mn} + \xi^i_{nm}\xi^j_{mn} \Big)
    \Bigg]
\end{split}
\end{equation}
and
\begin{equation}
\label{sigmaK}
\begin{split}
    \sigma^{iljk}_K(\omega) = \frac{1}{6}\sum_{\sigma(jlk)} \Bigg[ c\epsilon^{ija} \Big( \mathring{\beta}^{alk}(\omega) -\mathring{\gamma}^{alk}(\omega) \Big) + i\omega\Big( \mathring{\Pi}^{ijlk}(\omega) + \mathring{\Omega}^{ijlk}(\omega)
    -\mathring{\Sigma}^{ijlk}(\omega) \Big) + \bar{\sigma}^{iljk}_K(\omega)\Bigg],
\end{split}
\end{equation}
where
\begin{equation}
\begin{split}
    \bar{\sigma}^{iljk}_K(\omega) = \frac{e^2}{12\hbar} \sum_{n,m\neq n} f_{n} \int_{BZ} \frac{d\textbf{k}}{(2\pi)^3} \xi^j_{nm} \xi^k_{mn} \Bigg[ 
     \partial_l \xi^i_{nn} - \partial_i \xi^l_{nn}  
    \Bigg].
\end{split}
\end{equation}
\end{widetext}

Instructions for how to obtain the `purified' multipole response tensors are given in Appendix \ref{Appendix:GaugeInvariantTensors}. With $\sigma^{ilj}_L(\omega)$ and $\sigma^{iljk}_K(\omega)$ now obtained, the total conductivity can be calculated using eq. (\ref{eq:q2conductivity}). The magnetic field dependent contribution is multiplied by $\frac{ic}{\omega}\epsilon^{akl}$, since Faraday's law was used (eq. (\ref{FaradaysLaw})) to convert the derivatives of the magnetic field to double derivatives of the electric field. This appears to now be a divergent contribution to $\sigma^{iljk}(\omega)$ as $\omega\rightarrow 0$. However, if the magnetic field is time-independent the limit can be taken noting that $\frac{\epsilon^{akl}}{\omega} \frac{\partial E^k(\textbf{x},\omega)}{\partial x^l}$ is finite as $\omega\rightarrow 0$, so the induced current is not divergent. Here we see another benefit to the partitioning of the response into $\sigma^{iljk}_K(\omega)$ and $\sigma^{ilj}_L(\omega)$, they are both separately gauge-invariant and are finite as $\omega\rightarrow 0$. 

\section{Application to crystal system: Haldane Model}\label{sec:HaldaneModel}

\begin{figure}
\centering
\includegraphics[width=8cm]{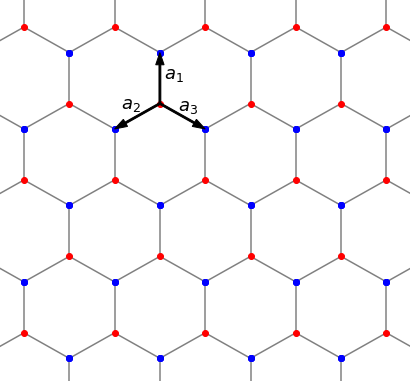}
\caption{Honeycomb lattice of the Haldane model. There are three nearest neighbour vectors connecting the sites that belong to sublattice A (in red) to those on sublattice B (in blue). The vectors are given in the text.}\label{fig:Lattice}
\end{figure}

To investigate the ``geometric" terms in more detail, test the sum rules that must be employed when performing these calculations, and see the effects of the ``crystalline" itinerant terms to the response tensors we consider the optical response of the Haldane model \cite{HaldaneModel}.

\begin{figure}
\centering
\includegraphics[width=7cm]{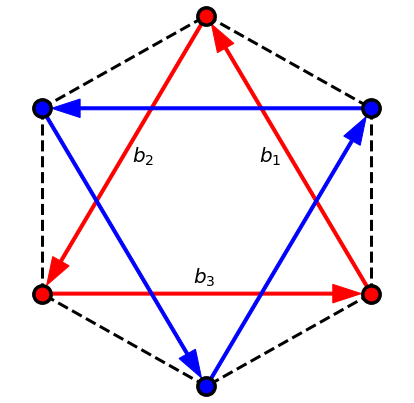}
\caption{Complex NNN hoppings that connect sites on the same sublattice. The direction indicates positive imaginary hopping of strength $t_2\sin(\phi)$}
\label{fig:HaldaneHopping}
\end{figure}

The Haldane model is a honeycomb lattice with real nearest neighbour (NN) hopping between A and B sites and complex next nearest neighbour (NNN) hoppings. The phase of the NNN hoppings are direction dependent, with positive imaginary hoppings if the direction is counterclockwise around a unit cell, as seen in Figure \ref{fig:HaldaneHopping}. This breaks the sublattice symmetry and time-reversal symmetry. A mass term to break inversion symmetry and to gap the system is also included.

The NN lattice vectors are $\textbf{a}_1 = (0,a)$, $\textbf{a}_2 = (-\frac{\sqrt{3}a}{2},-\frac{a}{2})$, and $\textbf{a}_3 = (\frac{\sqrt{3}a}{2},-\frac{a}{2})$, where $a$ is the distance between A and B sites. The NNN lattice vectors are $\textbf{b}_1 = (\frac{-\sqrt{3}a}{2},\frac{3a}{2})$, $\textbf{b}_2 = (-\frac{\sqrt{3}a}{2},-\frac{3a}{2})$, and $\textbf{b}_3 = (\sqrt{3}a,0)$. The high symmetry Dirac points are located at $\textbf{K} = \frac{2\pi}{3a}( \frac{\sqrt{3}}{3},1)$ and $\textbf{K}'= \frac{2\pi}{3a}(-\frac{\sqrt{3}}{3},1)$ in the first Brillouin zone.

In the basis of cell-periodic Bloch functions associated with A and B sites the Hamiltonian can be written with the use of pseudo-spin Pauli matrices

\begin{equation}
\label{HaldaneHamiltonian}
\begin{split}
    \mathcal{H}(\textbf{k}) = \textbf{f}(\textbf{k}) \cdot \boldsymbol\sigma,
\end{split}
\end{equation}
where $\textbf{f}(\textbf{k})$ is a four component vector whose components are 
\begin{equation}
\label{HaldaneFunctions}
\begin{split}
    &f_0(\textbf{k}) = 2t_2\cos(\phi) \sum_i \cos(\textbf{k}\cdot\textbf{b}_i),
    \\
    &f_x(\textbf{k}) = -t\sum_i \cos(\textbf{k}\cdot\textbf{a}_i),
    \\
    &f_y(\textbf{k}) = t\sum_i \sin(\textbf{k}\cdot\textbf{a}_i),
    \\
    &f_z(\textbf{k}) = M-2t_2\sin(\phi)\sum_i \sin(\textbf{k}\cdot\textbf{b}_i) ,
\end{split}
\end{equation}
and $\sigma^i$ the Pauli matrices; $M$ is the `mass' term, $t$ the NN hopping strength from site A to site B, and $t' = t_2 e^{i\phi}$ the complex NNN hopping strength. 

The Haldane model exhibits two topological phases with Chern number $C = \pm 1$ if the magnitude of the imaginary hopping $t_2 \sin(\phi)$ is greater than a critical value $t_c = \frac{M}{3\sqrt{3}}$. The real part of the NNN hopping serves only to break the particle hole symmetry. At the critical hopping strength the gap closes at the $\textbf{K}$ point for $t_2 = t_c$ and at the $\textbf{K}'$ point for $t_2 = -t_c$.

\begin{figure}[h]
\centering
\includegraphics[width=8.5cm]{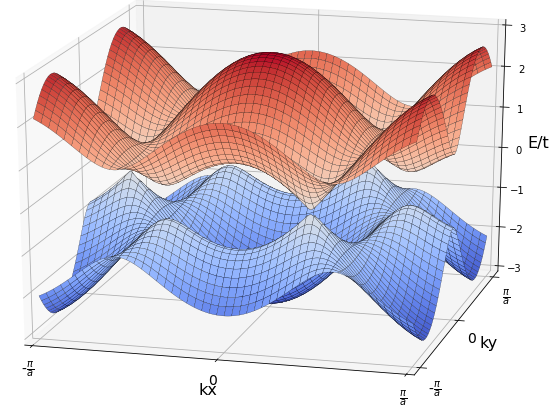}
\caption{Dispersion of the Haldane model given in eq. (\ref{HaldaneHamiltonian}) and eq. (\ref{HaldaneFunctions}). $M = 0.2t$, $\phi=\pi/2$, and $t_2 =0.04t$, which is above the critical value, thus the model is in the topological phase with $C=+1$. The gap is almost closed at the three equivalent $\textbf{K}$ points.}\label{fig:Dispersion}
\end{figure}

We consider the deep tight-binding regime where the atomic orbitals $\phi_\textbf{R}(\textbf{x})$ are well localized about their respective lattice sites \textbf{R}, and are orthonormal to orbitals on different lattice sites. The cell periodic Bloch functions associated with the A/B sites are then constructed via the prescription:
 \begin{equation}
\begin{split}\label{BlochFunctions}
    \phi^{A/B}_\textbf{k}(\textbf{x}) = \frac{1}{\sqrt{N}} \sum_{\textbf{R}_{A/B}} e^{i\textbf{k}\cdot(\textbf{x}-\textbf{R}_{A/B})} \phi_{\textbf{R}_{A/B}}(\textbf{x}). 
\end{split}
\end{equation}

The Haldane Hamiltonian can be diagonalized to obtain the dispersion relation for the two bands
\begin{equation}
\begin{split}\label{eq:Dispersion}
    E_{\pm}(\textbf{k}) = f_0(\textbf{k}) \pm \sqrt{ f_x(\textbf{k})^2 + f_y(\textbf{k})^2 + f_z(\textbf{k})^2 },
\end{split}
\end{equation}
and the associated eigenvectors are
\begin{equation}
    \phi^{\pm}_\textbf{k}(\textbf{x}) = c^{\pm}_A(\textbf{k})\phi^A_\textbf{k} + c^{\pm}_B(\textbf{k}) \phi^B_\textbf{k},
\end{equation}
with
\begin{equation}
\begin{split}
    c_A^{+} = \frac{1}{\sqrt{1+|\alpha^+(\textbf{k})|^2 }},
    \hspace{5pt}
    c_B^{+} = \alpha^+(\textbf{k}) c_A^{+},
\end{split}
\end{equation}
\begin{equation}
\begin{split}
    c_B^{-} = \frac{1}{\sqrt{1+|\alpha^-(\textbf{k})|^2 }},
    \hspace{5pt}
    c_A^{-} = \alpha^-(\textbf{k}) c_B^{-},
\end{split}
\end{equation}
where 
\begin{equation}
    \alpha^+(\textbf{k}) = \frac{ f_z -  \sqrt{ f_x(\textbf{k})^2 + f_y(\textbf{k})^2 + f_z(\textbf{k})^2 }}{ -f_x(\textbf{k}) + if_y(\textbf{k})  },
\end{equation}
and
\begin{equation}
    \alpha^-(\textbf{k}) = \frac{ -f_x(\textbf{k}) + if_y(\textbf{k})  }{f_z(\textbf{k}) + \sqrt{ f_x(\textbf{k})^2 + f_y(\textbf{k})^2 + f_z(\textbf{k})^2 } }.
\end{equation}
With the eigenvectors and eigenvalues (and their derivatives) in hand one can then begin to compute the myriad response tensors. 

\subsection{Frequency Dependent Response}

With all the response tensors derived we can now look at a theoretical calculation of the $q^2$ effective conductivity tensor. We will look in particular at the tensor $\sigma^{ilj}_L(\omega)$ that characterizes the current induced by a spatially varying magnetic field. This is because at zero frequency this tensor is completely described the the DC magnetic susceptibility. As we extend to finite frequency other magnetization contributions like $\gamma^{ijl}_\mathcal{M}(\omega)$ and $\beta^{ijl}_\mathcal{M}(\omega)$ must be included as well to describe the total response of the magnetization to a non-uniform electromagnetic field. Additionally, polarization contributions must be included to obtain the gauge-invariant tensor $\sigma^{ilj}_L(\omega)$. We will only consider the components $\sigma^{xzy}_L(\omega) = -\sigma^{yzx}_L(\omega)$ since for the 2 dimensional model we consider they are the only tensor components that are non-vanishing at zero frequency.

\begin{figure}[h]
\centering
\includegraphics[width=8.5cm]{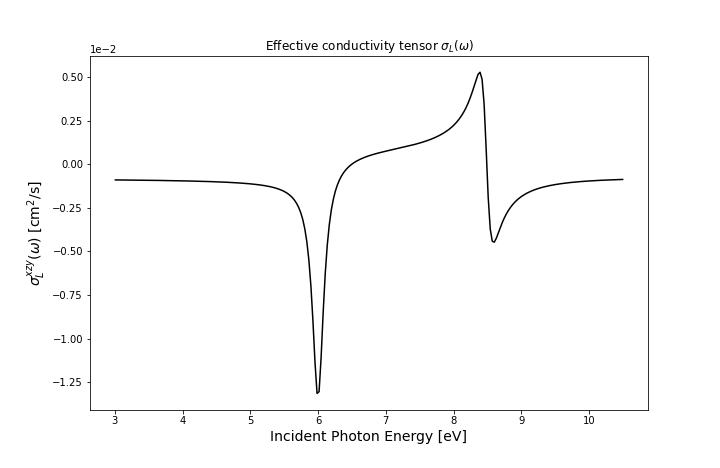}
\caption{The real part of the effective conductivity tensor $\sigma^{xzy}_L(\omega)$, plotted over a range of incident field photon energies.}
\label{fig:ACMagnetization}
\end{figure}

We set the parameters of the model to $t=M=3\text{eV}$, and $t_2 = 0$. The numerical integration is performed on a 800x800 grid of $\textbf{k}$-points over the Brillouin zone, and $0^+$ is taken to 100 meV to smooth out the peaks. As we see in Figure \ref{fig:ACMagnetization}, while the incident photon energy $\hbar\omega$ is less than the gap energy the response is almost a constant value of $c\chi_\text{static}$. There is then a sharp change in $\sigma^{xzy}_L(\omega)$ when $\hbar\omega$ approaches the energy gap which occurs at the $\textbf{K}$ and $\textbf{K}'$ points in the BZ, degenerate in this model for $t_2 = 0$. The response is then an order of magnitude larger and then switches to being ``paramagnetic" as the frequency increases to then match the energy gap at the $\textbf{M}$ point. We call this a ``paramagnetic switch" since, if we were to naively think about this response as purely being due to the induced magnetization, it has now switched signs. Away from any resonances the response returns to being dominated by the static susceptibility.

\subsection{DC Magnetic Susceptibility}

The derivation of the response tensors initially assume topologically trivial insulators. This assumption is made by the choice for the ground state SPDM in the Wannier function basis, eq. (\ref{SPDM0}), where one can relate a set of filled Wannier functions to a set of filled Bloch functions, which cannot be done in the case of a Chern Insulator. Under the assumption of eq. (\ref{SPDM0}) there is no static linear conductivity of an insulator and there is no free current at linear response. Work has been done to extend to the more general class of insulators that include Chern insulators \cite{PerryChern}, where the well known quantized anomalous Hall conductivity is proportional to the Chern number. This contribution can be identified as arising due to the so called ``free current", and not associated with the charges described by the polarization and magnetization. Thus, an examination of possible additional topological contributions to the $q^2$ effective conductivity is warranted since new ``free current" contributions could arise. With this caveat in mind we can still \textit{attempt} to evaluate all the polarization and magnetization response tensors in any phase of the Haldane model. 

\begin{figure}[h]
\centering
\includegraphics[width=8.5cm]{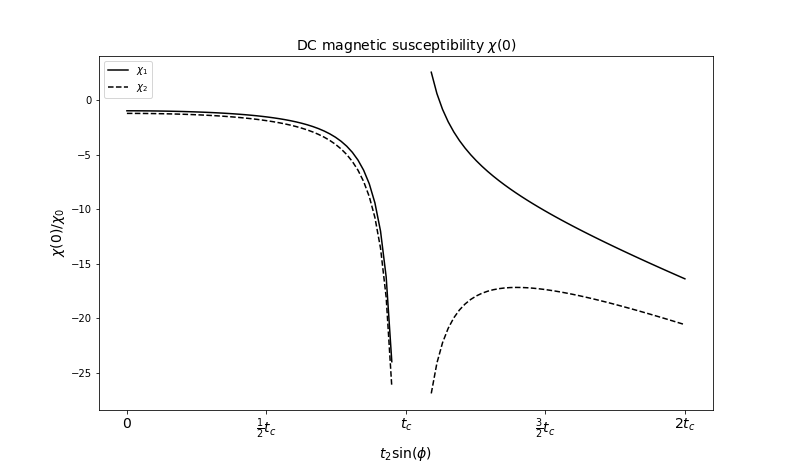}
\caption{The static magnetic susceptibility plotted as a function of the NNN imaginary hopping strength. We compare the use of two different choices for the definitions of the static magnetic susceptibility, $\chi_\text{1}$ and $\chi_\text{2}$. They are normalized by $\chi_0$, which here is the value of $|\chi_\text{1}|$ when $t_2 = 0$. }
\label{fig:staticMagSus}
\end{figure}

Here we evaluate the the static component of the magnetic susceptibility $\chi^{zz}_\text{static}$. It was previously shown that by employing various sum rules and integration by parts the static magnetic susceptibility could be expressed in multiple ways, but all equivalent \cite{DuffMagneticSusceptibility}. This was not originally appreciated and the expressions were considered different \cite{OgataMagSus2017,GaoGeometricalSus}. However, this equivalence was only rigorously proven for topologically trivial insulators. In Figure \ref{fig:staticMagSus} we plot $\chi_1 = \chi^{zz}_\text{inter}+\chi^{zz}_\text{occ}+\chi^{zz}_\text{occ2}$ which is the form identified by Ogata \cite{OgataMagSus2017}, and $\chi_2$ which is obtained by computing eq. (\ref{chiDC_ring}), in which the diagonal elements of the Berry connection have been removed by use of sum rules and integration by parts.

As we increase $t_2$, as long as we remain in the trivial phase of the Haldane model $\chi_1$ and $\chi_2$ are very nearly equal. Both diverge as we approach the critical value of $t_2 \sin(\phi)$, which indicates the change into the topological phase. This large diamagnetism right before the transition can be understood since the gap closes at either the $\textbf{K}$ or $\textbf{K}'$ point as one approaches the phase transition. When $t_2 = 0$ the diamagnetism also grows as the gap size shrinks and the Haldane model approaches a graphene model. This behavior is also independent of the sign of $t_2 \sin(\phi)$, the response is the same in the $C=\pm 1$ phases. However, once we enter the topological phase $\chi_1$ and $\chi_2$ become drastically different. An apparent flip from diamagnetic to paramagnetic behavior is predicted by $\chi_1$, however the model remains diamagnetic as predicted by $\chi_2$. This difference arises from $\chi_\text{geo}$ flipping sign across the phase transition, while $\mathring{\chi}_\text{geo}$ does not.     

This striking difference between the two numerical results that have been proven to be analytically equivalent in the topologically trivial phase indicate that the sum rules or the use of integration by parts is not valid in the topological phase. We consider this an indication that there may be a topological contribution to the DC magnetic susceptibility that is as of yet unknown. 

\subsection{Sum Rules}

In evaluating the expressions for the response tensors gauge covariant first and second derivatives appear. There are sum rules that can be employed to facilitate the computation of these derivatives, i.e. that avoid evaluating diagonal elements of the Berry connection. Since we are employing an effective tight-binding model the Hessian and \textit{Tressian} matrix corrections must be included in the sum rules. 

The gauge-covariant derivative sum rule is \cite{NonlinearOpticalSipe,ShiftCurrent,AversaSipe}

\begin{equation}\label{SingleGaugeCov}
\begin{split}
    \partial_{;a} \xi^b_{nm} = \hbar \frac{ (v^a_{nn}-v^a_{mm}) \xi^b_{nm} + (v^b_{nn}-v^b_{mm}) \xi^a_{nm} }{\Delta_{mn}} 
    \\
    + \hbar \sum_{l'} \frac{ v^b_{nl}\xi^a_{lm} - \xi^a_{nl}v^b_{lm}}{\Delta_{mn}} + i \frac{ w^{ab}_{nm}}{\Delta_{mn}},
\end{split}
\end{equation}
where $v^a_{nn}$ and $v^a_{mm}$ are diagonal matrix elements of the velocity. The sum $l'$ is restricted such that $l\neq n,m$, and $w^{ab}_{nm}$ is the Hessian matrix of the model Hamiltonian,

\begin{equation}
    w^{ab}_{nm} = \Big( u_{n\textbf{k} }\Big| \frac{\partial^2 \mathcal{H}(\textbf{k})}{\partial k_a \partial k_b} \Big|u_{m\textbf{k}}\Big).
\end{equation}
Equation (\ref{SingleGaugeCov}) can be obtained by taking double $\textbf{k}$ derivatives of the Hamiltonian matrix elements, comparing the result obtained by evaluating the matrix element then taking the derivatives, or using the product rule to apply the derivative on the bra, ket, and operator.  

The gauge-covariant double derivative sum rule, with the derivatives evaluated from right to left is

\begin{equation}\label{DoubleGaugeCov}
\begin{split}
    &\partial_{;c} \partial_{;b} \xi^a_{nm} = \frac{ig^{abc}_{nm}}{\Delta_{mn}} + \frac{(w^{ab}_{nn}-w^{ab}_{mm})\xi^c_{nm}}{\Delta_{mn}}
    \\
    &+\sum_{l'} \Bigg[\frac{\Big( w^{ab}_{nl}\xi^c_{lm} - \xi^c_{nl} w^{ab}_{lm} \Big) - \partial_{;c}\Big(\xi^b_{nl}v^a_{lm} - v^a_{nl}\xi^b_{lm}\Big) }{\Delta_{mn}}
    \Bigg]
    \\
    &-\frac{\Delta^a_{mn} \partial_{;c} \xi^b_{nm} 
    + \Delta^b_{mn} \partial_{;c} \xi^a_{nm} + \Delta^c_{mn} \partial_{;b} \xi^a_{nm}
    }{\Delta_{mn}}
    \\
    &-\frac{\Delta^{ac}_{mn}\xi^b_{nm} + \Delta^{bc}_{mn} \xi^a_{nm}}{\Delta_{mn}},
\end{split}
\end{equation}
where $g^{abc}_{nm}$ are the \textit{Tressian} matrix elements, 
\begin{equation}
\begin{split}
    g^{abc}_{nm} = \Big( u_{n\textbf{k}} \Big|  \frac{\partial^3 \mathcal{H}(\textbf{k})}{\partial k_a \partial k_b \partial k_c} \Big|u_{m\textbf{k}}\Big).
\end{split}
\end{equation}
Eq. (\ref{DoubleGaugeCov}) is obtained by taking double \textbf{k} derivatives of the velocity matrix elements, and exactly like what was done for equation (\ref{SingleGaugeCov}) comparing the result obtained by evaluating the matrix element then taking the derivatives, or using the product rule. 

These sum rules were tested by applying them to the Haldane model, the left hand side of the equation was evaluated by explicitly taking the gauge-covariant derivatives of the Berry connection, and the right side used the sum rule that only requires the dispersion and its derivatives and off-diagonal Berry connection matrix elements. Comparing the two methods we find exact agreement. Note that since this is only a two band model there is no band $l\neq n, m$ so those terms make no contribution. 

Additionally, the velocity matrix elements can be evaluated by either using equation (\ref{velocityEq}) or

\begin{equation}
\begin{split}
    v^a_{nm} = \Big( u_{n\textbf{k}} \Big| \frac{\partial \mathcal{H}(\textbf{k})}{\partial k_a} \Big| u_{m\textbf{k}} \Big),
\end{split}
\end{equation}
and likewise one can then define the off-diagonal elements of the Berry connection as 
\begin{equation}
    \xi^a_{nm} = \frac{\hbar v^a_{nm}}{i\Delta_{nm}},
\end{equation}
where one must take care at the set of $\textbf{k}$ points where band $n$ and $m$ are degenerate. This complication can be handled by choosing a gauge for which $v^a_{nm} = 0$ for these \textbf{k} points.

\section{Conclusion}\label{sec:Conclusion}

We have applied a microscopic theory of polarization and magnetization in crystals to determine the spatially dispersive contributions of the conductivity tensor that go beyond optical activity. By exploiting a kind of multipolar expansion valid in crystals we can attribute parts of the response to eight induced multipole response tensors, one of which being the generalization of the magnetic susceptibility to finite frequency. The expressions include both orbital and spin contributions to the susceptibilities, as well as itinerant contributions due to the overlap of Wannier functions localized about different sites.

The question of gauge dependence must be dealt with carefully in any multipolar expansion, and indeed the appropriate combination of tensors that describes the effective conductivity tensor is gauge-invariant. By writing expressions with diagonal elements of the Berry connection explicitly removed and gauge covariant derivatives of matrix elements an additional contribution must be added, the substitution cannot be made trivially.

Limiting to the trivial phase of the model, effectively the h-BN model we considered earlier \cite{DuffMagneticSusceptibility}, we determine the conductivity tensor component $\sigma^{xzy}_L(\omega)$ over the range of frequencies that lead to transitions between the two band model. We find the DC susceptibility approximately describes the response away from any resonances, but obtain large contributions around the \textbf{K}, \textbf{K}', and \textbf{M} points in the BZ from the other multipole tensors that contribute to the total induced current. 

We then consider the DC magnetic susceptibility of the Haldane model. It shows singular behavior about the topological phase transition, as the gap closes at the $\textbf{K}$ or $\textbf{K}'$ point. It appears that the equality between the different expressions for the DC magnetic susceptibility that we have identified in the literature does not hold in the topological phase of the model. This is perhaps not surprising as the equality relies upon sum rules and integration by parts that does not necessarily hold for Chern insulators.

The generalization of these results to Chern insulators and any ``surface/boundary effects" remains an open question. For example, if there is any global topological quantity that is related to the magnetic susceptibility -- as the Chern number is related to the anomalous Hall conductivity -- remains as of yet unknown.

\appendix

\section{Adjusted Wannier Functions}\label{AppendixA}

The unperturbed Wannier functions before any fields are applied are related to the unperturbed Bloch functions through a unitary transformation and Fourier transform, eq. (\ref{Wannier}). A common approach to including the effects of a magnetic field is to multiply the Wannier functions by a generalized Peierls phase to obtain a modified Wannier function, 

\begin{equation}
\begin{split}
    W'_{\alpha\textbf{R}}(\textbf{x},t) = e^{i\Phi(\textbf{x},\textbf{R};t)} W_{\alpha\textbf{R}}(\textbf{x}),
\end{split}
\end{equation}
where the phase factor is given in equation (\ref{phiphase}). However, these are not the adjusted Wannier functions that serve as the basis for expanding site quantities. In the presence of a non-uniform vector potential the $W'_{\alpha\textbf{R}}(\textbf{x},t)$ are in general not orthogonal. Therefore we use Lowdin's method of symmetric orthogonalization to obtain the adjusted Wannier functions from the $W'_{\alpha\textbf{R}}(\textbf{x},t)$ \cite{Lowdin}, these can be expressed as

\begin{equation}
    \bar{W}_{\alpha\textbf{R}}(\textbf{x},t) = e^{i\Phi(\textbf{x},\textbf{R};t)} \chi_{\alpha\textbf{R}}(\textbf{x},t),
\end{equation}
where the functions $\chi_{\alpha\textbf{R}}(\textbf{x},t)$ are expanded in a power series of the electromagnetic field. The first two terms are given by
\begin{equation}
\label{ChiExpansion}
\begin{split}
    \chi_{\alpha\textbf{R}}(\textbf{x},t) =& W_{\alpha\textbf{R}}(\textbf{x}) 
    -\frac{i}{2} \sum_{\beta\textbf{R}'} W_{\beta\textbf{R}'}(\textbf{x}) 
    \\
    &\times \int W^*_{\beta\textbf{R}'}(\textbf{x}) \Delta(\textbf{R}',\textbf{x},\textbf{R};t) W_{\alpha\textbf{R}}(\textbf{x},t),
\end{split}
\end{equation}
where $\Delta(\textbf{R}',\textbf{x},\textbf{R};t) = \Phi(\textbf{R}',\textbf{x};t)+\Phi(\textbf{x},\textbf{R};t)+\Phi(\textbf{R},\textbf{R}';t)$, so is a closed line integral of the vector potential - by Stokes theorem this can then be expressed as the magnetic flux passing through a surface bounded by the path $\textbf{R}'\rightarrow \textbf{R} \rightarrow \textbf{x} \rightarrow \textbf{R}'$. Since we are only concerned with the linear response in this manuscript it is sufficient to truncate the series after the above two terms in equation (\ref{ChiExpansion}).  

\begin{widetext}
\section{Gauge Invariance of Ground State Magnetization}\label{ExtraAppendix}

The magnetization is split into an atomic, itinerant, and spin contribution. While the ground state spin contribution is gauge-invariant this is not so for the ground state atomic and itinerant contributions. They depend on diagonal elements of the Berry connection and the $\mathcal{W}$ matrix elements. The ground state atomic magnetization written in the Bloch basis is obtained by using eq. (\ref{Wannier}) for the relation between the Wannier and Bloch basis in eq. (\ref{AtomicMagWannier})

\begin{equation}
\label{AtomicMagBloch}
\begin{split}
    \bar{M}^i = \epsilon^{iab} \frac{e}{4c} \sum_{mn} f_{n} \int_{BZ} \frac{d\textbf{k}}{(2\pi)^3} \Bigg[ 
    (\xi^a_{nm}+\mathcal{W}^a_{nm}) v^b_{mn} 
    + v^b_{nm} (\xi^a_{mn} + \mathcal{W}^a_{mn}) 
    \Bigg],
\end{split}
\end{equation}
and in eq. (\ref{ItinerantMagWannier}) for the itinerant contribution
\begin{equation}
\label{ItinerantMagBloch}
\begin{split}
    \tilde{M}^i = \epsilon^{iab} \frac{e}{4\hbar c} \sum_{mn} f_{n} \int_{BZ} \frac{d\textbf{k}}{(2\pi)^3} \Bigg[ 
    2\partial_b E_{n\textbf{k}} (\xi^a_{nn}+\mathcal{W}^a_{nn})
    +
    i(E_{m\textbf{k}}-E_{n\textbf{k}}) \Big( (\xi^b_{nm}+\mathcal{W}^b_{nm})\mathcal{W}^a_{mn}
    - \mathcal{W}^a_{nm}(\xi^b_{mn}+\mathcal{W}^b_{mn}) \Big) 
    \Bigg].
\end{split}
\end{equation}

Combining eq. (\ref{AtomicMagBloch}) and eq. (\ref{ItinerantMagBloch}) we find for the total orbital magnetization
\begin{equation}
\label{TotalMag_Pre}
\begin{split}
    \bar{M}^i + \tilde{M}^i = \epsilon^{iab} \frac{e}{2\hbar c} \sum_{mn} f_{n} \int_{BZ} \frac{d\textbf{k}}{(2\pi)^3} \Bigg[ &
    2 \partial_b E_{n\textbf{k}} \xi^a_{nn} 
    +i(E_{m\textbf{k}}-E_{n\textbf{k}}) \xi^a_{nm}\xi^b_{mn}
    \\
    &+ 2 \partial_b E_{n\textbf{k}} \mathcal{W}^a_{nn}
    +i(E_{n\textbf{k}}-E_{m\textbf{k}}) \mathcal{W}^a_{nm}\mathcal{W}^b_{mn}
    \Bigg].
\end{split}
\end{equation}

To then get eq. (\ref{TotalMag_Pre}) to match that obtained from the ``modern theory", eq. (\ref{GroundStateMag}), requires integration by parts to obtain contributions that have $\partial_b \xi^a_{nn}$ or $\partial_b \mathcal{W}^a_{nn}$. We then use the sum rules $\epsilon^{lab} \partial_b \xi^a_{nn} = i\epsilon^{lab} \sum_{m} \xi^b_{nm}\xi^a_{mn}$ and $\epsilon^{lab} \partial_b \mathcal{W}^a_{nn} = i\epsilon^{lab} \sum_{m} \mathcal{W}^a_{nm}\mathcal{W}^b_{mn}$ to process this expression to get
\begin{equation}
\begin{split}
    \bar{M}^i + \tilde{M}^i = \epsilon^{iab} \frac{ie}{2\hbar c} \sum_{mn} f_{n} \int_{BZ} \frac{d\textbf{k}}{(2\pi)^3} \Bigg[ 
    (E_{m\textbf{k}}+E_{n\textbf{k}}) \xi^a_{nm}\xi^b_{mn}
    -(E_{m\textbf{k}}+E_{n\textbf{k}}) \mathcal{W}^a_{nm}\mathcal{W}^b_{mn}
    \Bigg].
\end{split}
\end{equation}
\end{widetext} 
To then obtain the final expression, eq. (\ref{GroundStateMag}), requires performing a relabelling of indices and using the property that for a topologically trivial insulator $f_{nm} \mathcal{W}^a_{nm} = 0$. Thus the $\mathcal{W}$ dependence has vanished, and in this form it is clear that there is no contribution from the sum when $n=m$ due to the $\epsilon^{lab}$. 

\begin{widetext}

\section{Combination of Multipole Response Tensors in Coordinate Space}\label{AppendixB}

The current density depends on time and spatial derivatives of the polarization and magnetization fields respectively. Thus the induced current can be expressed in terms of various induced multipole contributions. The $\mathcal{O}(q^2)$  contribution to the effective conductivity tensor written in real space can be separated into an interaction with a spatially varying magnetic field ($\sigma^{ilj}_L(\omega)$) and with the symmetric double derivatives of the electric field ($\sigma^{iljk}_K(\omega)$). To form these response tensors from the constituent multipole contributions it is instructive to first see how a single derivative of the electric field can be expanded into the magnetic field $B^l(\textbf{x},\omega)$ and symmetric double derivatives of the electric field $F^{ij}(\textbf{x},\omega)$,
\begin{equation}
\label{B1}
\begin{split}
    \frac{\partial E^i(\textbf{x},\omega)}{\partial x^j} = \frac{1}{2}\Big( \frac{\partial E^i(\textbf{x},\omega)}{\partial x^j} + \frac{\partial E^j(\textbf{x},\omega)}{\partial x^i} \Big)
    + \frac{1}{2} \Big( \frac{\partial E^i(\textbf{x},\omega)}{\partial x^j} - \frac{\partial E^j(\textbf{x},\omega)}{\partial x^i} \Big)
    \\
    = F^{ij}(\textbf{x},\omega) +\frac{i\omega}{2c} \epsilon^{ilj} B^l(\textbf{x},\omega),
\end{split}
\end{equation}
where $F^{ij}(\textbf{x},\omega)$ has been defined in eq. (\ref{F-field}), and to identify the magnetic field we have used Faraday's law, eq. (\ref{FaradaysLaw}). 

Next, it is instructive to examine how a derivative of $F^{ij}(\textbf{x},\omega)$ can be expressed as symmetric double derivatives of the electric field $K^{ijk}(\textbf{x},\omega)$ and derivatives of the magnetic field $L^{jl}(\textbf{x},\omega)$,

\begin{equation}
\label{B2}
\begin{split}
    \frac{\partial F^{ij}(\textbf{x},\omega)}{\partial x^k} =& \frac{1}{3} \Big( \frac{\partial^2 E^i(\textbf{x},\omega)}{\partial x^j \partial x^k} + \frac{\partial^2 E^j(\textbf{x},\omega)}{\partial x^k \partial x^i} + \frac{\partial^2 E^k(\textbf{x},\omega)}{\partial x^i \partial x^j}\Big) 
    \\
    +&\frac{1}{6} \frac{\partial}{\partial x^j} \Big( \frac{\partial E^i(\textbf{x},\omega)}{\partial x^k} - \frac{\partial E^k(\textbf{x},\omega)}{\partial x^i}\Big)
    +\frac{1}{6} \frac{\partial}{\partial x^i} \Big( \frac{\partial E^j(\textbf{x},\omega)}{\partial x^k} - \frac{\partial E^k(\textbf{x},\omega)}{\partial x^j}\Big)
    \\
    =& K^{ijk}(\textbf{x},\omega) + \frac{i\omega}{6c} \epsilon^{ilk} L^{jl}(\textbf{x},\omega) + \frac{i\omega}{6c} \epsilon^{jlk} L^{il}(\textbf{x},\omega).
\end{split}
\end{equation}
Where $K^{ijk}(\textbf{x},\omega)$ and $L^{jl}(\textbf{x},\omega)$ have been defined in equations (\ref{K-field}) and (\ref{L-field}). Lastly, using equations (\ref{B1}) and (\ref{B2}) it follows that the double derivatives of the electric field can be written in terms of $K^{jlk}(\textbf{x},\omega)$ and ${L}^{jl}(\textbf{x},\omega)$ as
\begin{equation}
\label{B3}
\begin{split}
    \frac{\partial^2 E^k(\textbf{x},\omega)}{\partial x^j \partial x^l} &= \frac{1}{2} \Big( \frac{\partial}{\partial x^j} \frac{\partial E^k(\textbf{x},\omega)}{\partial x^l} + \frac{\partial}{\partial x^l} \frac{\partial E^k(\textbf{x},\omega)}{\partial x^j} \Big)
    \\
    &= \frac{1}{2} \frac{\partial}{\partial x^j} \Big( F^{lk}(\textbf{x},\omega) + \frac{i\omega}{2c} \epsilon^{alk} B^a(\textbf{x},\omega) \Big) + \frac{1}{2} \frac{\partial}{\partial x^l} \Big( F^{jk}(\textbf{x},\omega) + \frac{i\omega}{2c} \epsilon^{ajk} B^a(\textbf{x},\omega) \Big)
    \\
    &= K^{jlk}(\textbf{x},\omega) +\frac{i\omega}{12c} \Big( \epsilon^{ajl}\frac{\partial B^a(\textbf{x},\omega)}{\partial x^k} + \epsilon^{ajk} \frac{\partial B^a(\textbf{x},\omega)}{\partial x^l}\Big) + \frac{i\omega}{12c} \Big(\epsilon^{alj} \frac{\partial B^a(\textbf{x},\omega)}{\partial x^k} +  \epsilon^{alk} \frac{\partial B^a(\textbf{x},\omega)}{\partial x^j}\Big)
    \\
    & \hspace{10pt} 
    + \frac{i\omega}{4c} \epsilon^{alk} \frac{\partial B^a(\textbf{x},\omega)}{\partial x^j} 
    + \frac{i\omega}{4c} \epsilon^{ajk} \frac{\partial B^a(\textbf{x},\omega)}{\partial x^l} 
    \\
    &= K^{jlk}(\textbf{x},\omega) + \frac{i\omega}{3c} \epsilon^{alk} L^{ja}(\textbf{x},\omega) + \frac{i\omega}{3c} \epsilon^{ajk}L^{la}(\textbf{x},\omega)  .
\end{split}
\end{equation}

With equations (\ref{B1}), (\ref{B2}), and (\ref{B3}) in hand, when we plug equations (\ref{PolarizationExpansion}) and (\ref{MagnetizationExpansion}) into the second line of eq. (\ref{MacroJ}) we find
\begin{equation}
\label{B4}
\begin{split}
    J^{i}(\textbf{x},\omega) = &-i\omega P^i(\textbf{x},\omega) + c\epsilon^{ijk} \frac{\partial M^k(\textbf{x},\omega)}{\partial x^j}
    \\
    &=... + \Bigg( i\omega \Big(  \Gamma^{ijl}(\omega) -\Lambda^{ijl}(\omega) \Big) 
    +\frac{\omega^2}{3c} \epsilon^{lab} \Big( \Sigma^{ibaj}(\omega)
    +2\Omega^{ijab}  \Big) \Bigg) L^{jl}(\textbf{x},\omega)
    \\
    &+\Bigg( c\epsilon^{ijk} \Big(\chi^{kl}_{\mathcal{M}}
    -\frac{i\omega}{2c}\epsilon^{lab} \beta^{kab}_{\mathcal{M}}(\omega)\Big)
    +  \frac{i\omega}{3}\epsilon^{iak} \epsilon^{lab}\Big(\gamma^{kjb}_{\mathcal{M}}(\omega) - \frac{1}{2}\Big( \beta^{kjb}_{\mathcal{M}}(\omega) + \beta^{kbj}_{\mathcal{M}}(\omega) \Big)\Bigg)L^{jl}(\textbf{x},\omega) 
    \\
    &+i\omega\Bigg(
    \Sigma^{ijlk}(\omega)-\Pi^{ijlk}(\omega)-\Omega^{ijlk}(\omega) 
    \Bigg) K^{jlk}(\textbf{x},\omega)
    + c\epsilon^{ijb} \Big(\gamma^{blk}_{\mathcal{M}}
    -\beta^{blk}_{\mathcal{M}}(\omega) \Big) K^{jlk}(\textbf{x},\omega),
\end{split}
\end{equation}
where `...' are the long wavelength and $\mathcal{O}(q)$ contributions to the induced current. $\sigma^{ilj}_L(\omega)$ and $\sigma^{iljk}_K(\omega)$ can then be read off from equation (\ref{B4}). Note that since $\sigma^{iljk}_K(\omega)$ is contracted with the totally symmetric $K^{jlk}(\textbf{x},\omega)$ we enforce the symmetry by summing over all six permutations explicitly in equation (\ref{q2conductivityK}).

\section{Gauge-invariant Total Conductivity Tensors}\label{Appendix:GaugeInvariantTensors}

In writing the expressions in a gauge-invariant manner with the $\mathring{}$ accent we explicitly remove diagonal elements of the Berry connection and use gauge-covariant derivatives. After making these changes the multipole response tensors are written as

\begin{equation}
\begin{split}
    \mathring{\chi}^{kl}_{\mathcal{M}} = \mathring{\chi}^{kl}_\text{static} &+ \sum_{mn} f_{nm} \int_{BZ} \frac{d\textbf{k}}{(2\pi)^3} \frac{\hbar\omega}{\Delta_{mn}} \frac{\mathring{M}^k_{nm} }{\Delta_{mn}-\hbar(\omega+i0^+)}\Bigg[ 
    \mathring{M}^l_{mn} - \frac{e\omega}{4c} \epsilon^{lab} \frac{\partial_a (E_{m\textbf{k}}+E_{n\textbf{k}}) \xi^b_{mn}}{\Delta_{mn}-\hbar(\omega+i0^+)}
    \Bigg]
    \\
    &+\frac{\omega e^2}{16\hbar c^2} \sum_{nm} f_{n} \int_{BZ} \frac{d\textbf{k}}{(2\pi)^3} \Big( \mathring{\Omega}^k_{nm} \mathring{\Omega}^l_{mn} - \mathring{\Omega}^l_{nm} \mathring{\Omega}^k_{mn} \Big) 
    \\
    &-\frac{ie^2 \omega}{16 c^2} \epsilon^{kcd} \epsilon^{lab} \sum_{mn} f_{nm} \int_{BZ} \frac{d\textbf{k}}{(2\pi)^3} \frac{ \partial_{;a} \partial_{;c} v^d_{nm} \xi^b_{mn}}{\Delta_{mn}-\hbar\tilde\omega},
\end{split}
\end{equation}
where $\mathring{\chi}^{kl}_\text{static}$ is found in a previous manuscript \cite{DuffMagneticSusceptibility}. The `purified off-diagonal magnetization matrix elements' are
\begin{equation}
\begin{split}
    \mathring{M}^l_{mn} = \frac{e}{4c} \epsilon^{lab} \Bigg[ \sum_{l\neq m,n} \Big( \xi^a_{ml}v^b_{ln} + v^b_{ml}\xi^a_{ln} \Big)  + \frac{2}{\hbar} \partial_b(E_{n\textbf{k}}+E_{m\textbf{k}}) \xi^a_{mn} \Bigg] + \frac{e}{mc} S^l_{mn}.
\end{split}
\end{equation}
The non-Abelian Berry curvature must instead use the gauge covariant derivative
\begin{equation}
    \mathring{\Omega}^k_{nm} = \epsilon^{lab} \partial_{;a} \xi^b_{nm} = i\sum_{l\neq n,m} \xi^a_{nl}\xi^b_{lm},
\end{equation}
and lastly double gauge-covariant derivatives of the off-diagonal velocity matrix elements are
\begin{equation}
\begin{split}
    \partial_{;a}\partial_{;c} v^d_{nm} = i(E_{n\textbf{k}}-E_{m\textbf{k}}) \partial_{;a} \partial_{;c} \xi^d_{nm} 
    + i\partial_c (E_{n\textbf{k}}-E_{m\textbf{k}}) \partial_{;a} \xi^d_{nm}
    + i\partial_a (E_{n\textbf{k}}-E_{m\textbf{k}}) \partial_{;c} \xi^d_{nm}
    \\
    + i\partial_a \partial_c (E_{n\textbf{k}}-E_{m\textbf{k}}) \xi^d_{nm},
\end{split}
\end{equation}
where the single and double gauge-covariant derivatives of the Berry connection are defined in eq. (\ref{gaugecovariantderivative_eq}) and eq. (\ref{gaugecovariantdoublederivative_eq}) respectively. 

The `purified' magnetization response to F is
\begin{equation}
\begin{split}
    \mathring{\gamma}^{ijl}_\mathcal{M}(\omega) 
    =&\sum_{mn} f_{nm} \int_{BZ} \frac{d\textbf{k}}{(2\pi)^3} \frac{\mathring{M}^i_{nm} }{\Delta_{mn}-\hbar\tilde\omega}\Bigg[ \mathring{\mathcal{Q}}^{jl}_{\mathcal{P}:mn} 
    + \frac{ie}{4} \frac{\partial_j (E_{n\textbf{k}}+E_{m\textbf{k}}) \xi^l_{mn} + \partial_l (E_{m\textbf{k}}+E_{n\textbf{k}}) \xi^j_{mn} }{\Delta_{mn}-\hbar(\omega+i0^+)}
    \Bigg]
    \\
    &-\frac{e}{2\hbar c} \sum_{nm} f_{n} \text{Re}\Bigg[ 
    \int_{BZ} \frac{d\textbf{k}}{(2\pi)^3} \mathring{\Omega}^i_{nm} \Bigg( 
    \mathring{\mathcal{Q}}^{jl}_{\mathcal{P}:mn} + \frac{ie}{4} \Bigg( \partial_{;j} \xi^l_{mn} + \partial_{;l} \xi^j_{mn} \Bigg) 
    \Bigg)
    \Bigg]
    \\
    &-\frac{e^2}{16 c} \epsilon^{iab} \sum_{mn} f_{nm} \int_{BZ} \frac{d\textbf{k}}{(2\pi)^3} \frac{ \partial_{;j}\partial_{;a} v^b_{nm} \xi^l_{mn} + \partial_{;l}\partial_{;a} v^b_{nm}\xi^j_{mn}  }{\Delta_{mn}-\hbar\tilde\omega},
\end{split}
\end{equation}
where 
\begin{equation}
    \mathring{\mathcal{Q}}^{jl}_{\mathcal{P}:mn} = \frac{e}{4} \sum_{l\neq m,n} \Big( \xi^j_{ml}\xi^l_{ln} + \xi^l_{ml}\xi^j_{ln} \Big). 
\end{equation}

The `purified' magnetic quadrupolarization response to E is 

\begin{equation}
\begin{split}
    \mathring{\beta}^{ijl}_{\mathcal{M}}(\omega) = &e\sum_{mn} f_{nm} \int_{BZ} \frac{d\textbf{k}}{(2\pi)^3} \frac{ \mathring{\mathcal{Q}}^{ij}_{\mathcal{M}:nm} \xi^l_{mn}}{\Delta_{mn}(\textbf{k})-\hbar(\omega+i0^+)} 
    \\
    &-\frac{e^2}{24\hbar c} \epsilon^{iab} \sum_{mn} f_{nm} \int_{BZ} \frac{d\textbf{k}}{(2\pi)^3} \frac{ \xi^l_{mn}}{\Delta_{mn}(\textbf{k})-\hbar(\omega+i0^+)}  \Bigg[ 
    \partial_a (E_{n\textbf{k}}+E_{m\textbf{k}}) \sum_{l\neq n,m} \Big(\xi^b_{nl}\xi^j_{lm} + \xi^j_{nl}\xi^b_{lm}\Big)
    \Bigg] 
    \\
    &-\frac{e^2}{8 c} \epsilon^{iab} \sum_{mn} f_{nm} \int_{BZ} \frac{d\textbf{k}}{(2\pi)^3} \frac{ \xi^l_{mn}}{\Delta_{mn}(\textbf{k})-\hbar(\omega+i0^+)} \Bigg[ 
    \partial_{;j} \partial_{;a} v^b_{nm} + \frac{i}{3\hbar} (E_{m\textbf{k}}-E_{n\textbf{k}}) \partial_{;j}\partial_{;a} \xi^b_{nm} 
    \Bigg]
    \\
    &+\frac{e^2}{6\hbar c} \sum_{n,m\neq n} f_{n} \int_{BZ} \frac{d\textbf{k}}{(2\pi)^3} \text{Re}\Bigg[ 
    2i\partial_{;j} \xi^l_{nm} \mathring{\Omega}^i_{mn}
    + \partial_{;a} \xi^l_{nm} \sum_{l\neq m,n} \xi^j_{ml}\xi^b_{ln} 
    \Bigg],
\end{split}
\end{equation}
where 
\begin{equation}
\begin{split}
    \mathring{Q}^{ij}_{\mathcal{M}:nm} = &\frac{e}{12 c} \epsilon^{iab} \sum_{l's'} \Bigg[ \xi^j_{nl} \Big(\xi^a_{ls}v^b_{sm} + v^b_{ls}\xi^a_{sm} \Big) + \Big(\xi^a_{ns}v^b_{sl} + v^b_{ns}\xi^a_{sl} \Big) \xi^j_{lm} \Bigg] 
    \\
    &+ \frac{e}{12 \hbar c} \epsilon^{iab} \sum_{l'} \Bigg[ 
    \partial_b (E_{m\textbf{k}}+E_{l\textbf{k}}) \xi^j_{nl}\xi^a_{lm} 
    + \partial_b (E_{n\textbf{k}}+E_{l\textbf{k}}) \xi^a_{nl}\xi^j_{lm} 
    \Bigg]
    \\
    &+\frac{ie}{12 c} \epsilon^{iab} \Bigg[ 
    \sum_{l'} \Big( v^b_{nl} \partial_{;a} \xi^j_{lm} - \partial_{;a} \xi^j_{nl}v^b_{lm} \Big) 
    + \frac{1}{\hbar} \partial_b (E_{n\textbf{k}}-E_{m\textbf{k}}) \partial_{;a} \xi^j_{nm}
    + \frac{1}{\hbar} \partial_j \partial_a (E_{n\textbf{k}}-E_{m\textbf{k}}) \xi^b_{nm}
    \Bigg]
    \\
    &+\frac{e}{2mc} \sum_{l'} \Big( S^i_{nl}\xi^j_{lm} + \xi^j_{nl} S^i_{lm} \Big).
\end{split}
\end{equation}

The `purified' electric dipole response to derivatives of the magnetic field $\textbf{L}$ is

\begin{equation}
\begin{split}
    \mathring{\Lambda}^{ijl}(\omega) = &e\sum_{mn} f_{nm} \int_{BZ} \frac{d\textbf{k}}{(2\pi)^3} \frac{ \xi^i_{nm} \mathring{\mathcal{Q}}^{lj}_{\mathcal{M}:mn}}{\Delta_{mn}(\textbf{k})-\hbar(\omega+i0^+)}
    \\
    &+\frac{ie^2}{12 c} \epsilon^{lab} \sum_{mn} f_{nm} \int_{BZ} \frac{d\textbf{k}}{(2\pi)^3} \xi^i_{nm} \Bigg[  \sum_{l'} \Big( \xi^a_{ml}v^b_{ln} + v^b_{ml}\xi^a_{ln} \Big) + \frac{2}{\hbar} \partial_b (E_{n\textbf{k}}+E_{m\textbf{k}}) \xi^a_{mn} \Bigg] \frac{ \partial_j (E_{n\textbf{k}}+E_{m\textbf{k}})}{(\Delta_{mn}(\textbf{k})-\hbar(\omega+i0^+))^2}
    \\
    &
    +\frac{ie^2}{12 c} \epsilon^{lab} \sum_{mn} f_{nm} \int_{BZ} \frac{d\textbf{k}}{(2\pi)^3} \xi^i_{nm} \Bigg[\sum_{l'} \Big( \xi^j_{ml}v^b_{ln} + v^b_{ml}\xi^j_{ln} \Big)
    -\frac{1}{\hbar} \partial_j (E_{n\textbf{k}}+E_{m\textbf{k}}) \xi^b_{mn}
    \Bigg] \frac{ \partial_a (E_{n\textbf{k}}+E_{m\textbf{k}})}{(\Delta_{mn}(\textbf{k})-\hbar(\omega+i0^+))^2} 
    \\
    &+\frac{ie^2}{2mc} \sum_{mn} f_{nm} \int_{BZ} \frac{d\textbf{k}}{(2\pi)^3} \xi^i_{nm} S^l_{mn} \frac{ \partial_j (E_{n\textbf{k}}+E_{m\textbf{k}})}{(\Delta_{mn}(\textbf{k})-\hbar(\omega+i0^+))^2}
    \\
    &-\frac{i\omega e^2}{12 c} \epsilon^{lab} \sum_{mn} f_{nm} \int_{BZ} \frac{d\textbf{k}}{(2\pi)^3} \frac{ \partial_{;j} \partial_{;a} \xi^i_{nm} \xi^b_{mn}}{\Delta_{mn}(\textbf{k})-\hbar(\omega+i0^+)}
    \\  
    &-\frac{i\omega e^2}{12 c} \epsilon^{lab} \sum_{mn} f_{nm} \int_{BZ} \frac{d\textbf{k}}{(2\pi)^3} \xi^i_{nm} \xi^b_{mn} \Bigg[ 
    \frac{\partial_j \partial_a (E_{n\textbf{k}}-E_{m\textbf{k}})}{(\Delta_{mn}(\textbf{k})-\hbar(\omega+i0^+))^2} + 2 \frac{ \partial_j (E_{m\textbf{k}}+E_{n\textbf{k}}) \partial_a (E_{m\textbf{k}}+E_{n\textbf{k}}) }{(\Delta_{mn}(\textbf{k})-\hbar(\omega+i0^+))^3} 
    \Bigg]
    \\
    &+\frac{e^2}{12\hbar c} \epsilon^{lab} \sum_{n,m\neq n} f_{n} \text{Re}\Bigg[ 
    \int_{BZ} \frac{d\textbf{k}}{(2\pi)^3} \partial_{;a} \xi^i_{nm}\sum_{l'} \Big( \xi^j_{ml}\xi^b_{ln} + \xi^b_{ml}\xi^j_{ln}\Big)  + 2i \partial_{;j}\xi^i_{nm} \mathring{\Omega}^l_{mn} 
    \Bigg].
\end{split}
\end{equation}

The `purified' electric dipole response to symmetric double derivatives of the electric field $\textbf{K}$ is

\begin{equation}
\label{Pi_Response_GI}
\begin{split}
    \mathring{\Pi}^{ijlk}(\omega) &= \frac{e}{6}\sum_{mn} f_{nm} \sum_{ \{jlk\} }\int_{BZ} \frac{d\textbf{k}}{(2\pi)^3} \frac{\xi^i_{nm}}{\Delta_{mn}(\textbf{k})-\hbar(\omega+i0^+)} \Bigg[ 
    \mathring{\mathcal{O}}^{jlk}_{\mathcal{P}:mn}
    + \frac{i}{2} \frac{ \partial_j (E_{n\textbf{k}}+E_{m\textbf{k}}) \mathring{\mathcal{Q}}^{lk}_{\mathcal{P}:mn}}{\Delta_{mn}-\hbar(\omega+i0^+)}
    \\
    &+ \frac{e}{4} \frac{\partial_j \partial_l (E_{m\textbf{k}}-E_{n\textbf{k}}) \xi^k_{mn} }{\Delta_{mn}-\hbar(\omega+i0^+)}
    + \frac{e}{4} \frac{\partial_j (E_{m\textbf{k}}-E_{n\textbf{k}}) \partial_l \xi^k_{mn}}{\Delta_{mn}-\hbar(\omega+i0^+)}
    - \frac{e}{2} \frac{\partial_j E_{n\textbf{k}} \partial_l E_{n\textbf{k}} + \partial_j E_{m\textbf{k}} \partial_l E_{m\textbf{k}} }{ (\Delta_{mn}-\hbar(\omega+i0^+))^2 }
    \Bigg],
\end{split}
\end{equation}
where
\begin{equation}
\begin{split}
    \mathring{\mathcal{O}}^{jlk}_{\mathcal{P}:mn} = \frac{e}{36} \sum_{ \{jlk\} } \Bigg[ \sum_{s'} \Bigg( 
    \sum_{l'} \xi^i_{nl}\xi^j_{ls}\xi^l_{sm}
    +\frac{i}{2} \Big( \xi^i_{ns} \partial_{;j} \xi^l_{sm} - \partial_{;j} \xi^i_{ns}\xi^l_{sm} \Big) 
    \Bigg) 
    - \partial_{;j}\partial_{;l} \xi^k_{mn}\Bigg]
    .
\end{split}
\end{equation}

The `purified' quadrupolarization response to symmetric derivatives of the electric field (\textbf{F}) is

\begin{equation}
\begin{split}
    \mathring{\Sigma}^{ijlk}(\omega) = &\sum_{mn} f_{nm} \int_{BZ} \frac{d\textbf{k}}{(2\pi)^3} \frac{ \mathring{\mathcal{Q}}^{ij}_{\mathcal{P}:nm}  }{\Delta_{mn}(\textbf{k})-\hbar(\omega+i0^+)} \Bigg[ 
    \mathring{\mathcal{Q}}^{kl}_{\mathcal{P}:mn}
    + \frac{ie}{4}\frac{\partial_k (E_{n\textbf{k}}+E_{m\textbf{k}}) \xi^l_{mn} + \partial_l (E_{n\textbf{k}}+E_{m\textbf{k}}) \xi^k_{mn}}{\Delta_{mn}(\textbf{k})-\hbar(\omega+i0^+)}
    \Bigg] 
    \\
    &-\frac{e^2}{16} \sum_{mn} f_{nm} \int_{BZ} \frac{d\textbf{k}}{(2\pi)^3} \frac{\partial_{;l}\Big( \partial_{;i} \xi^j_{nm} + \partial_{;j} \xi^i_{nm} \Big)   \xi^k_{mn} + \partial_{;k} \Big( \partial_{;i} \xi^j_{nm} + \partial_{;j} \xi^i_{nm}\Big)\xi^l_{mn}}{\Delta_{mn}(\textbf{k})-\hbar(\omega+i0^+)} .
\end{split}
\end{equation}

The `purified' quadrupolarization response to the magnetic field (\textbf{B}) is
\begin{equation}
\begin{split}
    \mathring{\Gamma}^{ijl}(\omega) = &\sum_{mn} f_{nm} \int_{BZ} \frac{d\textbf{k}}{(2\pi)^3} \frac{ \mathring{\mathcal{Q}}^{ij}_{\mathcal{P}:nm} }{\Delta_{mn}(\textbf{k})-\hbar(\omega+i0^+)} \Bigg[ 
    \mathring{M}^l_{mn}-\frac{e\omega}{4c} \epsilon^{lab} \frac{ \partial_a (E_{n\textbf{k}}+E_{m\textbf{k}})\xi^b_{mn} }{\Delta_{mn}(\textbf{k})-\hbar(\omega+i0^+)}
    \Bigg]
    \\
    &-\frac{i\omega e^2}{16 c} \epsilon^{lab} \sum_{mn} f_{nm} \int_{BZ} \frac{d\textbf{k}}{(2\pi)^3} \frac{ \partial_{;a} \Big(  \partial_{;j} \xi^i_{nm} + \partial_{;i} \xi^j_{nm}  \Big) \xi^b_{mn}}{\Delta_{mn}(\textbf{k})-\hbar(\omega+i0^+)}
    \\
    &-\frac{e}{2\hbar c} \sum_{n, m\neq n} f_{n} \int_{BZ} \frac{d\textbf{k}}{(2\pi)^3} \text{Re}\Bigg[ \mathring{\Omega}^l_{nm} \Big( \mathring{\mathcal{Q}}^{ij}_{\mathcal{P}:mn} + \frac{ie}{4} \Big(\partial_{;i} \xi^j_{mn} + \partial_{;j} \xi^i_{mn}\Big) \Big) \Bigg] .
\end{split}
\end{equation}

Lastly, the `purified' octupolarization response to the electric field (\textbf{E}) is

\begin{equation}
\begin{split}
    \mathring{\Omega}^{ijlk}(\omega) = e \sum_{mn} f_{nm} \int_{BZ} \frac{d\textbf{k}}{(2\pi)^3} \frac{ \mathring{\mathcal{O}}^{ijl}_{\mathcal{P}:nm}\xi^k_{mn}}{\Delta_{mn}(\textbf{k})-\hbar(\omega+i0^+)} .
\end{split}
\end{equation}

\end{widetext}

\section{Relator Expansion - Spatially Varying Fields}\label{AppendixC}

The relators $s^i(\textbf{w};\textbf{x},\textbf{y})$ and $\alpha^{jk}(\textbf{w};\textbf{x},\textbf{y})$ have been defined in eq. (\ref{s-relator}) and eq. (\ref{alpha-relator}) respectively in the main text. Some of the properties of the relators are \textit{path independent}, they only depend on the endpoints of the integration, however when looking at the multipole expansion it is helpful to make a particular choice of path to simplify the expansion. Here we consider the straight line path parametrized by $u$, 
\begin{equation}
    z^i(u) = y^i + u(x^i-y^i),
\end{equation}
where $u$ ranges from 0 to 1, $\textbf{z}(0) = \textbf{y}$, and $\textbf{z}(1) = \textbf{x}$. We can then write the relators as an integration over the parameter $u$
\begin{equation}
\begin{split}
    s^i(\textbf{w};\textbf{x},\textbf{y}) = \int_0^{1} du (x^i-y^i) \delta(\textbf{w}-\textbf{y} - u(\textbf{x}-\textbf{y})),
\end{split}
\end{equation}
and
\begin{equation}
\begin{split}
    \alpha^{jk}(\textbf{w};\textbf{x},\textbf{y}) = \epsilon^{jmk} \int_0^1 du (x^m-y^m) u \delta(\textbf{w}-\textbf{y}-u(\textbf{x}-\textbf{y})).
\end{split}
\end{equation}

Two quantities that depend on these relators and the electromagnetic fields are
\begin{equation}\label{OmegaR}
    \Omega^j_\textbf{R}(\textbf{x},t) = \int \alpha^{lj}(\textbf{w};\textbf{x},\textbf{R}) B^l(\textbf{w},t) d\textbf{w},
\end{equation}
\begin{equation}\label{Omega0}
    \Omega^{0}_\textbf{R}(\textbf{x},t) = \int s^i(\textbf{w};\textbf{x},\textbf{R}) E^i(\textbf{w},t) d\textbf{w},
\end{equation}
as well as $\Phi(\textbf{R},\textbf{R}';t)$, given in the main text at eq. (\ref{phiphase}), and $\Delta(\textbf{R}',\textbf{x},\textbf{R};t)$. These quantities appear when considering the perturbed Hamiltonian, see earlier work \cite{Notes_on_EMHamiltonian} as well as Appendix \ref{AppendixD} of the current manuscript to see how they appear when treating the linear response. 

Considering nearly uniform fields we perform a Taylor expansion about the lattice point \textbf{R} of the magnetic field up to first order and the electric field up to second order. Then equation (\ref{OmegaR}) is written
\begin{equation}
\begin{split}
    \Omega^{b}_\textbf{R}(\textbf{x},t) \approx &B^l(\textbf{R},t) \int d\textbf{w} \alpha^{lb}(\textbf{w};\textbf{x},\textbf{R})
    \\
    &+ L^{kl}(\textbf{R},t) \int d\textbf{w} (w^k-R^k) \alpha^{lb}(\textbf{w};\textbf{x},\textbf{R}).
\end{split}
\end{equation}
\begin{widetext}
Using the straight line path one can then show that this is
\begin{equation}
\begin{split}\label{C1}
    \Omega^{b}_\textbf{R}(\textbf{x},t) \approx &B^l(\textbf{R},t) \frac{\epsilon^{lab}}{2} (x^a-R^a) 
    +L^{jl}(\textbf{R},t) \frac{\epsilon^{lab}}{3} (x^a-R^a) (x^j-R^j).
\end{split}
\end{equation}

Equation (\ref{Omega0}) is written
\begin{equation}
\begin{split}
    \Omega^0_\textbf{R}(\textbf{x},t) &\approx  E^i(\textbf{R},t) \int s^i(\textbf{w};\textbf{x},\textbf{R}) d\textbf{w} 
    + \frac{\partial E^i(\textbf{x},t)}{\partial x^j}\Big|_{\textbf{x}\rightarrow\textbf{R}} \int d\textbf{w} (w^j-R^j) s^i(\textbf{w};\textbf{x},\textbf{R}) 
    \\
    &+ \frac{1}{2} \frac{\partial^2 E^i(\textbf{x},t)}{\partial x^j \partial x^k}\Big|_{\textbf{x}\rightarrow\textbf{R}} \int d\textbf{w} (w^j-R^j)(w^k-R^k) s^i(\textbf{w};\textbf{x},\textbf{R}),
\end{split}
\end{equation}
where upon implementing the straight line path we can write 
\begin{equation}
\begin{split}\label{C2}
    \Omega^0_\textbf{R}(\textbf{x},t) &\approx  E^i(\textbf{R},t) (x^i-R^i) 
    + \frac{1}{2} (x^i-R^i)(x^j-R^j) F^{ij}(\textbf{R},t) 
    \\
    &+ \frac{1}{6} (x^i-R^i)(x^j-R^j)(x^l-R^l) K^{ijl}(\textbf{R},t). 
\end{split}
\end{equation}

The magnetic flux through the surface bounded by the paths $\textbf{x}\rightarrow \textbf{R}$, $\textbf{R}\rightarrow \textbf{R}'$, and $\textbf{R}'\rightarrow \textbf{x}$ is given by the expression $\Delta(\textbf{x},\textbf{R},\textbf{R}';t)$, which is the sum of line integrals of the vector potential. This requires expanding the vector potential up to double derivatives to account for a spatially varying magnetic field. The details are somewhat tedious, but one can push through the calculation and find

\begin{equation}\label{C3}
\begin{split}
    \Delta(\textbf{x},\textbf{z},\textbf{y};t) = &-\frac{e}{2\hbar c} \epsilon^{lab} B^l(\textbf{y},t) \Big( (x^a-y^a)(z^b-y^b) \Big) 
    \\
    &- \frac{e}{6\hbar c} \epsilon^{lab} L^{jl}(\textbf{y},t) \Big( (x^j-y^j)+(z^j-y^j) \Big)(x^a-y^a)(z^b-y^b).
\end{split}
\end{equation}
Equations (\ref{C1}), (\ref{C2}) and (\ref{C3}) are then used in the linear response calculations. 

These relators are also used to expand the microscopic polarization and magnetization fields. Formally, employing the straight line path this leads to the following for the polarization expansion

\begin{equation}
\begin{split}
    p^i_\textbf{R}(\textbf{x},t) =& \int d\textbf{w}  \rho_\textbf{R}(\textbf{w},t) (w^i-R^i)  \int_{0}^{1} du \delta(\textbf{x}-\textbf{R}-\textbf{u}(\textbf{w}-\textbf{R})) 
    \\
    =& \delta(\textbf{x}-\textbf{R}) \int d\textbf{w} (w^i-R^i) \rho_\textbf{R}(\textbf{w},t)
    -\frac{1}{2} \frac{\partial \delta(\textbf{x}-\textbf{R})}{\partial x^j} \int d\textbf{w} (w^i-R^i)(w^j-R^j) \rho_\textbf{R}(\textbf{w},t) 
    +...,
\end{split}
\end{equation}
where the electric dipole and quadrupole moment can be read off. 

The atomic and itinerant magnetization are defined in terms of the relator $\alpha(\textbf{x};\textbf{y},\textbf{R})$ and the atomic and itinerant site currents respectively;
\begin{equation}
\begin{split}
    \bar{m}^j_\textbf{R}(\textbf{x},t) + \tilde{m}^j_\textbf{R}(\textbf{x},t) \equiv \frac{1}{c} \int d\textbf{y} \alpha^{jk}(\textbf{x};\textbf{y},\textbf{R}) \Big( j_\textbf{R}^{\mathfrak{p},k}(\textbf{y},t) 
    + \tilde{j}_\textbf{R}^{k}(\textbf{y},t) \Big).
\end{split}
\end{equation}
The expressions for the momentum current $j^{\mathfrak{p},k}_\textbf{R}(\textbf{y},t)$ and the itinerant current $\tilde{j}^k_{\textbf{R}}(\textbf{y},t)$ are found in earlier work \cite{Perry_Sipe,DuffOpticalActivity,DuffMagneticSusceptibility}. Expanding the relator by employing the straight line path yields the magnetic dipole and quadrupole terms
\begin{equation}
\label{ObtainingMagneticMultipole}
\begin{split}
    \bar{m}^j_\textbf{R}(\textbf{x},t) + \tilde{m}^j_\textbf{R}(\textbf{x},t) &=  \frac{\epsilon^{jab}}{2c} \int d\textbf{y} (y^a-R^a) \Big( j_\textbf{R}^{\mathfrak{p},b}(\textbf{y},t) 
    + \tilde{j}_\textbf{R}^{b}(\textbf{y},t) \Big)   
    \delta(\textbf{x}-\textbf{R})
    \\
    &- \frac{\epsilon^{jab}}{3c} \int d\textbf{y} (y^k-R^k) (y^a-R^a) \Big( j_\textbf{R}^{\mathfrak{p},b}(\textbf{y},t) 
    + \tilde{j}_\textbf{R}^{b}(\textbf{y},t) \Big)  \frac{\partial \delta(\textbf{x}-\textbf{R})}{\partial x^k}
    \\
    &+ ..., 
\end{split}
\end{equation}
where the magnetic dipole and quadrupole moment can be read off from eq. (\ref{ObtainingMagneticMultipole}). 

The spin magnetization does not involve a relator,
\begin{equation}
\begin{split}
    \breve{m}^j_\textbf{R}(\textbf{x},t) = \frac{e\hbar}{4mc} \sum_{\alpha\beta\textbf{R}'\textbf{R}''} \Big( \delta_{\textbf{R}\textbf{R}'} + \delta_{\textbf{R}\textbf{R}''}\Big) e^{i\Delta(\textbf{R}',\textbf{x},\textbf{R}'';t)} \chi^\dag_{\beta\textbf{R}'}(\textbf{x},t) \sigma^j \chi_{\alpha\textbf{R}''}(\textbf{x},t) \eta_{\alpha\textbf{R}'';\beta\textbf{R}'}(t). 
\end{split}
\end{equation}
Instead we perform an ad-hoc moment expansion to obtain
\begin{equation}
\label{ObtainingSpinMultipoles}
\begin{split}
    \breve{m}_\textbf{R}^j(\textbf{x},t) = \delta(\textbf{x}-\textbf{R}) \int d\textbf{y} \breve{m}^j_\textbf{R}(\textbf{y},t) - \int d\textbf{y} (y^i-R^i) \breve{m}^j_\textbf{R}(\textbf{y},t) \frac{\partial \delta(\textbf{x}-\textbf{R})}{\partial x^i} + ...
\end{split}
\end{equation}
The spin magnetic dipole and quadrupole can then be read off from eq. (\ref{ObtainingSpinMultipoles}).

\section{Linear Response of SPDM to Spatially Varying Electromagnetic Fields}\label{AppendixD}

Using the equation of motion for the SPDM in the adjusted Wannier function basis, the expansion of the relator dependent quantities that appear in the dynamical equation, we find the response of the SPDM to the applied electromagnetic fields evaluated at an arbitrary lattice site $\textbf{R}_a$. The dynamical equation for the SPDM is derived in an earlier work \cite{DuffOpticalActivity}, we quote the result here

\begin{equation}
\label{DynamicalEquation_eta}
\begin{split}
    i\hbar \frac{\partial \eta_{\alpha\textbf{R};\beta\textbf{R}'}(t)}{\partial t} = \sum_{\lambda\textbf{R}''} \Bigg[ e^{i\Delta(\textbf{R},\textbf{R}_a,\textbf{R}'',\textbf{R}';t)}\bar{H}_{\alpha\textbf{R};\lambda\textbf{R}''}(\textbf{R}_a,t) \eta_{\lambda\textbf{R}'';\beta\textbf{R}'}(t) -e^{i\Delta(\textbf{R}'',\textbf{R}_a,\textbf{R}',\textbf{R};t)} \eta_{\alpha\textbf{R};\lambda\textbf{R}''}(t) \bar{H}_{\lambda\textbf{R}'';\beta\textbf{R}'}(\textbf{R}_a,t) \Bigg]
    \\
    -\hbar \frac{\partial \Delta(\textbf{R},\textbf{R}_a,\textbf{R}';t)}{\partial t} \eta_{\alpha\textbf{R};\beta\textbf{R}'}(t), 
\end{split}
\end{equation}
where the modified Hamiltonian matrix elements in the adjusted Wannier function basis are
\begin{equation}
\label{HamiltonianMatrixElements}
\begin{split}
    \bar{H}_{\alpha\textbf{R};\lambda\textbf{R}''}(\textbf{R}_a,t) =& \int d\textbf{x} \chi^\dag_{\alpha\textbf{R}}(\textbf{x},t) e^{i\Delta(\textbf{R},\textbf{x},\textbf{R}_a;t)} \mathcal{H}_{\textbf{R}_a}(\textbf{x},t) e^{i\Delta(\textbf{R}_a,\textbf{x},\textbf{R}'';t)} \chi_{\beta\textbf{R}''}(\textbf{x},t) 
    \\
    &-\frac{i\hbar}{2} \int d\textbf{x} \Bigg[ e^{i\Delta(\textbf{R},\textbf{x},\textbf{R}_a;t)} \chi^\dag_{\alpha\textbf{R}}(\textbf{x},t) \frac{\partial}{\partial t} \Big(e^{i\Delta(\textbf{R}_a,\textbf{x},\textbf{R}'';t)} \chi_{\beta\textbf{R}''}(\textbf{x},t) \Big)
    \\
    &\hspace{50pt}- \frac{\partial }{\partial t} \Big(
    e^{i\Delta(\textbf{R},\textbf{x},\textbf{R}_a;t)} \chi^\dag_{\alpha\textbf{R}}(\textbf{x},t)
    \Big) \chi_{\beta\textbf{R}''}(\textbf{x},t) e^{i\Delta(\textbf{R}_a,\textbf{x},\textbf{R}'';t)} \Bigg]
\end{split}
\end{equation}
where the differential operator is defined as
\begin{equation}
\begin{split}
    \mathcal{H}_{\textbf{R}_a} (\textbf{x},t) = \frac{(\mathfrak{p}(\textbf{x}) - \frac{e}{c}\boldsymbol\Omega_{\textbf{R}_a}(\textbf{x},t) )^2 }{2m} + \text{V}(\textbf{x}) + \frac{\hbar}{4m^2 c^2} \boldsymbol\sigma \cdot \Big( \nabla\text{V}(\textbf{x}) \times \Big( \mathfrak{p}(\textbf{x}) - \frac{e}{c}\boldsymbol\Omega_{\textbf{R}_a}(\textbf{x},t) \Big) \Big)
    \\
    - \frac{e\hbar}{2mc} \boldsymbol\sigma \cdot \textbf{B}(\textbf{x},t) - \frac{e\hbar}{2mc} \boldsymbol\sigma \cdot \textbf{B}_\text{static}(\textbf{x})
    -e \Omega^{0}_{\textbf{R}_a}(\textbf{x},t) 
    , 
\end{split}
\end{equation}
where $\text{V}(\textbf{x})$ is the lattice potential. We have also included the possibility of a static vector potential $\textbf{A}_\text{static}(\textbf{x})$ and magnetic field $\textbf{B}_\text{static}(\textbf{x}) = \nabla\times\textbf{A}_\text{static}(\textbf{x})$ to break time-reversal symmetry in the ground state within the independent particle approximation we adopt. This then alters the canonical momentum, so we define $\mathfrak{p}(\textbf{x}) \equiv -i\hbar\nabla - \frac{e}{c} \textbf{A}_\text{static}(\textbf{x})$ in the above. More details on this choice of Hamiltonian are discussed in an earlier work \cite{Notes_on_EMHamiltonian}. Then solving this dynamical equation perturbatively leads to the linear response of the SPDM to applied spatially varying fields.

The SPDM response to an electric field is
\begin{equation}
\begin{split}
    \eta^{(E)}_{\alpha\textbf{R};\beta\textbf{R}'}(\omega) = e E^l(\textbf{R}_a,\omega) \mathcal{V}_{uc} \sum_{mn} f_{nm} \int_{BZ} \frac{d\textbf{k}}{(2\pi)^3} e^{i\textbf{k}\cdot(\textbf{R}-\textbf{R}')} \frac{ U^\dag_{\alpha m} \xi^l_{mn}  U_{n\beta}}{\Delta_{mn}(\textbf{k})-\hbar(\omega+i0^+)}.
\end{split}
\end{equation}
The SPDM response to symmetric derivatives of the electric field is
\begin{equation}
\begin{split}
    \eta^{(F)}_{\alpha\textbf{R};\beta\textbf{R}'}(\omega) = &\mathcal{V}_{uc} F^{jl}(\textbf{R}_a,\omega) \sum_{mn} f_{nm} \int_{BZ} \frac{d\textbf{k}}{(2\pi)^3} e^{i\textbf{k}\cdot(\textbf{R}-\textbf{R}')}  \Bigg[ 
    \frac{U^\dag_{\alpha m}U_{n\beta}}{\Delta_{mn}(\textbf{k})-\hbar(\omega+i0^+)} \Bigg[ \mathcal{Q}^{jl}_{\mathcal{P}:mn}
    \\
    &\hspace{200pt}+ \frac{ie}{4} \frac{\partial_j (E_{n\textbf{k}}+E_{m\textbf{k}})\xi^l_{mn} + \partial_l (E_{n\textbf{k}}+E_{m\textbf{k}})\xi^j_{mn}}{\Delta_{mn}(\textbf{k})-\hbar(\omega+i0^+)}   
    \Bigg]  \Bigg] 
    \\
    &+\frac{e\mathcal{V}_{uc}}{2} F^{jl}(\textbf{R}_a,\omega) \sum_{mn} f_{nm} \int_{BZ} \frac{d\textbf{k}}{(2\pi)^3} e^{i\textbf{k}\cdot(\textbf{R}-\textbf{R}')} \frac{ U^\dag_{\alpha m} \xi^l_{mn} U_{n\beta}}{\Delta_{mn}(\textbf{k})-\hbar(\omega+i0^+)} \Bigg( (R^j-R^j_a) + (R'^j-R^j_a) \Bigg)
    \\
    &+\frac{ie\mathcal{V}_{uc}}{2} F^{jl}(\textbf{R}_a,\omega) \sum_{mn} f_{nm} \int_{BZ} \frac{d\textbf{k}}{(2\pi)^3} e^{i\textbf{k}\cdot(\textbf{R}-\textbf{R}')} \frac{\xi^l_{mn}}{\Delta_{mn}(\textbf{k})-\hbar(\omega+i0^+)} \Bigg( U^\dag_{\alpha m} \partial_j U_{n\beta} - \partial_j U^\dag_{\alpha m} U_{n\beta} \Bigg) .
\end{split}
\end{equation}
The SPDM response to a magnetic field is
\begin{equation}
\begin{split}
    \eta^{(B)}_{\alpha\textbf{R};\beta\textbf{R}'}(\omega) = &B^l(\textbf{R}_a,\omega) \mathcal{V}_{uc} \sum_{mn} f_{nm} \int_{BZ} \frac{d\textbf{k}}{(2\pi)^3} e^{i\textbf{k}\cdot(\textbf{R}-\textbf{R}')} 
    \frac{U^\dag_{\alpha m} M^l_{mn} U_{n\beta}}{\Delta_{mn}(\textbf{k})-\hbar(\omega+i0^+)} 
    \\
    &+ \frac{e\omega }{4c} \epsilon^{lab} B^l(\textbf{R}_a,\omega) \mathcal{V}_{uc}  \sum_{mn} f_{nm} \int_{BZ} \frac{d\textbf{k}}{(2\pi)^3} e^{i\textbf{k}\cdot(\textbf{R}-\textbf{R}')} \frac{ \partial_b (E_{n\textbf{k}}+E_{m\textbf{k}}) \xi^a_{mn}}{(\Delta_{mn}(\textbf{k})-\hbar(\omega+i0^+))^2} 
    \\
    &+\frac{ie\omega }{4c} \epsilon^{lab} B^l(\textbf{R}_a,\omega) \mathcal{V}_{uc}  \sum_{mn} f_{nm} \int_{BZ} \frac{d\textbf{k}}{(2\pi)^3} e^{i\textbf{k}\cdot(\textbf{R}-\textbf{R}')} \frac{ U^\dag_{\alpha m} \xi^b_{mn} U_{n\beta}}{\Delta_{mn}(\textbf{k})-\hbar(\omega+i0^+)} \Bigg( (R^a-R^a_a) + (R'^a-R^a_a)\Bigg)
    \\
    &+\frac{e }{4\hbar c} \epsilon^{lab} B^l(\textbf{R}_a,\omega) \mathcal{V}_{uc} \sum_{mn} f_{nm} \int_{BZ} \frac{d\textbf{k}}{(2\pi)^3} e^{i\textbf{k}\cdot(\textbf{R}-\textbf{R}')}  \frac{ \Delta_{mn}(\textbf{k}) \xi^b_{mn}}{\Delta_{mn}(\textbf{k})-\hbar(\omega+i0^+)} \Bigg( \partial_a U^\dag_{\alpha m} U_{n\beta} - U^\dag_{\alpha m} \partial_a U_{n\beta} \Bigg).
\end{split}
\end{equation}
The SPDM response to derivatives of the magnetic field is
\begin{equation}
\begin{split}
    \eta^{(L)}_{\alpha\textbf{R};\beta\textbf{R}'}(\omega) = &\mathcal{V}_{uc} L^{jl}(\textbf{R}_a,\omega) \sum_{mn} f_{nm} \int_{BZ} \frac{d\textbf{k}}{(2\pi)^3} e^{i\textbf{k}\cdot(\textbf{R}-\textbf{R}')} \frac{ U^\dag_{\alpha m} \mathcal{A}^{jl}_{mn}(\textbf{k},\omega) U_{n\beta} }{\Delta_{mn}(\textbf{k}) - \hbar(\omega+i0^+)}
    \\
    &+\mathcal{V}_{uc} L^{jl}(\textbf{R}_a,\omega) \sum_{mn} f_{nm} \int_{BZ} \frac{d\textbf{k}}{(2\pi)^3} e^{i\textbf{k}\cdot(\textbf{R}-\textbf{R}')} \frac{U^\dag_{\alpha m} \mathcal{B}^{jl}_{mn}(\textbf{R},\textbf{R}';\textbf{k},\omega) U_{n\beta} }{\Delta_{mn}(\textbf{k})-\hbar(\omega+i0^+)} 
    \\
    &+ \mathcal{V}_{uc} L^{jl}(\textbf{R}_a,\omega) \sum_{mn} f_{nm} \int_{BZ} \frac{d\textbf{k}}{(2\pi)^3} e^{i\textbf{k}\cdot(\textbf{R}-\textbf{R}')} U^\dag_{\alpha m} \mathcal{C}^{jl}_{mn}(\textbf{R},\textbf{R}';\textbf{k}) U_{n\beta} 
    \\
    &+ \mathcal{V}_{uc} L^{jl}(\textbf{R}_a,\omega) \int_{BZ} \frac{d\textbf{k}}{(2\pi)^3} e^{i\textbf{k}\cdot(\textbf{R}-\textbf{R}')} \mathcal{X}^{jl}_{\alpha\beta}(\textbf{R},\textbf{R}';\textbf{k},\omega)
\end{split}
\end{equation}
where we define 
\begin{equation}
\begin{split}
    \mathcal{A}^{jl}_{mn}(\textbf{k},\omega) \equiv &\mathcal{Q}^{lj}_{\mathcal{M}:mn}
    +\Big[\frac{i}{3} M^l_{\text{Orb}:mn} 
    + \frac{i}{2} M^l_{\text{Spin}:mn} \Big]\frac{ \partial_j (E_{n\textbf{k}}+E_{m\textbf{k}}) }{\Delta_{mn}(\textbf{k})-\hbar(\omega+i0^+)} 
    \\
    &+\frac{ie}{12 c} \epsilon^{lab} \sum_{l} (\xi^j_{ml}v^b_{ln} + v^b_{ml}\xi^j_{ln} -\frac{1}{\hbar} \partial_j (E_{n\textbf{k}}+E_{m\textbf{k}}) \xi^b_{mn} ) \frac{ \partial_a (E_{n\textbf{k}}+E_{m\textbf{k}}) }{\Delta_{mn}(\textbf{k})-\hbar(\omega+i0^+)}
    \\
    &-\frac{ie \omega}{12 c} \epsilon^{lab} 
    \frac{\partial_j \partial_a (E_{n\textbf{k}}-E_{m\textbf{k}}) \xi^b_{mn} }{\Delta_{mn}(\textbf{k})-\hbar(\omega+i0^+)}
    -\frac{ie \omega}{6 c} \epsilon^{lab} \xi^b_{mn}\frac{ \partial_j (E_{m\textbf{k}}+E_{n\textbf{k}}) \partial_a (E_{m\textbf{k}}+E_{n\textbf{k}}) }{(\Delta_{mn}(\textbf{k})-\hbar(\omega+i0^+))^2},
\end{split}
\end{equation}
where $M^l_{\text{Orb}:mn}$ and $M^{l}_{\text{Spin}:mn}$ are the orbital and spin contributions to the magnetization matrix element eq. (\ref{SpontaneousMagnetization}) respectively,

\begin{equation}
\begin{split}
    M^l_{\text{Orb}:mn} &= \frac{e}{4c} \epsilon^{lab}\Bigg[ \sum_{s} \Big(\xi^a_{ms}v^b_{sn} + v^b_{ms}\xi^a_{sn} \Big) - \frac{1}{\hbar} \partial_a (E_{n\textbf{k}}+E_{m\textbf{k}}) \xi^b_{mn} \Bigg],
    \\
    M^{l}_{\text{Spin}:mn} &= \frac{e}{mc} S^l_{mn}.
\end{split}
\end{equation}

\begin{equation}
\begin{split}
    \mathcal{B}^{jl}_{mn}(\textbf{R},\textbf{R}';\textbf{k},\omega) \equiv &\Big[ (R^j-R^j_a) + (R'^j-R^j_a)\Big] \Bigg[\frac{1}{3} M^l_{\text{orb}:mn} + \frac{1}{2} M^l_{\text{spin}:mn} + \frac{e\omega}{12 c} \epsilon^{lab} \frac{\partial_b (E_{n\textbf{k}}+E_{m\textbf{k}}) \xi^a_{mn}}{\Delta_{mn}(\textbf{k})-\hbar(\omega+i0^+)}\Bigg]  
    \\
    &+ \Big[ (R^a-R^a_a) + (R'^a-R^a_a) \Big] \frac{e}{12c} \epsilon^{lab} \Bigg[ \sum_{l} (\xi^j_{ml}v^b_{ln} + v^b_{ml}\xi^j_{ln}) +i v^b_{mn}\frac{ \partial_j (E_{n\textbf{k}}+E_{m\textbf{k}}) }{\Delta_{mn}(\textbf{k})-\hbar(\omega+i0^+)} \Bigg] 
    \\
    &+\frac{e}{12 c} \epsilon^{lab} \Big( (R^j-R^j_a) + (R'^j-R^j_a) \Big)\Big( (R^a-R^a_a) + (R'^a-R^a_a) \Big) v^b_{mn},
\end{split}
\end{equation}
and
\begin{equation}
\begin{split}
    \mathcal{C}^{jl}_{mn}(\textbf{R},\textbf{R}';\textbf{k}) \equiv &-\frac{ie}{24 \hbar c} \epsilon^{lab} \Big( (R^a-R^a_a) + (R'^a-R^a_a) \Big) \Big[ \xi^b_{ml}\xi^j_{ln} + \xi^j_{ml}\xi^b_{ln} \Big]
    \\
    &-\frac{ie}{24\hbar c} \Big( (R'^a-R^a_a) + (R^a - R^a_a) \Big)\Big( (R'^j-R^j_a) + (R^j-R^j_a) \Big)\xi^b_{mn}
    \\
    &-\frac{ie}{12 \hbar c} \Big( (R^a-R^a_a)(R^j-R^j_a) + (R'^a-R^a_a)(R'^j-R^j_a)  \Big)  \xi^b_{mn},
\end{split}
\end{equation}
and $\mathcal{X}^{jl}_{\alpha\beta}(\textbf{R},\textbf{R}';\textbf{k},\omega)$ depends on the gauge-choice of the Wannier functions through the $\mathcal{W}(\textbf{k})$ matrix elements. It is not relevant for the purposes of this manuscript since we only consider the set of gauge-transformations that alter the phase of the Bloch functions, and so only show that the total response does not depend on the diagonal elements of the Berry connection, and so prove the result is gauge-invariant in this more restrictive sense.  

Lastly, the response to symmetric double derivatives of the electric field is
\begin{equation}
\begin{split}
    \eta^{(K)}_{\alpha\textbf{R};\beta\textbf{R}'}(\omega) = &\mathcal{V}_{uc} K^{ijl}(\textbf{R}_a,\omega) \sum_{mn} f_{nm} \int_{BZ} \frac{d\textbf{k}}{(2\pi)^3} e^{i\textbf{k}\cdot(\textbf{R}-\textbf{R}')} \frac{U^\dag_{\alpha m} \mathcal{D}^{ijl}_{mn}(\textbf{k},\omega) U_{n\beta}}{\Delta_{mn}(\textbf{k})-\hbar(\omega+i0^+)}
    \\
    &+ \mathcal{V}_{uc} K^{ijl}(\textbf{R}_a,\omega) \sum_{mn} f_{nm} \int_{BZ} \frac{d\textbf{k}}{(2\pi)^3} e^{i\textbf{k}\cdot(\textbf{R}-\textbf{R}')} \frac{U^\dag_{\alpha m} \mathcal{E}^{ijl}_{mn}(\textbf{R},\textbf{R}';\textbf{k},\omega) U_{n\beta}}{\Delta_{mn}(\textbf{k})-\hbar(\omega+i0^+)}
    \\
    &+ \mathcal{V}_{uc} K^{ijl}(\textbf{R}_a,\omega) \int_{BZ} \frac{d\textbf{k}}{(2\pi)^3} e^{i\textbf{k}\cdot(\textbf{R}-\textbf{R}')} \mathcal{Y}^{ijl}_{\alpha\beta}(\textbf{R},\textbf{R}';\textbf{k}),
\end{split}
\end{equation}
where
\begin{equation}
\begin{split}
    \mathcal{D}^{ijl}_{mn}(\textbf{k},\omega) = \mathcal{O}^{ijl}_{\mathcal{P}:mn}  
    +\frac{e}{8} \partial_i \partial_j \xi^l_{mn}
    +\frac{i}{2} \frac{\partial_i (E_{m\textbf{k}}+E_{n\textbf{k}}) \mathcal{Q}^{jl}_{\mathcal{P}:mn}}{\Delta_{mn}(\textbf{k})-\hbar(\omega+i0^+)} 
    +\frac{e}{8} \frac{ \partial_i \partial_j (E_{m\textbf{k}}-E_{n\textbf{k}}) \xi^l_{mn} }{\Delta_{mn}(\textbf{k})-\hbar(\omega+i0^+)} 
    \\
    -\frac{e}{4} \xi^i_{mn}\frac{\partial_j (E_{n\textbf{k}}+E_{m\textbf{k}})\partial_l (E_{n\textbf{k}}+E_{m\textbf{k}})}{(\Delta_{mn}(\textbf{k})-\hbar(\omega+i0^+))^2},
\end{split}
\end{equation}
and
\begin{equation}
\begin{split}
    \mathcal{E}^{ijl}_{mn}(\textbf{R},\textbf{R}';\textbf{k},\omega) 
    = 
    \frac{1}{2}\Big( (R'^i-R^i_a) + (R^i-R^i_a) \Big) \Bigg[ \mathcal{Q}^{jl}_{\mathcal{P}:mn}
    +\frac{ie}{2} \frac{ \xi^l_{mn} \partial_j (E_{n\textbf{k}}+E_{m\textbf{k}})}{\Delta_{mn}(\textbf{k})-\hbar(\omega+i0^+)} \Bigg] 
    \\
    +\frac{e}{8} \Big( (R'^i-R^i_a) + (R^i-R^i_a) \Big) \Big( (R'^j-R^j_a) + (R^j-R^j_a) \Big) \xi^l_{mn}, 
\end{split}
\end{equation}
and $\mathcal{Y}^{ijl}_{\alpha\beta}(\textbf{R},\textbf{R}';\textbf{k},\omega)$ depends on the gauge-choice of the Wannier functions through the $\mathcal{W}$ matrix elements and their derivatives.  

\section{Long Wavelength Conductivity Tensor and Optical Activity}\label{AppendixE}

In the long wavelength limit, for topologically trivial insulators the conductivity depends only on the electric susceptibility, 

\begin{equation}
\begin{split}
    \chi^{il}_{\mathcal{P}}(\omega) = e^2 \sum_{mn} f_{nm} \int_{BZ} \frac{d\textbf{k}}{(2\pi)^3} \frac{\xi^i_{nm}\xi^l_{mn}}{\Delta_{mn}(\textbf{k})-\hbar(\omega+i0^+)}. 
\end{split}
\end{equation}

Extending to treat spatial dispersion requires going beyond the dipole approximation and considering the magnetic dipole moment and the electric quadrupole moment. The four tensors required to describe optical activity, as listed in Table \ref{tab:1} are

\begin{equation}
\begin{split}
    \gamma^{ijl}_{\mathcal{P}}(\omega) &= e \sum_{mn} f_{nm} \int_{BZ} \frac{d\textbf{k}}{(2\pi)^3} \frac{ \xi^i_{nm} Q^{jl}_{\mathcal{P}:mn}  }{\Delta_{mn}(\textbf{k})-\hbar(\omega+i0^+)} 
    \\
    &+ \frac{ie^2}{4} \sum_{mn} f_{nm} \int_{BZ} \frac{d\textbf{k}}{(2\pi)^3} \frac{  \xi^i_{nm} \Big( \partial_j (E_{n\textbf{k}}+E_{m\textbf{k}})\xi^l_{mn} + \partial_l (E_{m\textbf{k}}+E_{n\textbf{k}}) \xi^j_{mn}\Big) }{(\Delta_{mn}(\textbf{k})-\hbar(\omega+i0^+))^2}
\end{split}
\end{equation}

\begin{equation}
\begin{split}
    \chi^{ijl}_\mathcal{Q}(\omega) = e\sum_{mns} f_{nm} \int_{BZ} \frac{d\textbf{k}}{(2\pi)^3} \frac{ \mathcal{Q}^{ij}_{\mathcal{P}:nm}\xi^l_{mn} }{\Delta_{mn}(\textbf{k})-\hbar(\omega+i0^+)},
\end{split}
\end{equation}

\begin{equation}
\begin{split}
    \alpha^{ia}_{\mathcal{P}}(\omega) = \alpha^{ia}_{G} + \alpha^{ia}_{S;\mathcal{P}}(\omega) + \frac{\omega e^2}{4c} \epsilon^{abc} \sum_{mn} f_{nm} \int_{BZ} \frac{d\textbf{k}}{(2\pi)^3} \frac{ \xi^i_{nm}\mathcal{B}^{bc}_{mn}(\textbf{k},\omega) }{\Delta_{mn}(\textbf{k})-\hbar(\omega+i0^+)}, 
\end{split}
\end{equation}
and
\begin{equation}
\begin{split}
    \alpha^{lb}_{\mathcal{M}}(\omega) = \alpha^{lb}_G + \alpha^{lb}_{S;\mathcal{M}}(\omega) - \frac{\omega e^2}{4c} \epsilon^{abc} \sum_{mn} f_{nm} \int_{BZ} \frac{d\textbf{k}}{(2\pi)^3} \frac{\xi^l_{mn}}{\Delta_{mn}(\textbf{k})-\hbar(\omega+i0^+)}
    \\
    \times \Bigg[ 
    2 \frac{\partial_c (E_{m\textbf{k}}+E_{n\textbf{k}})}{\Delta_{mn}(\textbf{k})} \xi^a_{nm} + i\sum_{s} \Bigg[ 
    \frac{\Delta_{sm}(\textbf{k})}{\Delta_{mn}(\textbf{k})} - \frac{\Delta_{ns}(\textbf{k})}{\Delta_{mn}(\textbf{k})}
    \Bigg] \xi^a_{ns} \xi^c_{sm}
    \Bigg].
\end{split}
\end{equation}
Where
\begin{equation}
\begin{split}
    B^{bc}_{mn}(\textbf{k},\omega) \equiv i\sum_{s} \Bigg[ 
    \frac{\Delta_{sn}(\textbf{k})}{\Delta_{mn}(\textbf{k})} \xi^b_{ms}\xi^c_{sn} + \frac{\Delta_{sm}(\textbf{k})}{\Delta_{mn}(\textbf{k})}\xi^b_{ms}\xi^c_{sn}
    \Bigg] - \Bigg[3 + \frac{\hbar\omega}{\Delta_{mn}(\textbf{k})-\hbar(\omega+i0^+)} \Bigg] \frac{\partial_b (E_{m\textbf{k}}+E_{n\textbf{k}})}{\Delta_{mn}(\textbf{k})} \xi^c_{mn},
\end{split}
\end{equation}
and the spin and ``cross-gap" ME effects are
\begin{equation}
\begin{split}
    \alpha^{ia}_{S;\mathcal{P}}(\omega) = \frac{e^2}{mc} \sum_{mn} f_{nm} \int_{BZ} \frac{d\textbf{k}}{(2\pi)^3} \frac{\xi^i_{nm}S^a_{mn}}{\Delta_{mn}(\textbf{k})-\hbar(\omega+i0^+)},
\end{split}
\end{equation}

\begin{equation}
\begin{split}
    \alpha^{lb}_{S;\mathcal{M}}(\omega) = \frac{e^2}{mc} \sum_{mn} f_{nm} \int_{BZ} \frac{d\textbf{k}}{(2\pi)^3} \frac{S^b_{nm}\xi^l_{mn}}{\Delta_{mn}(\textbf{k})-\hbar(\omega+i0^+)},
\end{split}
\end{equation}

\begin{equation}
\label{CrossGap}
\begin{split}
    \alpha^{ia}_{G} = &\frac{e^2}{\hbar c} \epsilon^{lab} \int_{BZ} \frac{d\textbf{k}}{(2\pi)^3} \Bigg\{ 
    \sum_{vcc'} \frac{E_{c\textbf{k}}-E_{c'\textbf{k}}}{E_{v\textbf{k}}-E_{c\textbf{k}}} \text{Re}\Bigg[ (\partial_b v_{\textbf{k}}|c'_\textbf{k})(c'_\textbf{k}|\partial_a c_\textbf{k})(c_\textbf{k}|\partial_i v_\textbf{k}) \Bigg]
    \\
    &-\sum_{cv} \frac{\partial_b (E_{c\textbf{k}}+E_{v\textbf{k}})}{E_{v\textbf{k}}-E_{c\textbf{k}}} \text{Re}\Bigg[ (\partial_a v_\textbf{k}|c_\textbf{k})(c_\textbf{k}|\partial_i v_\textbf{k})\Bigg] 
    - \sum_{vv'c} \frac{E_{v\textbf{k}}-E_{v'\textbf{k}}}{E_{v\textbf{k}}-E_{c\textbf{k}}} \text{Re}\Bigg[ 
    (\partial_b v_\textbf{k}|v'_\textbf{k})(\partial_a v'_\textbf{k}|c_\textbf{k})(c_\textbf{k}|\partial_i v_{\textbf{k}}) 
    \Bigg]
    \Bigg\},
\end{split}
\end{equation}
where $v$ labels an occupied state and $c$ labels an unoccupied state in equation (\ref{CrossGap}),

\end{widetext}

\bibliographystyle{apsrev4-1}
\bibliography{references.bib}

\end{document}